\documentclass[9pt,twocolumn,twoside]{pnas-new}
\setboolean{displaywatermark}{false}

\articletype{CLASSIFICATION}%%%% Article topic/classification

\templatetype{pnasresearcharticle} % Choose template
\usepackage{tabularx}
\usepackage{xltabular}
\usepackage{array}
\usepackage{booktabs}
\usepackage{caption}
\usepackage{placeins} % optional, for \FloatBarrier
\usepackage{float}
\usepackage[normalem]{ulem}
\usepackage{xcolor}
\usepackage{eso-pic}
\usepackage{fancyhdr}

\AtBeginDocument{%
  \fancypagestyle{firststyle}{%
    \fancyhf{}
    \fancyfoot[C]{\thepage}

  }

  \pagestyle{fancy}
  \fancyhf{}
  \fancyfoot[C]{\thepage}

}
\AddToShipoutPictureFG*{%
  \AtPageUpperLeft{%
    \raisebox{-0.75in}{%
      \makebox[\paperwidth]{\color{white}\rule{\paperwidth}{0.45in}}%
    }%
  }%
}

\begin{document}

%\title{Symmetry Protected Curvature-Free Folding in Periodic Origami}
\title{Nonlinear Curvature-Free Periodic Folding in Origami via Symmetry Classification of Linear Modes}

% Use letters for affiliations, numbers to show equal authorship (if applicable) and to indicate the corresponding author
\author[a]{Nan Cheng}
\author[b]{Leon M. Kamp}
\author[a]{Yanxin Feng}
\author[a]{Wenqian Sun}
\author[b]{Katia Bertoldi}
\author[a,1]{D. Zeb Rocklin}

\affil[a]{School of Physics, Georgia Institute of Technology, Atlanta, GA 30332, USA}
\affil[b]{School of Engineering and Applied Science, Harvard University, Cambridge, MA 02138}

% Please give the surname of the lead author for the running footer
\leadauthor{Cheng}

\significancestatement{Origami-inspired structures can transform shape dramatically, but periodic folding usually induces unwanted global curvature. Why a few patterns, such as the Miura-ori, avoid this has remained unresolved. We show that the answer lies in symmetry: the spatial symmetries of a periodic crease pattern determine which collective folding modes are allowed and when curvature-free motions are protected, even in over-constrained sheets. Using layer groups, we derive counting rules for these modes, explain the Miura mechanism, and classify and clarify the relationship between the bending and stretching responses of periodic origami. These results establish symmetry as a practical design principle for deployable and reconfigurable mechanical sheets with targeted shape-change properties.}

\authorcontributions{Nan Cheng and D. Zeb Rocklin conceived the project. Nan Cheng formulated the problem in group-theoretic terms, developed the resummation and reduction trick, carried out the group-theoretical analysis for all layer groups, proved the nonlinear extension theorem. Nan Cheng and Yanxin Feng developed algorithms for numerical validation. Leon M. Kamp performed the fabrication and experimental validation. Wenqian Sun and Katia Bertoldi contributed to discussions. Nan Cheng, Leon M. Kamp and D. Zeb Rocklin wrote the manuscript.}
\authordeclaration{The authors declare no conflicts of interest.}
%\equalauthors{\textsuperscript{1}A.O.(Author One) contributed equally to this work with A.T. (Author Two) (remove if not applicable).}
\correspondingauthor{\textsuperscript{1}To whom correspondence should be addressed. E-mail: zebrocklin@gatech.edu}


\begin{abstract}
    Broadly studied origami crease patterns such as the Miura-ori are capable of rigid folding that maintains their spatial periodicity.
    However, periodic crease patterns generically require additional creases and fold into quasi-cylindrical shapes.
    Here, using group theory, we show that the folding properties of such sheets are largely determined by their spatial symmetries, known as layer groups.
    Within each symmetry class, the allowed linear shape-periodic modes of sheets are constrained and classified by the irreducible representation they belong to. 
    Such classification further gives a sufficient condition, containing a symmetry constraint and a topological constraint, for the existence of a nonlinear strain mode. 
    In this way, familiar special cases are placed within a unified framework that connects symmetry, network topology, and deployability. 
    Finally, we turn our classification results into a predictive design principle that enables the rapid design of novel sheets with prescribed properties.
\end{abstract}

\dates{This manuscript was compiled on \today}
% Submission: uncomment the line below
%\doi{\url{www.pnas.org/cgi/doi/10.1073/pnas.XXXXXXXXXX}}

\maketitle
\thispagestyle{firststyle}
\ifthenelse{\boolean{shortarticle}}{\ifthenelse{\boolean{singlecolumn}}{\abscontentformatted}{\abscontent}}{}

\Firstpage
\Endparasplit

\dropcap{O}rigami turns geometry into mechanical function. By prescribing a network of creases, a thin sheet can be compacted, deployed, and reconfigured through motions encoded primarily by the crease pattern rather than by the material itself. This geometric control has made origami a versatile platform for deployable structures~\cite{zirbel2013accommodating,filipov2015origami}, soft and self-folding robots~\cite{felton2014method,li2017fluid}, medical devices~\cite{kuribayashi2006self,johnson2017fabricating}, flexible energy-storage devices~\cite{cheng2013folding,song2014origami}, programmable matter~\cite{hawkes2010programmable,dudte2016programming,karami2024curved}, and mechanical metamaterials~\cite{silverberg2014using,overvelde2016three,waitukaitis2015origami,bertoldi2017flexible,zhai2021mechanical}. Because the underlying kinematics can be realized across materials and length scales~\cite{lebee2015folds,callens2018flat,Peraza-Hernandez_2014,shenoy2012self,misseroni2024origami}, origami offers a versatile route to large, reversible shape change from simple local rules.

For many applications, however, it is not enough for a finite crease pattern such as an origami crane to possess a folding motion. One would like the mechanism to be scalable: it should persist as the number of unit cells increases, rather than rely on boundary effects or finite-size flexibility~\cite{tachi2015rigid}. Periodic origami provides a natural setting for this requirement, since a spatially periodic folding mode reduces the global compatibility problem to the compatibility of a repeating unit cell. This reduction links microscopic crease geometry and fold angles to continuum-scale quantities such as the effective metric strain and curvature, as well as to the isometries and macroscopic compatibility conditions that govern the folded sheet~\cite{nassar2017curvature,nassar2022strain,xu2024derivation,nassar2024periodic,nassar2024effective,sun2025geometric}. In general, such periodic folding can produce macroscopic bending. 
One remarkable exception is the Miura-ori: its nonlinear local rigid-folding branch is compatible across the entire sheet, producing a global shape-changing mechanism that preserves spatial periodicity and remains free of macroscopic curvature~\cite{schenk2013geometry,wei2013geometric}. This scalable curvature-free mechanism has made the Miura-ori a model system for origami metamaterials, motivating extensive work on negative and tunable Poisson response, Poisson-ratio switching, and coupled stretching--bending behavior~\cite{yasuda2015reentrant,pratapa2019geometric,misseroni2022experimental}.

The success of the Miura-ori suggests a broader design question: 
how can one design new periodic origami patterns that admit scalable rigid-folding mechanisms that change the macroscopic metric without generating macroscopic curvature? Several families with this property have been obtained by generalizing the Miura-ori using wallpaper-group symmetries~\cite{sareh2015isomorphic,SarehMiura2015,sareh2015framework}. These constructions demonstrate the usefulness of symmetry, but they remain tied to Miura-based design principles and therefore sample only a restricted portion of the possible design space. A general theory should instead identify the symmetry conditions under which curvature-free folding is protected, and should do so without assuming a Miura-like crease geometry in advance.

The main difficulty is that curvature-free folding is a nonlinear property. A linear analysis about the flat state can identify infinitesimal zero modes, but such modes may be spurious: they 
may not integrate to finite-amplitude rigid-folding mechanisms. Conversely, directly solving the nonlinear compatibility equations for an arbitrary periodic crease pattern is often analytically opaque and computationally costly. 

Here we address this difficulty by shifting the viewpoint from the flat crease pattern to a generic folded state, where the relevant three-dimensional symmetries of the periodic origami are explicit. In this setting, layer groups~\cite{kopsky2002international}, which generalize the concept of wallpaper groups by including point group operations that reverse the orientation of the third dimension, provide the natural symmetry language of origami sheets.
We solve this nonlinear curvature-free folding problem by classifying the symmetry-allowed linear zero modes in each layer group according to their macroscopic strain and curvature responses, and by proving a symmetry-based extension theorem identifying linear modes admitting nonlinear curvature-free continuations.

Our central results are twofold. First, we give an exhaustive table of the symmetry-allowed linear zero modes for each relevant layer group, organized by their macroscopic strain and curvature responses. Second, we identify symmetries that protect nonlinear curvature-free folding by combining this exhaustive table with a nonlinear extension theorem. We show that such protection arises only from certain symmetries that reverse the out-of-plane orientation, namely inversion, glide rotation, and glide-$z$-reflection. Glide rotation and glide-$z$-reflection protect curvature-free mechanisms directly, whereas inversion does so only when an additional topological occupation rule is satisfied.

The paper is organized as follows. We first introduce layer groups as the natural framework for describing the symmetries of periodic origami in a folded state. We then use group representation theory to develop a symmetry-adapted Maxwell counting for the relevant deformations. This counting classifies linear zero modes by their macroscopic strain and curvature responses. We further present an exhaustive table summarizing our classification. Finally, we establish a criterion for the existence of finite curvature-free folding mechanisms by developing a nonlinear extension theorem and combining it with the exhaustive table, and turn this criterion into a design principle.

\section*{Symmetry-Adapted Maxwell Counting}
\subsection*{Layer Groups}\label{sec:layergroups}
\begin{figure}[ht]
    \centering
    \includegraphics[width=1.0\linewidth]{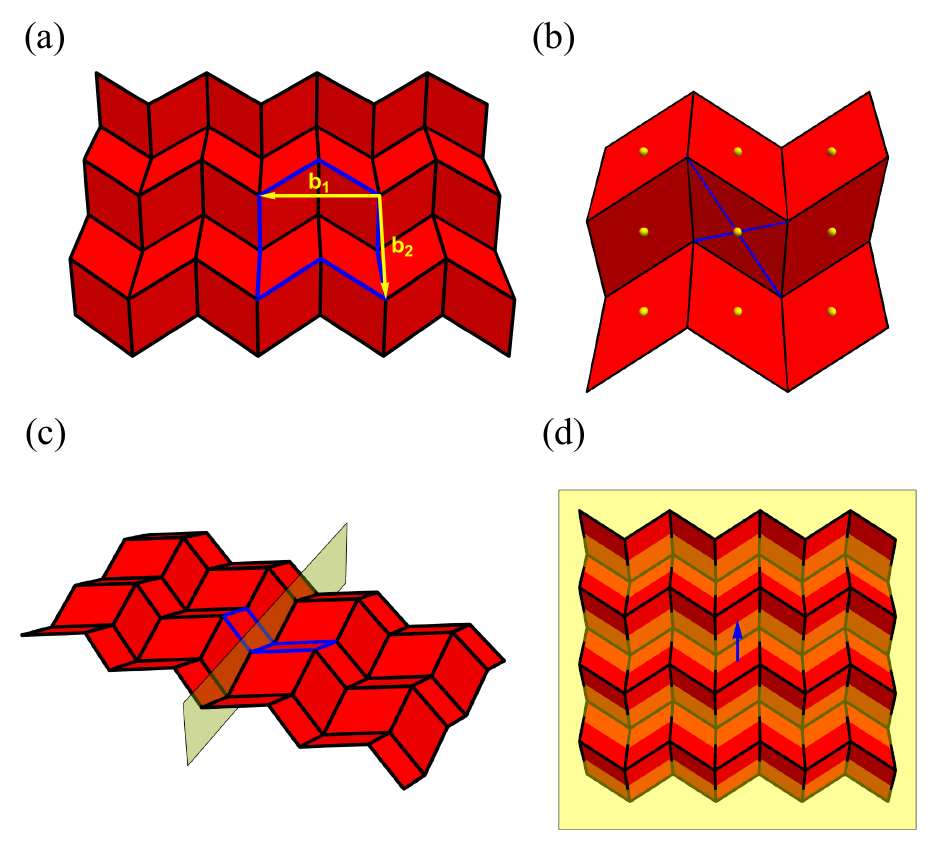}
    \caption{Miura fold and generators of its layer group. (a) The Miura fold in its folded configuration. The yellow vectors indicate the lattice vectors and the blue lines outline one unit cell. (b) Inversion symmetry: the yellow dots mark inversion centers, and the blue lines connect pairs of vertices that are related by an inversion. (c) Mirror symmetry: the green plane denotes the mirror plane, and the two parallelograms outlined in blue are mirror images of each other. (d) Glide-$z$-reflection symmetry: the glide-$z$-reflection operation is the composition of reflection across the yellow mirror plane and a half-lattice translation indicated by the blue arrow.}
    \label{fig:miurafoldsymmetry}
\end{figure}

Wallpaper groups, crystallographic objects  describing the symmetries of periodic crease patterns in their unfolded states, are used in the literature to characterize the folding of flat origami~\cite{sareh2015isomorphic}. Periodic origami sheets in their folded states, however, extend into the third spatial dimension. The appropriate crystallographic objects describing their symmetries are the layer groups~\cite{kopsky2002international}. Fig.~\ref{fig:miurafoldsymmetry} provides an example analyzing the layer group of the Miura fold (Fig.~\ref{fig:miurafoldsymmetry} (a)), which 
is generated by an inversion (Fig.~\ref{fig:miurafoldsymmetry} (b)), a reflection (Fig.~\ref{fig:miurafoldsymmetry} (c)), and a glide-$z$-reflection (Fig.~\ref{fig:miurafoldsymmetry} (d)).

A further reason for introducing the more refined framework of layer groups in origami folding is that they provide a natural way to distinguish curvature-free folding (strain modes) from curved folding (bending modes). This distinction arises because layer groups can include operations that reverse the out-of-plane direction, such as inversion, glide-$z$-reflection, and glide-rotation (The formal definitions of these operations can be found in Ref.~\cite{kopsky2002international}). In-plane strain is unchanged under such a reversal, whereas curvature changes sign. Consequently, symmetry operations that reverse the out-of-plane direction can place strain-like and bending-like foldings in different symmetry sectors in the group-theoretic sense~\cite{serre2012linear}.

With this observation in mind, we study curvature-free folding by an idea of nonlinear extension initially proposed in Ref.~\cite{KANGWAI19995507}. Starting from a folded periodic structure, we use its layer group to determine whether it can continue to fold infinitesimally without developing curvature. If this infinitesimal deformation further preserves the original layer group, then this infinitesimal folding extends to a finite mechanism, and the corresponding structure can fold nonlinearly in a curvature-free manner.

\subsection*{Symmetry-Adapted Counting for Infinitesimal Zero Modes}\label{sec:symmetrycounting}

The low-energy deformations of origami sheets are well-modeled by treating faces as rigid objects that rotate freely at their shared hinges. In the case of triangulated origami, this is equivalent to treating each vertex as a point mass and each crease as an inextensible rod~\cite{ChenPRX2018}. An extensive literature dating back to Maxwell demonstrates how a system of solid elements achieves flexibility when its number of degrees of freedom, $N_{\rm dof}$, exceeds its number of constraints, $N_{\rm con}$~\cite{Maxwell01041864,Maoannreview2018}. Calladine refined this counting argument by showing that adding a constraint need not always eliminate a zero mode, but may instead generate a \emph{state of self stress}: a pattern of bond tensions that exerts no net force on the sites~\cite{CALLADINE1978161}. Guest and collaborators further combined this framework with group theory to account for structures with spatial symmetries~\cite{FOWLER20001793,connelly2022frameworks}. While group theory is used extensively in atomic, condensed matter and particle physics~\cite{hamermesh2012group}, it has heightened utility in the present case because constraints, unlike forces, lie in an entirely distinct vector space from displacements, and may thus contain irreducible components with different multiplicities than the vector space of displacements. 
The result is  symmetry-adapted Maxwell counting (See SI, Section 1.3 for details):

\begin{equation}
\label{eq:generalizedMaxwellcounting}
    N_{\rm dof}^{(j)}-N_{\rm con}^{(j)}=N_{\rm ZM}^{(j)}-N_{\rm SSS}^{(j)},
\end{equation}
where $N_{\rm dof}^{(j)}$, $N_{\rm con}^{(j)}$, $N_{\rm ZM}^{(j)}$ and $N_{\rm SSS}^{(j)}$ are, respectively, the numbers of degrees of freedom, constraints, zero modes and states of self-stress in sector $j$. As a consequence, a zero mode necessarily exists when 
\begin{equation}
\label{eq:ZMcriterion}
    N_{\rm dof}^{(j)}-N_{\rm con}^{(j)}>0.
\end{equation}

\begin{figure*}[ht]
    \centering
    \includegraphics[width=1.0\linewidth]{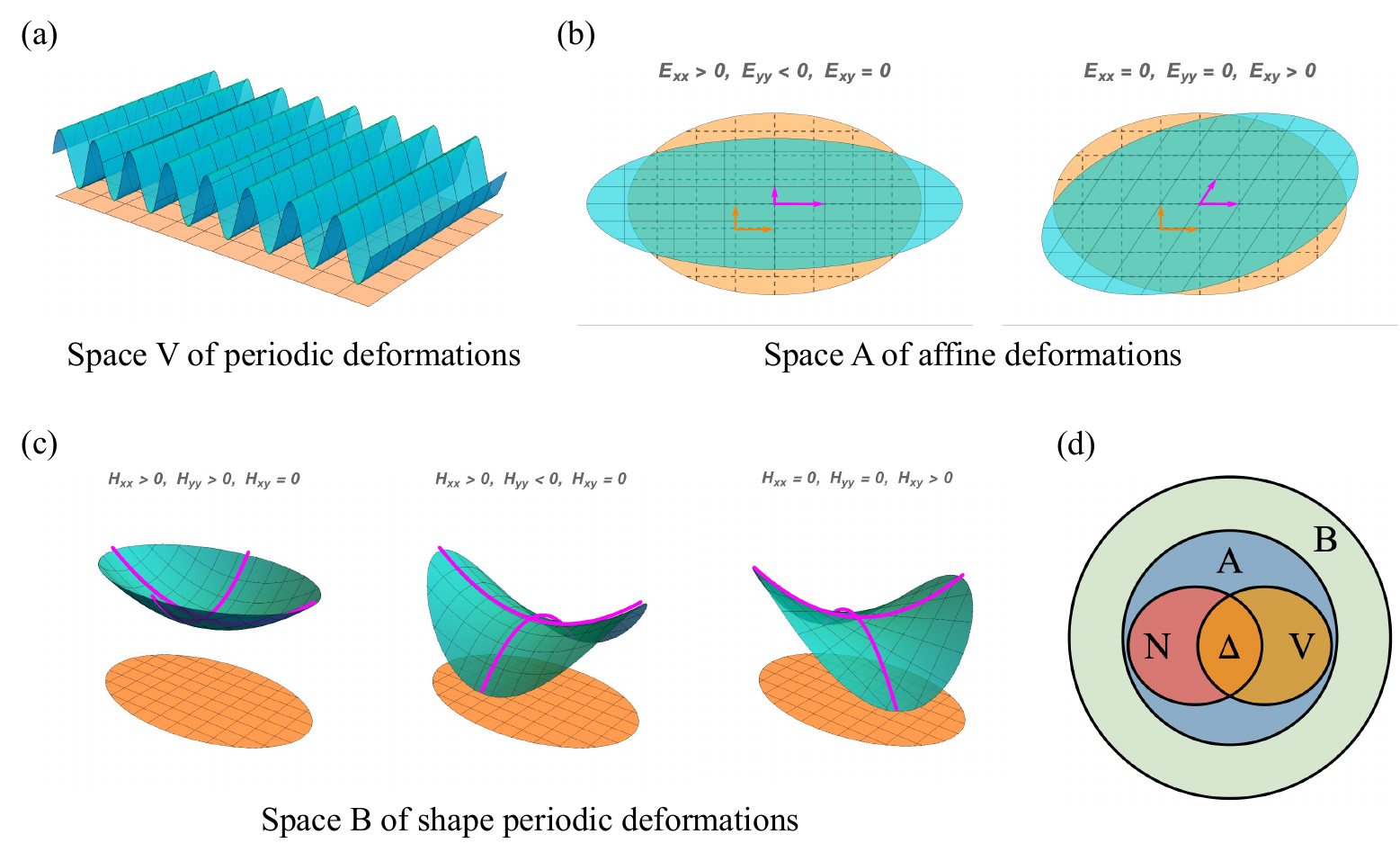}
    \caption{Schematic illustration of deformations in the spaces $V, A, B$ and their relations between $\Delta$ and $N$. (a) Periodic deformation. The deformed sheet (cyan) preserves periodicity. The undeformed sheet (orange) has the same lattice vectors as the deformed one (cyan). (b) Strain deformation. The deformation (cyan) preserves periodicity but the lattice vectors (orange vectors for the undeformed sheet and magenta for the deformed sheet) are changed. (c) Shape-periodic deformation. Each unit cell is bent in the same way. The deformed system is no longer spatially periodic. The magenta lines mark directions of principal curvatures. (d) The Venn diagram demonstrating the relation between the spaces $V, A, B, \Delta, N$.}
    \label{fig:spaces}
\end{figure*}

We use this idea to study the folding of triangulated origami.
The folding of origami involves not only the space $V$ of periodic vertex motions (Fig.~\ref{fig:spaces} (a)),
but also two more spaces $A$ and $B$. The space $A$ of affine deformations (Fig.~\ref{fig:spaces} (b), see SI, Section 1.1 for detailed definitions) contains curvature-free foldings where spatial periodicity is preserved but lattice vectors may change. The space $B$ of shape-periodic deformations (Fig.~\ref{fig:spaces} (c), see SI, Section 1.1 for detailed definitions) contains foldings where different cells have the same shape but potentially different orientations. Equivalently, space $B$ contains all deformations with periodic folding angles.
These three types of deformation spaces are organized into nested spaces: $V\subsetneq A\subsetneq B$, where $\subsetneq$ denotes strict inclusion.

Let $\Delta\subsetneq V $ be the space of rigid translations and $N\subsetneq A$ be the space of all rigid-body motions, which includes rotations as well as translations. Fig.~\ref{fig:spaces} (d) shows the relations between the spaces $V, A, B, \Delta, N$. 
Since the rigid body motions are not of interest, we factor out the corresponding rigid-body motions from the relevant spaces.
As a result, $V/\Delta$ contains purely periodic internal motions, $A/N$ adds strain modes, and $B/N$ further adds bending modes. 
Thus, the strain modes are the new zero modes that appear when passing from $V/\Delta$ to $A/N$, and the bending modes are the new zero modes that appear when passing from $A/N$ to $B/N$.

Let $W$ be the space of constraints, the idea of decomposing the relevant spaces $V/\Delta$, $A/N$, $B/N$ and $W$ into irreducible sectors in \eqref{eq:generalizedMaxwellcounting} applies similarly here (See SI, Section 1.3 for details). This gives us $N_{\rm dof}^{(j)}(V/\Delta)$, $N_{\rm dof}^{(j)}(A/N)$, $N_{\rm dof}^{(j)}(B/N)$ and $N_{\rm con}^{(j)}(W)$. Using \eqref{eq:ZMcriterion}, we obtain the following sufficient condition: an infinitesimal curvature-free folding mode exists in irreducible sector $j$, when
\begin{equation}
\label{eq:infinitesimalstrainmodecriterion}
   N_{\rm dof}^{(j)}(A/N)>\max\{N_{\rm dof}^{(j)}(V/\Delta), N_{\rm con}^{(j)}(W)\}.
\end{equation}

\noindent $N_{\rm dof}^{(j)}(A/N)> N_{\rm con}^{(j)}(W)$ ensures that the sector contains at least one zero mode that does not violate any constraints. $N_{\rm dof}^{(j)}(A/N)>N_{\rm dof}^{(j)}(V/\Delta)$ ensures that this zero mode involves non-zero strain. In principle, there could be two or more strain modes if the difference between the quantities were greater, but this does not occur for any layer group. By the same logic, infinitesimal curved folding modes exist when we have:
\begin{equation}\label{eq:infinitesimalbendingmodecriterion}
    N_{\rm dof}^{(j)}(B/N)>\max\{N_{\rm dof}^{(j)}(A/N), N_{\rm con}^{(j)}(W)\}.
\end{equation}

The final ingredient for this symmetry-adapted counting is a dimension counting formula from the canonical decomposition~\cite{serre2012linear}. For any space $S=W,V/\Delta,A/N,B/N$ of degrees of freedom or constraints, let $\rho_S(g)$ be the induced action of $g\in G/T$ on $S$, where $G$ is the layer group and $T$ is the translation subgroup. Let $\chi_j$ be the character of $j^{\textrm{th}}$ irreducible representation of $G/T$, the $j^{\rm th}$ canonical component of $S$ has dimension
\begin{equation}
\label{eq:dimensioncountingformula}
\begin{aligned}
    \dim S^{(j)}&=\frac{\chi_j(\mathrm{id})}{|G/T|}\sum_{g\in G/T}\chi_j(g)^*\,\operatorname{Tr}\rho_S(g),\\
    S&=W,V/\Delta,A/N,B/N,
\end{aligned}
\end{equation}
where $|G/T|$ is the number of elements in the group $G/T$, and $\mathrm{id}\in G/T$ is the identity element.
Once $\operatorname{Tr}\rho_S(g)$ is obtained for each space $S$ (See SI, Chapter 1 for details), we can compute $N_{\rm dof}^{(j)}(V/\Delta)$, $N_{\rm dof}^{(j)}(A/N)$, $N_{\rm dof}^{(j)}(B/N)$ and $N_{\rm con}^{(j)}(W)$ using \eqref{eq:dimensioncountingformula}. Combining \eqref{eq:infinitesimalstrainmodecriterion} and \eqref{eq:infinitesimalbendingmodecriterion}, the number of infinitesimal zero modes can be determined. For example, Table~\ref{table:sectordimensionsandfundamentalformsforgliderotationsymmetry} summarizes the sector dimensions of relevant spaces of the glide-rotation symmetric origami in Fig.~\ref{fig:nonlinearfoldinggliderotandglideref} (a). From the table, that structure has one pure strain mode in the symmetric sector, one bending mode in the symmetric sector, and one bending mode in the antisymmetric sector.

\begin{table}[h!]
\centering
\caption{Sector dimensions of spaces $V/\Delta, A/N, B/N$, $W$, and the fundamental forms of the resulting zero modes in the glide-rotation symmetric origami (group $p2_{1}11$) in Fig.~\ref{fig:nonlinearfoldinggliderotandglideref} (a).}
\label{table:sectordimensionsandfundamentalformsforgliderotationsymmetry}
\footnotesize
\setlength{\tabcolsep}{2.5pt}
\renewcommand{\arraystretch}{1.15}
\begin{tabularx}{\columnwidth}{
    @{}l
    cccc
    >{\raggedright\arraybackslash}X@{}
}
\toprule
\textbf{Sector}
& \(V/\Delta\)
& \(A/N\)
& \(B/N\)
& \(W\)
& \textbf{Zero modes and their fundamental forms} \\
\midrule
Symmetric & 5 & 7 & 8 & 6 & One strain mode with $E_{12}=0$. One shape-periodic mode with $E_{12}=0$ and $H_{11}=H_{22}=0$.
\\[2pt]
Antisymmetric & 4 & 5 & 7 & 6 & One shape-periodic mode with $E_{11}=E_{22}=0$ and $H_{12}=0$.\\
\bottomrule
\end{tabularx}
\end{table}

\section*{Infinitesimal Shape-Periodic Zero Modes}

\subsection*{The Symmetry-Determined Macroscopic Response}
The macroscopic response of a sheet to its linear folding mode $v$ can be characterized by the strain $E(v)$  (Fig.~\ref{fig:spaces} (b)) and by the induced curvature $H(v)$ (Fig.~\ref{fig:spaces} (c)), which are respectively the changes to its first and second fundamental forms.
Symmetry constrains the possible forms of $E(v)$ and $H(v)$ of shape-periodic zero modes. Let $b_{1}$ and $b_{2}$ be the lattice vectors of the periodic triangulated origami, $\eta(g)$ be the matrix of an operation $g\in G/T$ under the basis $\{b_{1},b_{2}\}$, $\lambda(g)=\pm 1$ indicating if the operation preserves ($+1$) or reverses ($-1$) the out-of-plane direction. From the geometric perspective, $E(v)$ and $H(v)$ transform under the operation $g\in G/T$ by
\begin{equation}\label{eq:EHtransform}
\begin{aligned}
E(v)&\to\eta(g^{-1})^T E(v)\eta(g^{-1}),\\
H(v)&\to\lambda(g)\eta(g^{-1})^T H(v)\eta(g^{-1}).
\end{aligned}
\end{equation}

\onecolumn

\begingroup
\small
\renewcommand{\arraystretch}{1.2}
\setlength{\LTcapwidth}{\textwidth}

\begin{xltabular}{\textwidth}{
    >{\raggedright\arraybackslash}p{0.16\textwidth}
    >{\centering\arraybackslash}p{0.12\textwidth}
    >{\raggedright\arraybackslash}X
}

\caption{Mode structures for different layer groups.}
\label{tab:modestructures}\\

\toprule
Lattice System & Group Name & Mode Structures \\
\midrule
\endfirsthead

Triclinic 
& $p1$ 
& This group has trivial point group. \\

& $p\bar{1}$ 
& When all four inversion centers are occupied by edge centers, there is one symmetry-preserving nonlinear pure $E$ mode in the $\chi_{1}$ sector and two pure $H$ modes in the $\chi_{2}$ sector. Otherwise, there are three pure $H$ modes in the $\chi_{2}$ sector. \\

Monoclinic (Inclined)
& $p112$ 
& Three $(E,H)$ modes, all of them lie in the $\chi_{1}$ sector. \\
& $p11a$ 
& One symmetry-preserving nonlinear  pure $E$ mode in the $\chi_{1}$ sector, two pure $H$ modes  in the $\chi_{2}$ sector.\\
& $p112/a$
& When all 4 inversion centers are occupied by edge centers: one symmetry-preserving nonlinear pure $E$ mode in the $\chi_{1}$ sector, two pure $H$ modes in the $\chi_{3}$ sector. Otherwise, three pure $H$ modes in the $\chi_{3}$ sector.\\

Monoclinic (Orthogonal)
& $p2_{1}11$
& One symmetry-preserving nonlinear  pure $E$ mode in the $\chi_{1}$ sector, $E$ is diagonal. One $(E,H)$ mode in sector $\chi_{1}$, with diagonal $E$ and off-diagonal $H$. One $(E,H)$ mode in sector $\chi_{2}$, with off-diagonal $E$ and diagonal $H$.\\
& $pm11$
& Two linear $(E,H)$ modes in sector $\chi_{1}$ with diagonal $E$ and diagonal $H$. One $(E,H)$ mode in sector $\chi_{2}$, with off-diagonal $E$ and off-diagonal $H$. \\
& $pb11$
& Two linear $(E,H)$ modes in the $\chi_{1}$ sector with diagonal $E$ and diagonal $H$. One bending mode $(E,H)$ in sector $\chi_{2}$ with off-diagonal $E$ and off-diagonal $H$.\\
& $cm11$
& Same as $pm11$.\\
& $p2_{1}/m11$
& When all 4 inversion centers are occupied by edge centers: one symmetry-preserving nonlinear  pure $E$ mode in the $\chi_{1}$ sector with diagonal $E$. One pure $H$ mode in the $\chi_{3}$ sector with off-diagonal $H$. One pure $H$ mode  in the $\chi_{4}$ sector, with diagonal $H$. Otherwise, one pure $H$ mode in the $\chi_{3}$ sector with off-diagonal $H$. Two pure $H$ modes in the $\chi_{4}$ sector with diagonal $H$.\\
& $p2_{1}/b11$
& Same as $p2_{1}/m11$.\\

Orthorhombic
& $p2_{1}2_{1}2$
& One symmetry-preserving nonlinear  pure $E$ mode in the $\chi_{1}$ sector with diagonal $E$, one $(E,H)$ mode in the $\chi_{1}$ sector with diagonal $E$ and off-diagonal $H$. One $(E,H)$ mode in the $\chi_{4}$ sector with off-diagonal $E$ and diagonal $H$.\\
& $pmm2$
& Two linear $(E,H)$ modes in the $\chi_{1}$ sector with diagonal $E$ and diagonal $H$. One $(E,H)$ mode with off-diagonal $E$, off-diagonal $H$ in the $\chi_{4}$ sector.\\
& $cmm2$
& Same as $pmm2$\\
& $pma2$
& Two linear $(E,H)$ modes in the $\chi_{1}$ sector with diagonal $E$ and diagonal $H$. One $(E,H)$ mode with off-diagonal $E$, off-diagonal $H$ in the $\chi_{3}$ sector.\\
& $pm2_{1}b$
& One symmetry-preserving nonlinear  pure $E$ mode with diagonal $E$ in the $\chi_{1}$ sector. One pure $H$ mode with off-diagonal $H$ in the $\chi_{2}$ sector. One pure $H$ mode with diagonal $H$ in the $\chi_{3}$ sector.\\
& $pm2_{1}n$
& One symmetry-preserving nonlinear  pure $E$ mode with diagonal $E$ in the $\chi_{1}$ sector. One pure $H$ mode with off-diagonal $H$ in the $\chi_{2}$ sector. One pure $H$ mode with diagonal $H$ in the $\chi_{3}$ sector.\\
& $pba2$
& Two linear $(E,H)$ modes in the $\chi_{1}$ sector, with diagonal $E$ and diagonal $H$. One $(E,H)$ mode in the $\chi_{3}$ sector, with off-diagonal $E$ and off-diagonal $H$.\\
& $pb2_{1}a$
& One symmetry-preserving nonlinear pure $E$ mode in the $\chi_{1}$ sector with diagonal $E$. One pure $H$ mode in the $\chi_{2}$ sector with off-diagonal $H$. One pure $H$ mode in the $\chi_{3}$ sector with diagonal $H$.\\
& $pbma$
& When all 4 inversion centers are occupied by edge centers: one symmetry-preserving nonlinear  pure $E$ mode in the $\chi_{1}$ sector with diagonal $E$. One pure $H$ mode  in the $\chi_{5}$ sector with diagonal $H$. One pure $H$ mode in the $\chi_{8}$ sector, with off-diagonal $H$. Otherwise, two pure $H$ modes  in the $\chi_{5}$ sector with diagonal $H$. One pure $H$ mode  in the $\chi_{8}$ sector, with off-diagonal $H$.\\
& $pmmn$
& When all 4 inversion centers are occupied by edge centers: one symmetry-preserving nonlinear  pure $E$ mode in the $\chi_{1}$ sector with diagonal $E$. One pure $H$ mode in the $\chi_{5}$ sector with diagonal $H$. One pure $H$ mode  in the $\chi_{8}$ sector, with off-diagonal $H$. Otherwise, two pure $H$ modes  in the $\chi_{5}$ sector with diagonal $H$. One pure $H$ mode in the $\chi_{8}$ sector, with off-diagonal $H$.\\

Tetragonal
& $p4$
& One $(E,H)$ mode in the $\chi_{1}$ sector, with $E_{11}-E_{22}=E_{12}=0$ and $H_{11}-H_{22}=H_{12}=0$. Two $(E,H)$ modes in the $\chi_{3}$ sector, with $E_{11}+E_{22}=0$ and $ H_{11}+H_{22}=0$.\\
& $p\bar{4}$
& One pure $E$ mode in the $\chi_{3}$ sector, with $E_{11}+E_{22}=0$. One $(E,H)$ mode in the $\chi_{1}$ sector, with $E_{11}-E_{22}=E_{12}=0$ and $H_{11}+H_{22}=0$. One $(E,H)$ mode in the $\chi_{3}$ sector, with $E_{11}+E_{22}=0$ and $H_{11}-H_{22}=H_{12}=0$.\\
& $p4/n$
& When all 4 inversion centers are occupied by edge centers: one pure $E$ mode in the $\chi_{3}$ sector with $E_{11}+E_{22}=0$. One pure $H$ mode in the $\chi_{5}$ sector with $H_{11}-H_{22}=H_{12}=0$. One pure $H$ mode in the $\chi_{7}$ sector, with $H_{11}+H_{22}=0$. Otherwise, one pure $H$ mode in the $\chi_{5}$ sector with $H_{11}-H_{22}=H_{12}=0$. Two pure $H$ modes in the $\chi_{7}$ sector, with $H_{11}+H_{22}=0$.\\
& $p4mm$
& One $(E,H)$ mode in the $\chi_{1}$ sector with $E_{11}-E_{22}=E_{12}=0$ and $H_{11}-H_{22}=H_{12}=0$. One $(E,H)$ mode in the $\chi_{3}$ sector with $E_{11}+E_{22}=E_{12}=0$ and $H_{11}+H_{22}=H_{12}=0$. One $(E,H)$ mode in the $\chi_{4}$ sector with $E_{11}=E_{22}=0$ and $H_{11}=H_{22}=0$.\\
& $p4bm$
& One $(E,H)$ mode in the $\chi_{1}$ sector with $E_{11}-E_{22}=E_{12}=0$ and $H_{11}-H_{22}=H_{12}=0$. One $(E,H)$ mode in the $\chi_{3}$ sector with $E_{11}+E_{22}=E_{12}=0$ and $H_{11}+H_{22}=H_{12}=0$. One $(E,H)$ mode in the $\chi_{4}$ sector with $E_{11}=E_{22}=0$ and $H_{11}=H_{22}=0$.\\
& $p\bar{4}2_{1}m$
& One pure $E$ mode in the $\chi_{4}$ sector with $E_{11}=E_{22}=0$. One pure $H$ mode in the $\chi_{2}$ sector with $H_{11}+H_{22}=H_{12}=0$. One $(E,H)$ mode in the $\chi_{4}$ sector with $E_{11}=E_{22}=0$ and $H_{11}-H_{22}=H_{12}=0$.\\

Trigonal
& $p3$
& One $(E,H)$ mode in the $\chi_{1}$ sector with $2E_{12}=E_{11}=E_{22}$ and $2H_{12}=H_{11}=H_{22}$. One $(E,H)$ mode in the $\chi_{2}$ sector. One $(E,H)$ mode in the $\chi_{3}$ sector. They together can be recombined to a mode with $E_{11}=-E_{12},\;E_{22}=2E_{12}$ and $H_{11}=-H_{12},\;H_{22}=2H_{12}$, and a mode with $E_{11}=E_{12},\;E_{22}=0$  and $H_{11}=H_{12},\;H_{22}=0$.\\
& $p\bar{3}$
& One pure $H$ mode in the $\chi_{4}$ sector with $2H_{12}=H_{11}=H_{22}$. One pure $H$ mode in the $\chi_{5}$ sector, and one pure $H$ mode in the $\chi_{6}$ sector. They can be recombined into two real modes: one with $H_{11}-H_{12}=H_{22}=0$, and one with $H_{11}+H_{12}=0,\;  H_{22}=2H_{12}$.\\
& $p3m1$
& One $(E,H)$ mode in the $\chi_{1}$ sector with $2E_{12}=E_{11}=E_{22}$ and $2H_{12}=H_{11}=H_{22}$. Two $(E,H)$ modes in the $\chi_{3}$ sector. These two modes together can be recombined to a mode with $E_{11}=-E_{12},\;E_{22}=2E_{12}$ and $H_{11}=-H_{12},\;H_{22}=2H_{12}$, and a mode with $E_{11}=E_{12},\;E_{22}=0$  and $H_{11}=H_{12},\;H_{22}=0$.\\

Hexagonal
& $p6$
& One $(E,H)$ mode in the $\chi_{1}$ sector with $2E_{12}=E_{11}=E_{22}$ and $2H_{12}=H_{11}=H_{22}$. One $(E,H)$ mode in the $\chi_{3}$ sector and one $(E,H)$ mode in the $\chi_{5}$ sector. They together can be recombined to a mode with $E_{11}=-E_{12},\;E_{22}=2E_{12}$ and $H_{11}=-H_{12},\;H_{22}=2H_{12}$, and a mode with $E_{11}=E_{12},\;E_{22}=0$  and $H_{11}=H_{12},\;H_{22}=0$.\\
& $p6mm$
& One $(E,H)$ mode in the $\chi_{1}$ sector with $2E_{12}=E_{11}=E_{22}$ and $2H_{12}=H_{11}=H_{22}$. Two $(E,H)$ modes in the $\chi_{6}$ sector. They satisfy $E_{11}+E_{22}-E_{12}=0$, $H_{11}+H_{22}-H_{12}=0$.\\
\hline
\end{xltabular}
\endgroup

\begin{figure*}[ht]
    \centering
    \includegraphics[width=1.0\linewidth]{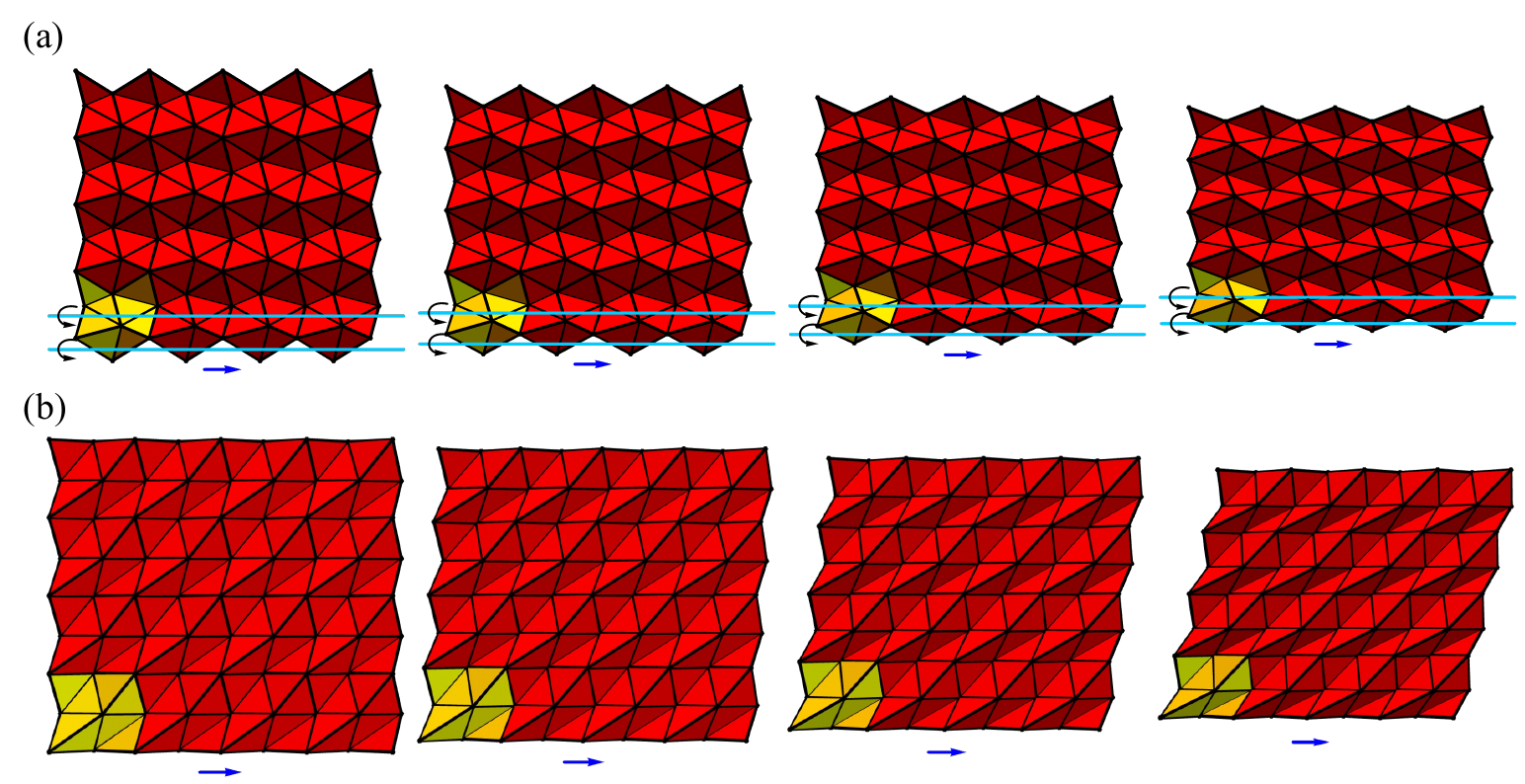}
    \caption{Symmetry-protected nonlinear curvature-free folding from glide rotation symmetry (group $p2_{1}11$) and glide-$z$-reflection symmetry (group $p11a$). (a) Glide–rotation-symmetric triangulated origami and its curvature-free folding sequence (left to right).  We highlight one unit cell in yellow. The blue arrow is the glide vector, and the solid light blue lines are the two inequivalent rotation axes. (b) Glide–z-reflection-symmetric triangulated origami. We highlight one unit cell in yellow. The blue arrow is the glide vector and the z-reflection reverses the out-of-plane direction.}
    \label{fig:nonlinearfoldinggliderotandglideref}
\end{figure*}

From the representation theory perspective, when the zero mode $v$ lies in a real one-dimensional representation sector $j$ (the case of sectors in real two-dimensional irreducible representations is handled in the SI, Chapter 3), we have $gv=\chi_j(g)v$, where $\chi_{j}(g)$ is the character of $g$ in the representation $j$. Since $E(v)$ and $H(v)$ are linear in $v$ we have the consistency equations 

\begin{equation}\label{eq:EHconstraints}
\begin{aligned}
    \eta(g^{-1})^T E(v)\eta(g^{-1})&=\chi_j(g)E(v),\\
    \lambda(g)\eta(g^{-1})^T H(v)\eta(g^{-1})&=\chi_j(g)H(v).
\end{aligned}
\end{equation}
\eqref{eq:EHconstraints} relates the macroscopic response of a zero mode to the irreducible sector it lies in. Thus, to determine what macroscopic response a periodic triangulated origami with layer group $G$ could possibly have, we only need to determine the distribution of its zero modes among the irreducible representations.
Returning to our example in Table~\ref{table:sectordimensionsandfundamentalformsforgliderotationsymmetry}, the glide rotation $g$ has $\eta(g)=\mathrm{diag}\{1,-1\}$ and $\lambda(g)=-1$, ~\eqref{eq:EHconstraints} then tells us that the two modes in the symmetric sector ($\chi_{1}(g)=1$) have only $E_{11}, E_{22}, H_{12}$ among their nonzero components, whereas the mode in the antisymmetric sector ($\chi_{2}(g)=-1$) contains components with nonzero $E_{12}, H_{11}, H_{22}$.

\twocolumn
These symmetry-based restrictions combine with a more fundamental relationship between the changes to the fundamental forms given by two distinct modes indexed by $a,b$:

\begin{align}
    \epsilon^{\alpha\mu}\epsilon^{\beta\nu}\Big(H^a_{\mu\nu}E^b_{\alpha\beta}-H^b_{\mu\nu}E^a_{\alpha\beta}\Big)=0
\end{align}

\noindent This equation, first derived in~\cite{sun2025geometric} and rederived for triangulated surfaces in the SI, indicates that the form of one of a sheet's deformation modes places a sharp restriction on the sheet's two other modes. The present work establishes which origami sheets, by virtue of symmetry, possess pure strain modes. Each pure strain mode has a well-defined in-plane Poisson ratio proportional to the ratio of its two principal strains. Such sheets also possess out-of-plane Poisson ratios, the ratio of their two principal curvatures. The above relationship establishes that for a sheet with a protected strain mode, the in-plane Poisson ratio of the pure strain mode and the out-of-plane Poisson ratios of the two other modes are equal in magnitude and opposite in sign. 
Previously, such equal and opposite Poisson ratios were observed for specific but broadening classes of origami where the in-plane Poisson's ratio can be defined, including the Miura-ori~\cite{schenk2013geometry, wei2013geometric}, the eggbox~\cite{nassar2017curvature}, morph pattern~\cite{pratapa2019geometric} and the four-parallelogram~\cite{mcinerney2022discrete}. The present result demonstrates that this is true for a large class of origami, including the inversion-symmetric origami depicted in Fig.~\ref{fig:nonlinearfoldinginversion} (e).

\subsection*{Towards the Classification: Resummation and Reduction}\label{sec:RRforclassification}
Given a periodic triangulated origami $X$ with layer group $G$, to classify all possible distributions among irreducible sectors of shape-periodic zero modes in $X$, we need to face the challenge that each layer group admits infinitely many possible triangulations. We now develop techniques to identify the essential features of a triangulation that determine its folding properties.

The first technique is a resummation trick. \eqref{eq:dimensioncountingformula} counts $\dim S^{(j)}$ from the contribution of each $g\in G/T$, where the trace of $\rho_S(g)$ is related to the objects fixed by symmetry $g$: vertices for displacement degrees of freedom, or edges for constraints. The key observation is that we can count how many degrees of freedom or constraints a vertex or edge contributes to each symmetry sector based purely on how symmetric its location is in the unit cell (e.g., lying at an inversion center or in a mirror plane). 

After this resummation, for a given layer group, $\dim S^{(j)}$ depends only on the counting of high-symmetry vertices and high-symmetry edges (See SI, Chapter 1 for details). The theory developed in the previous section provides a general procedure determining which irreducible sector must support linear zero modes when the entire structure is known. This resummation, however, identifies the contribution of the dimensions of the irreducible sectors from the local properties of the structure, and is useful for computing the change of irreducible sector dimensions when the structure is only changed locally.

The second technique is a reduction trick. Using the resummation trick, we can identify symmetry-preserving operations on the structure preserving the number of symmetry protected modes in every irreducible sector. Together, these two techniques (See SI, Chapter 2 for details) reduce the infinitely many triangulations associated with each layer group into finitely many equivalence classes such that all triangulations within the same class have identical counts of strain modes and bending modes.

\subsection*{The Classification}
Using the tricks of resummation and reduction, we classify the symmetry-protected zero modes in terms of their macroscopic response in Table~\ref{tab:modestructures} (See SI, Chapter 3 for details) without solving the compatibility matrix. For example, when $g$ is an inversion, $\eta(g)=-I$ and $\lambda(g)=-1$. An inversion-even mode can have nonzero $E$ but must have $H=0$, while an inversion-odd mode can have nonzero $H$ but must have $E=0$. 
In other words, inversion symmetry separates the linear zero modes into pure strain modes and pure bending modes. Glide-$z$-reflection produces the same even/odd separation. Glide rotation produces a complementary selection rule: its even sector contains a pure strain mode with diagonal $E$ (Table~\ref{table:sectordimensionsandfundamentalformsforgliderotationsymmetry} and Fig.~\ref{fig:nonlinearfoldinggliderotandglideref} (a), the lattice vectors are always perpendicular along the mechanism so $E$ is always diagonal) and a mixed mode with diagonal $E$ and off-diagonal $H$ (Table~\ref{table:sectordimensionsandfundamentalformsforgliderotationsymmetry}); its odd sector contains a mixed mode with off-diagonal $E$ and diagonal $H$ (Table~\ref{table:sectordimensionsandfundamentalformsforgliderotationsymmetry}).

We emphasize that the classification presented here is restricted to layer groups that contain neither the pure reflection about the $xy$-plane nor any in-plane $\pi$-rotation. The exclusion of the former symmetry is straightforward: folded origami configurations cannot realize it. The exclusion of in-plane $\pi$-rotations is subtle: such a rotation necessarily gives rise to additional states of self stress and therefore to  zero modes in $V/\Delta$.
These zero modes give rise to trivial behavior, such as a straight line of creases that permits the origami to fold up along that single line, breaking spatial periodicity in a quasi-one-dimensional way rather than true two-dimensional origami folding.

\section*{Designing Nonlinear Curvature-Free Folding}

\begin{figure*}[ht!]
    \centering
    \includegraphics[width=1.0\linewidth]{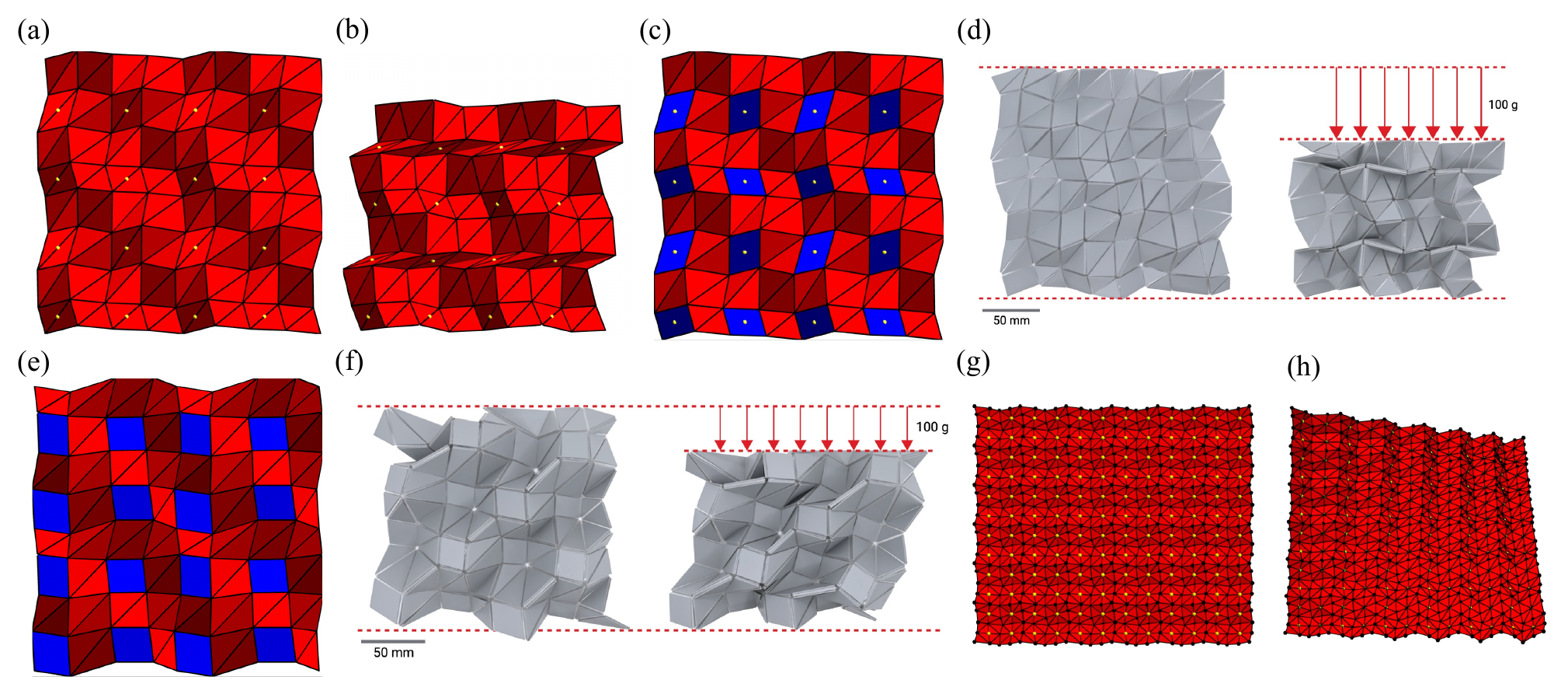}
    \caption{The topological occupation rule in inversion symmetric structures. (a) Inversion-symmetric triangulated origami satisfying the topological occupation rule. This figure contains $2\times 2$ unit cells. The inversion centers, highlighted in yellow, are all located at edge centers. (b) The resulting structure after folding the origami in (a) along the nonlinear inversion-symmetry-preserving strain mode. This configuration is obtained by numerically continuing the linear inversion-symmetric strain mode at each time step. (c) Merging the triangles containing the inversion center in inversion symmetric structures in (a), which satisfies the topological occupation rule. The merged triangles are highlighted in blue. The structure is over-constrained but still admits a nonlinear strain mode preserving the inversion symmetry. (d) Experimental demonstration of the nonlinear strain mode in (c). 
    (e) Periodic origami breaking inversion symmetry. This structure has the same number of triangles and parallelograms as that of (c) in each unit cell and they are connected in the same way. (f) Experimental demonstration of the nonexistence of nonlinear strain mode in (e). 
    (g) Inversion-symmetric triangulated origami that violates the topological occupation rule. Some inversion centers do not coincide with edge centers. (h) The folding of (g) induces curvature. Here we only showed one of its three linear zero modes.}
    \label{fig:nonlinearfoldinginversion}
\end{figure*}

\subsection*{Existence and Classification of Nonlinear Strain Modes}
A linear strain mode in the trivial ($\chi_{1}$) sector of a sheet preserves all of its symmetries. When such a mode is guaranteed to exist by the symmetry-adapted counting, this strain mode can be extended nonlinearly in a way that preserves all of the symmetries. The intuitive reason is that deforming a small step along this symmetry-protected mode preserves all symmetries.
Therefore, the deformed state, by the symmetry-adapted counting, also has a symmetry-protected mode and the procedure may continue. 
The only obstacle for this nonlinear extension argument is the possibility of a fully symmetric state of self-stress~\cite{KANGWAI19995507}.

The classification established in the previous section shows that only layer groups containing operations reversing the out-of-plane direction can possibly guarantee a strain mode in the trivial sector by the symmetry-adapted counting. For layer groups containing an inversion or a glide-$z$-reflection, if the structure considered further has no zero modes in $V/\Delta$, the symmetry-protected linear strain mode in the $\chi_{1}$ sector extends nonlinearly into a symmetry-preserving curvature-free folding (See SI, Chapter 1 for the proof). For layer groups whose only symmetry reversing the out-of-plane direction is the glide-rotation, the assumption of no zero modes in $V/\Delta$ implies that the symmetry-protected linear strain mode in the $\chi_{1}$ sector extends nonlinearly into a symmetry-preserving curvature-free folding unless the structure is fine-tuned (See SI, Chapter 1 for the exact statement and proof).
Flat sheets don't fall into this category. Additionally, the absence of zero modes in $V/\Delta$ is a weak condition, equivalent to saying that the structure has only the three states of self-stress guaranteed by the usual Maxwell counting~\cite{JamesPNAS2020}.

The results of this program for all layer groups associated with periodic folded origami are presented exhaustively in Table~\ref{tab:modestructures} (See SI, Chapter 3 for details). From this table, we conclude that any layer group admitting curvature-free nonlinear folding must contain an inversion, or a glide-rotation, or a glide-$z$-reflection. Glide-rotation and glide-$z$-reflection symmetries guarantee a curvature-free nonlinear folding mode without further conditions.
In contrast, such a mode exists in inversion-symmetric systems only when a certain topological occupation rule (discussed below) is satisfied. We provide examples illustrating how a single glide-rotation, glide-$z$-reflection, or inversion symmetry protects a curvature-free nonlinear folding mode.

\subsection*{Glide-Rotation and Glide-$z$-Reflection Symmetry}

Fig.~\ref{fig:nonlinearfoldinggliderotandglideref} (a) shows a periodic triangulated origami that is symmetric only under a glide-rotation operation. The glide axis and the rotation axis coincide with the horizontal axis. Direct numerical zero mode continuation confirms that the structure can fold without generating curvature (Fig.~\ref{fig:nonlinearfoldinggliderotandglideref} (a) from left to right).
Fig.~\ref{fig:nonlinearfoldinggliderotandglideref} (b) shows a periodic triangulated origami that is symmetric only under a glide-$z$-reflection operation. The glide axis is horizontal, and the z-reflection reverses the out-of-plane direction. The structure can also fold without generating curvature. 

\subsection*{Inversion Symmetry and the Topological Occupation Rule}

While there are intuitive reasons that symmetries involving $z \rightarrow -z$, which reverse curvature while leaving strain invariant, are crucial to generating strain modes, as we will see the group-theoretic approach reveals specific criteria that are also necessary.
An important example is the inversion-symmetric periodic triangulated origami (well-known origami patterns, such as the Miura-ori, satisfy both inversion symmetry and other symmetries). Such origami admits curvature-free folding only when all four inversion centers in a unit cell lie at edge centers after the reduction.
Fig.~\ref{fig:nonlinearfoldinginversion} (a) is one example of such a structure, and we verify 
with numerical continuation of the linear inversion-symmetric strain mode at each time step that this structure folds nonlinearly in a way that generates no curvature and preserves the inversion symmetry, as shown in  Fig.~\ref{fig:nonlinearfoldinginversion} (b).

However, if this occupation rule is violated by having vertices rather than edges lie on these inversion centers (Fig.~\ref{fig:nonlinearfoldinginversion} (g)), the structure generally cannot fold without curvature, as shown in Fig.~\ref{fig:nonlinearfoldinginversion} (h), where bending necessarily occurs as the folding mode is activated.
The rule is topological because, in an inversion-symmetric triangulation, each inversion center must be occupied by a symmetry-invariant element of the triangulation: either a vertex or the midpoint of an edge. Thus the distinction between vertex-occupied and edge-center-occupied inversion centers is a discrete combinatorial property of the triangulation, rather than a metric property that can be changed by continuously moving vertices while preserving the same connectivity and symmetry.

In addition to determining the deformations of triangulated origami, symmetry can also be used to design over-constrained sheets that can fold. 
Notably, when an inversion-symmetric periodic triangulated origami satisfies the topological occupation rule, the two triangles sharing a common inversion center must be coplanar in the entire nonlinear folding pathway preserving the inversion symmetry and can therefore be merged into a single rigid parallelogram face.
After this operation, the system becomes over-constrained, and the excess floppy modes in periodic triangulated origami can be removed. The Miura-ori, for example, is an over-constrained sheet that can be obtained using such an operation, and its nonlinear folding is a consequence of both inversion symmetry and the topological occupation rule.

We further check experimentally the validity of the above argument on creating over-constrained origami that can fold curvature-freely. The origami in Fig.~\ref{fig:nonlinearfoldinginversion} (a), after this operation, becomes an over-constrained origami (Fig.~\ref{fig:nonlinearfoldinginversion} (c)). We experimentally assembled the resulting structure (the left panel in Fig.~\ref{fig:nonlinearfoldinginversion} (d)) and confirmed experimentally that the assembled origami can fold in an inversion-symmetry-preserving way that generates no curvature when the sample is compressed between two plates with a mass of 100 g (the right panel in Fig.~\ref{fig:nonlinearfoldinginversion} (d) and SI Chapter 4).

To see why this nonlinear curvature-free folding is also protected by the inversion symmetry, we created a periodic origami that breaks the inversion symmetry (Fig.~\ref{fig:nonlinearfoldinginversion} (e)). This new origami has the same number of triangles and parallelograms as that in Fig.~\ref{fig:nonlinearfoldinginversion} (c), and their connections are the same. However, the new structure does not fold curvature-freely in either numerical simulations or experiments. As can be seen in Fig.~\ref{fig:nonlinearfoldinginversion} (f), a sample with the same surface area and loading conditions as the sample in Fig.~\ref{fig:nonlinearfoldinginversion} (d) bends and compresses non-uniformly.

\subsection*{Design Nonlinear Curvature-Free Folding}\label{sec:nonlineardesign}
With the exhaustive table, the design of triangulated origami that folds nonlinearly and curvature-freely is divided into two steps. 
First, choose a layer group $G$ that admits nonlinear, curvature-free folding from Table~\ref{tab:modestructures}, and read out the rules needed for a triangulation symmetric under $G$ to support such a mode. This step fixes the occupation of high-symmetry positions within the unit cell and the connectivity between vertices. The vertices can still move along the high-symmetry positions.
Second, vary the vertex coordinates to optimize the mode properties we are looking for, such as the exact values of the entries of the strain tensor $E$, and whether the folded origami is developable (can be unfolded into a flat sheet) or flat-foldable. For developability, we need only to reduce the angle deficit at each vertex to zero under the symmetry constraints determined in the first step and to make sure the optimized structure is not flat.

\section*{Conclusion}
We have developed a symmetry-adapted Maxwell counting for periodic, triangulated sheets that classifies their linear zero modes and nonlinear strain modes, identifies the symmetries responsible for nonlinear strain modes, and permits the design of new sheets with prescribed properties. For a sheet with layer group $G$, the relevant counting depends only on the placement of high-symmetry vertices and edges. 
Using a reduction trick, we reduce, for each layer group, the infinite space of possible patterns into a finite set of counting-equivalent triangulations, and give an exhaustive table classifying the linear periodic foldings in terms of their macroscopic response for each layer group. Combining this table with a new nonlinear extension theorem, we identify inversion, glide-rotation, and glide-$z$-reflection as the key symmetries protecting nonlinear curvature-free folding. Glide-rotation and glide-$z$-reflection protect curvature-free mechanisms directly, whereas inversion requires an additional topological occupation rule: inversion centers must be occupied by edge centers. This rule explains why the Miura-ori supports a finite curvature-free mechanism despite being over-constrained by conventional Maxwell counting.
Together, these results recast curvature-free deployability as a symmetry-protected property of periodic mechanical sheets. They provide design rules for origami metamaterials whose global shape change is selected by symmetry, topology, and connectivity, rather than by fine-tuning of local geometry.

Several questions remain for future work. One is how to extend the present formulation to periodically curved origami~\cite{JamesPNAS2020}, where periodicity, symmetry, and reference curvature must be treated simultaneously. 
A second question concerns the existence of nonlinear modes for flat (developed) sheets. For example, our technique correctly predicts that the Miura-ori has a nonlinear mode that extends to such a flat state, but beginning from the flat state it is less clear whether such modes extend nonlinearly, due to the obstruction of states of self stress.

\section*{Acknowledgments}

We gratefully acknowledge support from the ARO MURI program (W911NF-22-1-0219) (DZR, KB), and from 
 the NSF CAREER program (2338492) (DZR).

\bibliography{pnas}
\end{document}

\maketitle
\renewcommand\contentsname{} % the empty name
\begingroup
\let\clearpage\relax
\let\cleardoublepage\relax
\vspace{-4cm} % the removed space. Set as appropriate
\tableofcontents
\endgroup

\chapter{Shape Periodic Deformations and The Generalized Maxwell Counting}
\section{The Important Deformation Spaces}
\subsection{The Space of Shape Periodic Deformations}
A shape periodic (SP) mode is a mode where each unit cell is deforming in the same way. 
We formulate this idea in the following. Let $X$ be a doubly periodic triangulated origami, $K$ be a unit cell of $X$, $p_{1},...,p_{h}$ be nodes inside $K$ whose positions are $r_{1},...,r_{h}$, $b_{1}, b_{2}$ be the lattice vectors (taking their plane to be the $xy$ plane), $v_{i}(n_{1},n_{2})$ be the displacement of the node located at $r_{i}+n_{1}b_{1}+n_{2}b_{2}$. 
If the infinitesimal displacement field $\epsilon v_{i}(n_{1},n_{2})$ (here $\epsilon\ll 1$ is infinitesimal) is a shape periodic mode, then after this displacement, we should be able to find infinitesimal rotations $1+\epsilon J_{1}$, $1+\epsilon J_{2}$ and translations $b_{1}+\epsilon t_{1}$, $b_{2}+\epsilon t_{2}$ such that first, after rotating by $1+\epsilon J_{1}$ and then translating by $b_{1}+\epsilon t_{1}$, the system is invariant up to $\epsilon^{2}$ and cell $(n_{1},n_{2})$ is moved to cell $(n_{1}+1,n_{2})$ for all $(n_{1},n_{2})$. Then, after rotating by $1+\epsilon J_{2}$ and then translating by $b_{2}+\epsilon t_{2}$, the system is invariant up to $\epsilon^{2}$ and cell $(n_{1},n_{2})$ is moved to cell $(n_{1},n_{2}+1)$ for all $(n_{1},n_{2})$. Unwrapping the description and assuming $J_{k}r=m_{k}\times r$ we have
\begin{equation}\label{eq:shapeperiodiccondition}
\begin{aligned}
    (1+\epsilon J_{1})(\epsilon v_{i}(n_{1},n_{2})+r_{i}+n_{1}b_{1}+n_{2}b_{2})+b_{1}+\epsilon t_{1}&=\epsilon v_{i}(n_{1}+1,n_{2})+r_{i}+(n_{1}+1)b_{1}+n_{2}b_{2}\\
    (1+\epsilon J_{2})(\epsilon v_{i}(n_{1},n_{2})+r_{i}+n_{1}b_{1}+n_{2}b_{2})+b_{2}+\epsilon t_{2}&=\epsilon v_{i}(n_{1},n_{2}+1)+r_{i}+n_{1}b_{1}+(n_{2}+1)b_{2}
\end{aligned}
\end{equation}
Dropping the $\epsilon^{2}$ terms in Eq.~\eqref{eq:shapeperiodiccondition}, we have the solution
\begin{equation}\label{eq:shapeperiodicansatz}
\begin{aligned}
    v_i(n_1,n_2)&=v_i+n_1(m_1\times r_i)+n_2(m_2\times r_i)+n_1t_1+n_2t_2\\
    &\quad+\frac{n_1(n_1-1)}2(m_1\times b_1)+\frac{n_2(n_2-1)}2(m_2\times b_2)+\frac{n_1n_2}{2}\big(m_2\times b_1+m_1\times b_2\big),
\end{aligned}
\end{equation}
together with a compatibility condition
\begin{equation}\label{eq:rotationcompatibilitycondition}
    m_{1}\times b_{2}=m_{2}\times b_{1}.
\end{equation}
\begin{definition}
    The set of shape periodic deformations forms a vector space under the natural addition of vector fields and scalar multiplication. This space, denoted by $B$, is called the space of shape periodic deformation.
\end{definition}
\subsection{The Space of Affine Deformations and The Space of Periodic Deformations}
\begin{definition}
    The subspace $A\subset B$ consisting of shape periodic deformations with $m_{1}=m_{2}=0$ is called the space of affine deformations.
\end{definition}
These affine deformations are characterized by the fact that a periodic system remains periodic with possibly different lattice vectors under any affine deformation.
\begin{definition}
    The subspace $V\subset A$ consisting of shape periodic deformations with $m_{1}=m_{2}=t_{1}=t_{2}=0$ is called the space of periodic deformations. 
\end{definition}
These periodic deformations are characterized by the fact that a periodic system remains periodic with the same lattice vectors under any periodic deformation.
\subsection{The Space of Translation Modes and The Space of Rigid Body Modes}
\begin{definition}
    The subspace $\Delta\subset V$ consisting of constant displacement vectors over all lattice sites is called the space of translation modes.
\end{definition}
\begin{definition}
    The subspace $N\subset A$ consisting of displacement vectors satisfying $v(r_{i})=x+m\times r_{i}$ for all sites $r_{i}$ is called the space of rigid body modes.
\end{definition}
\subsection{The Space of Periodic Edge Extensions}
\begin{definition}
    An edge-extension field is periodic if edges that differ by lattice translations are extended by the same amount. The space of periodic edge extensions is denoted by $W$.
\end{definition}

\section{The Compatibility Matrix and Its Induced Maps}
\subsection{The Compatibility Matrix}
\begin{definition}
    The map mapping (generic) node displacements to (generic) edge extensions is called the compatibility matrix. The compatibility matrix is denoted by $C$.
\end{definition}
\begin{prop}
    The compatibility matrix maps $B$ to $W$.
\end{prop}
\begin{definition}
    Elements in the space $C^{-1}(0)\cap B$ are called shape periodic zero modes.
\end{definition}
\begin{prop}\label{prop:rigidbodymodes}
    $N\subset C^{-1}(0)\cap B$.
\end{prop}
Because of Proposition~\ref{prop:rigidbodymodes}, we should ignore the modes in $N$ when studying shape periodic zero modes.
\subsection{Internal Modes, Strain Modes and Bending Modes}
\begin{definition}
    $C$ induces a map $C_{0}: V/\Delta\to W$. The elements in the null space of $C_{0}$ are called internal modes.
\end{definition}
\begin{definition}
    There is a natural embedding $V/\Delta\to A/N$ and $C$ induces a map $C_{1}: A/N\to W$. Elements in the null space of $C_{1}$ that are not in the null space of $C_{0}$ are called strain modes.
\end{definition}
\begin{definition}
    There is a natural embedding $A/N\to B/N$ and $C$ induces a map $C_{2}: B/N\to W$. Elements in the null space of $C_{2}$ that are not in the null space of $C_{1}$ are called bending modes.
\end{definition}

\section{The Generalized Maxwell Counting}
\subsection{The Canonical Decomposition}
\begin{prop}
    Let $X$ be a non-planar periodic triangulated origami, $G$ be its space group and $T$ be its translation group, then $T\subset G$ is a normal subgroup of $G$. $P=G/T$ is called the point group of $X$.
\end{prop}
\begin{prop}
    Define the action of $G$ on the vector field $v$ by
    \begin{equation}
        (gv)(p)=R(g)v(g^{-1}p), \text{ where } R(g)\in O(3) \text{ and } gp=R(g)p+r \text{ for some } r\in\mathbb{R}^{3}.
    \end{equation}
    Then $T$ acts trivially on $W$, $V/\Delta$, $A/N$ and $B/N$. Therefore $W$, $V/\Delta$, $A/N$ and $B/N$ descend to representations of $P=G/T$.
\end{prop}
\begin{proof}
    This can be done by brute force computation.
\end{proof}
\begin{prop}\label{prop:canonicaldecomposition}
    Let the irreducible characters of $P$ be $\{\chi_{i}\}_{i=1}^{h}$, with $\chi_{1}\equiv 1$ be the trivial character. Then for all $S=W, V/\Delta, A/N, B/N$, we have
    \begin{equation}
        S=\oplus_{j=1}^{h} Q^{S}_{j}S,
    \end{equation}
    where the projection operators $Q^{S}_{j}$ are
    \begin{equation}
        Q^{S}_{j}=\frac{\chi_{j}(\mathrm{id})}{|P|}\sum_{g\in P}\chi_{j}(g)^{*}\rho_{S}(g),
    \end{equation}
    where $\rho_{S}$ is the representation of $P$ on $S$.
    We have the identity
    \begin{equation}
        \dim Q^{S}_{j}S=\mathrm{Tr}(Q^{S}_{j})=\frac{\chi_{j}(\mathrm{id})}{|P|}\sum_{g\in P}\chi_{j}(g)^{*}\mathrm{Tr}(\rho_{S}(g)),
    \end{equation}
    $C_{0}$ maps $(V/\Delta)_{j}\equiv Q_{j}^{V/\Delta}(V/\Delta)$ to $W_{j}\equiv Q_{j}^{W}W$, $C_{1}$ maps $(A/N)_{j}\equiv Q_{j}^{A/N}(A/N)$ to $W_{j}$ and $C_{2}$ maps $(B/N)_{j}\equiv Q_{j}^{B/N}(B/N)$ to $W_{j}$.
\end{prop}
\subsection{Trace Computation}
To use the generalized Maxwell counting, we need to compute $\dim Q^{S}_{j}S$ for all $S=W, V/\Delta, A/N, B/N$. It is enough to compute $\mathrm{Tr}(\rho_{S}(g))$ for all $g$. For our origami, we assume the lattice vectors $b_{1}$ and $b_{2}$ are in the $xy$ plane. The matrix $R(g)$ of $g$ under basis $b_{1}, b_{2}, b_{1}\times b_{2}$ (assume they form a right-handed system) is
\begin{equation}\label{eq:operationmatrixdecomposition}
\boxed{
    R(g)=\begin{pmatrix}
        \eta(g)_{11} & \eta(g)_{12} & 0\\
        \eta(g)_{21} & \eta(g)_{22} & 0\\
        0 & 0 & \lambda(g)
    \end{pmatrix}.
}
\end{equation}
\subsubsection{Trace in \texorpdfstring{$\Delta$}{translationmode}}
brute force computation gives
\begin{equation}
\boxed{
\mathrm{Tr}(\rho_{\Delta}(g))=\mathrm{Tr}(R(g))
}
\end{equation}
\subsubsection{Trace in \texorpdfstring{$N$}{rigidbodymode}}
brute force computation gives
\begin{equation}
\boxed{
\mathrm{Tr}(\rho_{N}(g))=(1+\det(R(g)))\mathrm{Tr}(R(g))
}
\end{equation}
\subsubsection{Trace in \texorpdfstring{W}{PeriodicEdgeExtension}}
Let $n_{g}^{e}$ be the number of edges within a unit cell (ignore their orientations) fixed by $g$ up to lattice translations, then 
\begin{equation}
\boxed{
\mathrm{Tr}(\rho_{W}(g))=n^{e}_{g}.
}
\end{equation}
\subsubsection{Trace in \texorpdfstring{$V/\Delta$}{VoverDelta}}
Let $n_{g}^{v}$ be the number of vertices within a unit cell fixed by $g$ up to lattice translations, then 
\begin{equation}
    \mathrm{Tr}(\rho_{V}(g))=\mathrm{Tr}(R(g))n_{g}^{v}.
\end{equation}
Hence, on $V/\Delta$ we have
\begin{equation}
\boxed{
\mathrm{Tr}(\rho_{V/\Delta}(g))=\mathrm{Tr}(R(g))n_{g}^{v}-\mathrm{Tr}(R(g)).
}
\end{equation}
\subsubsection{Trace in \texorpdfstring{$A/N$}{AoverN}}
We have
\begin{equation}
\begin{aligned}
    \mathrm{Tr}(\rho_{A/\Delta}(g))&=\mathrm{Tr}(\rho_{A/V}(g))+\mathrm{Tr}(\rho_{V/\Delta}(g))\\
    \mathrm{Tr}(\rho_{A/N}(g))&=\mathrm{Tr}(\rho_{A/\Delta}(g))-\mathrm{Tr}(\rho_{N/\Delta}(g))=\mathrm{Tr}(\rho_{A/\Delta}(g))-\mathrm{Tr}(\rho_{N}(g))+\mathrm{Tr}(\rho_{\Delta}(g))
\end{aligned}
\end{equation}
The representation $\rho_{A/V}$ is isomorphic to the representation $g\mapsto \eta(g)\otimes R(g)$ as there are two strain vectors $t_{1}$ and $t_{2}$.
Hence we have
\begin{equation}
    \mathrm{Tr}(\rho_{A/V}(g))=\mathrm{Tr}(\eta(g))\mathrm{Tr}(R(g)).
\end{equation}
As a consequence,
\begin{equation}
\boxed{
\mathrm{Tr}(\rho_{A/N}(g))=\mathrm{Tr}(R(g))(\mathrm{Tr}(\eta(g))-\det(R(g))-1)+\mathrm{Tr}(R(g))n_{g}^{v}
}
\end{equation}
\subsubsection{Trace in \texorpdfstring{$B/N$}{BoverN}}
We have
\begin{equation}
    \mathrm{Tr}(\rho_{B/N}(g))=\mathrm{Tr}(\rho_{B/A}(g))+\mathrm{Tr}(\rho_{A/N}(g)).
\end{equation}
brute force computation tells us $\rho_{B/A}$ is isomorphic to the representation $\gamma$ defined by
\begin{equation}
    \gamma(g): H\mapsto\lambda(g)\eta(g^{-1})^{T}H\,\eta(g^{-1}), \text{ where } H \text{ is a } 2\times 2 \text{ symmetric matrix.}
\end{equation}
The character of $\gamma$ is then
\begin{equation}
    \chi_{\gamma}(g)=\lambda(g)(\mathrm{Tr}(\eta(g^{-1}))^{2}-\det(\eta(g^{-1})))
\end{equation}
Since $\eta(g^{-1})=\eta(g)^{-1}$ is similar to some orthogonal matrix, we have 
\begin{equation}
\begin{aligned}
    \mathrm{Tr}(\eta(g^{-1}))&=\mathrm{Tr}(\eta(g)),\\
    \det(\eta(g^{-1}))&=\det(\eta(g))
\end{aligned}
\end{equation}
Hence we have
\begin{equation}
\boxed{
\mathrm{Tr}(\rho_{B/N}(g))=\lambda(g)(\mathrm{Tr}(\eta(g))^{2}-\det(\eta(g)))+\mathrm{Tr}(R(g))(\mathrm{Tr}(\eta(g))-\det(R(g))-1)+\mathrm{Tr}(R(g))n_{g}^{v}.
}
\end{equation}
\subsection{Resummation}\label{sec:designformula}
Now we give different expressions for $\dim(B/N)_{j}$, $\dim(A/N)_{j}$, $\dim(V/\Delta)_{j}$ and $\dim W_{j}$ that are suitable for design. We first introduce some notation.
\begin{definition}\label{def:nSv}
    Let $q\in X$ be a point, the stabilizer $S$ of the point $q$ is defined to be the subgroup of $P$ that only changes the position of $q$ by a translation (including the zero translation) in the translation group. $n_{S}^{v}$ is the number of inequivalent (in the sense of not differing by some translation $t\in T$) vertices with stabilizer $S$.
\end{definition}
\begin{definition}\label{def:nSe}
    Let $e\in E$ be an edge, the stabilizer $S$ of the edge $e$ is defined to be the subgroup of $P$ that changes the position of $e$ (viewed as a non-oriented edge) by a translation (including the zero translation) in the translation group. $n_{S}^{e}$ is the number of inequivalent (in the sense of not differing by some translation $t\in T$) edges with stabilizer $S$.
\end{definition}
With this notation, we first recompute $\dim(V/\Delta)_{j}$, we have
\begin{equation}\label{eq:designprinciple0}
\boxed{ 
\dim(V/\Delta)_{j}=-\frac{\chi_{j}(\mathrm{id})}{|P|}\sum_{s\in P}\chi_{j}(s)^{*}\mathrm{Tr}(R(s))+\frac{\chi_{j}(\mathrm{id})}{|P|}\sum_{S\leq P}n_{S}^{v}\sum_{s\in S}\chi_{j}(s)^{*}\mathrm{Tr}(R(s)).
}
\end{equation}
We then recompute $\dim(A/N)_{j}$,
\begin{equation}\label{eq:designprinciple1}
\boxed{
\begin{aligned}
    \dim(A/N)_{j}=&\frac{\chi_{j}(\mathrm{id})}{|P|}\sum_{s\in P}\chi_{j}(s)^{*}\mathrm{Tr}(R(s))(\mathrm{Tr}(\eta(s))-\det(R(s))-1)\\
    &+\frac{\chi_{j}(\mathrm{id})}{|P|}\sum_{S\leq P}n_{S}^{v}\sum_{s\in S}\chi_{j}(s)^{*}\mathrm{Tr}(R(s))
\end{aligned}
}
\end{equation}
We can also recompute $\dim(B/N)_{j}$,
\begin{equation}\label{eq:designprinciple2}
\boxed{
\begin{aligned}
    \dim(B/N)_{j}=&\frac{\chi_{j}(\mathrm{id})}{|P|}\sum_{s\in P}\chi_{j}(s)^{*}\lambda(s)(\mathrm{Tr}(\eta(s))^{2}-\det(\eta(s)))\\
    &+\frac{\chi_{j}(\mathrm{id})}{|P|}\sum_{s\in P}\chi_{j}(s)^{*}\mathrm{Tr}(R(s))(\mathrm{Tr}(\eta(s))-\det(R(s))-1)\\
    &+\frac{\chi_{j}(\mathrm{id})}{|P|}\sum_{S\leq P}n_{S}^{v}\sum_{s\in S}\chi_{j}(s)^{*}\mathrm{Tr}(R(s))
\end{aligned}
}
\end{equation}
Finally, we recompute $\dim W_{j}$,
\begin{equation}\label{eq:designprinciple3}
\boxed{
\dim W_{j}=\frac{\chi_{j}(\mathrm{id})}{|P|}\sum_{S\leq P}n_{S}^{e}\sum_{s\in S}\chi_{j}(s)^{*}.
}
\end{equation}
To capture the simple periodic local state of stress and the simple periodic local zero modes, we introduce the following two definitions.
\begin{definition}\label{def:nSz}
    Let $q\in X$ be a point supporting local trivial zero modes, the stabilizer $S$ of the point $q$ is defined to be the subgroup of $P$ that only changes the position of $q$ by a translation (including the zero translation) in the translation group. $n_{S}^{z}$ is the number of inequivalent (in the sense of not differing by some translation $t\in T$) points supporting local trivial zero modes with stabilizer $S$.
\end{definition}
\begin{definition}\label{def:nSy}
    Let $J$ be a local coplanar connected substructure with $u$ nodes and $2u-2$ edges. The stabilizer $S$ of the substructure $J$ is defined to be the subgroup of $P$ that only changes the position of $J$ by a translation (including the zero translation) in the translation group. $n_{S}^{y}$ is the number of inequivalent (in the sense of not differing by some translation $t\in T$) substructures supporting local trivial state of self-stress with stabilizer $S$
\end{definition}
The dimension for periodic local zero modes in the $j^{\textrm{th}}$ sector is given by
\begin{equation}\label{eq:designprinciple4}
\boxed{
\dim Z_{j}=\frac{\chi_{j}(\mathrm{id})}{|P|}\sum_{S\leq P}n_{S}^{z}\sum_{s\in S}\chi_{j}(s)^{*}\lambda(s).
}
\end{equation}
The dimension for periodic local state of self-stress in the $j^{\textrm{th}}$ sector is given by
\begin{equation}\label{eq:designprinciple5}
\boxed{
\dim Y_{j}=\frac{\chi_{j}(\mathrm{id})}{|P|}\sum_{S\leq P}n_{S}^{y}\sum_{s\in S}\chi_{j}(s)^{*}.
}
\end{equation}

\subsection{Mode Counting Formulas}
The number of strain modes for generic configurations in the $j^{\textrm{th}}$ sector is given by
\begin{equation}\label{eq:symmetryprotectedstrainmodes}
\boxed{
\begin{aligned}
    N_{j}^{\text{strain}}=&\max\{\dim (A/N)_{j}-\dim Z_{j}-\dim W_{j}+\dim Y_{j},0\}\\&
    -\max\{\dim (V/\Delta)_{j}-\dim Z_{j}-\dim W_{j}+\dim Y_{j},0\}.
\end{aligned}
}
\end{equation}
The number of bending modes for generic configurations in the $j^{\textrm{th}}$ sector, is given by
\begin{equation}\label{eq:symmetryprotectedbendingmodes}
\boxed{
\begin{aligned}
N_{j}^{\text{bending}}=&\max\{\dim (B/N)_{j}-\dim Z_{j}-\dim W_{j}+\dim Y_{j},0\}\\
&-\max\{\dim (A/N)_{j}-\dim Z_{j}-\dim W_{j}+\dim Y_{j},0\}.
\end{aligned}
}
\end{equation}

\section{Nonlinear Strain Modes}
We prove a theorem on the nonlinear extension of symmetry-protected strain modes in this section.

\subsection{The Duality Map}
\begin{prop}\label{prop:dualitymap}
    Define $\kappa: \ker(C_{2})\to W$ to be a map defined by the following property: for all $v\in \ker(C_{2})$, $\kappa(v)$ is the family of bond extensions such that the extension on the $n^{\textrm{th}}$ bond equals the change of dihedral angle associated to that bond when the system undergoes the deformation $v$. Then this map $\kappa$, called the duality map, is an injective linear map, and $\ker(C_{0}^{T})=\mathrm{Im}(\kappa)$.
\end{prop}
\begin{proof}
    This is a standard result~\cite{JamesPNAS2020}. The key point is that the inverse of $\kappa$ from $\mathrm{Im}(\kappa)$ to $\ker(C_{0}^{T})$ can be defined by converting the state of self-stress into the folding angles in a similar way. The resulting folding angle lies in $\ker(C_{2})$ and the statements follow.
\end{proof}

\begin{prop}\label{prop:equavarienceofthedualitymap}
    $\kappa$ is a $\lambda$-twisted intertwiner of representations between $\ker(C_{0}^{T})$ and $\ker(C_{2})$. In other words, $\kappa$ is an isomorphism as linear map between $\ker(C_{0}^{T})$ and $\ker(C_{2})$ and
    \begin{equation}\label{eq:equavarienceofthedualitymap}
        \rho_{\ker(C_{0}^{T})}(g)\circ\kappa=\lambda(g)\kappa\circ\rho_{\ker(C_{2})}(g), \forall g\in G/T,
    \end{equation}
    where $\lambda(g)=\pm 1$ is defined in Eq.\eqref{eq:operationmatrixdecomposition}.
\end{prop}
\begin{proof}
    $\kappa$ is clearly a linear isomorphism between $\ker(C_{0}^{T})$ and $\ker(C_{2})$. To prove Eq.~\eqref{eq:equavarienceofthedualitymap}, we only need to prove $\rho_{W}(g)(\kappa(v))=\lambda(g)\kappa(\rho_{B/N}(g)(v))$ for all $v\in \ker(C_{2})$. The equation manifests itself when tracing back to the physical meaning of its left- and right- hand sides.
\end{proof}

\noindent Proposition~\ref{prop:equavarienceofthedualitymap} tells us that the irreducible decomposition of the state of self-stress space $\ker(C_{0}^{T})$ and the shape periodic zero modes $\ker(C_{2})$ with respect to the point group $G/T$ are related by the irreducible representation $\lambda$. The duality map maps a shape periodic zero mode in $\chi_{j}$ sector to a state of self-stress in the irreducible sector whose character maps $g\in G/T$ to $\lambda(g)\chi_{j}(g)$.

\subsection{A Nonlinear Extension Theorem}
Here we give some sufficient conditions that guarantee the nonlinear extension of strain modes.
\begin{theorem}\label{thm:nonlinearstrainmodes1}
    Let $G$ be a layer group, $X$ be a periodic triangulated origami symmetric under $G$. Let $\chi_{1}$ be the trivial representation of $G/T$. If $\dim(A/N)_{1}-\dim W_{1}\geq 1$ equals the number of fully symmetric zero modes in the $(A/N)_{1}$ sector of $X$, then $X$ has at least one strain mode and its strain modes extend nonlinearly to at least second order.
\end{theorem}

\begin{proof}
    Since $\dim(A/N)_{1}-\dim W_{1}\geq 1$ equals the number of fully symmetric strain modes in the $\chi_{1}$ sector, $C_{1}$ has at least a linear zero mode $v\in (A/N)_{1}$. We move $X$ infinitesimally along $v$, since $v$ is a linear zero mode of $C_{1}$ and $v\in (A/N)_{1}$, this infinitesimal motion can at most induce a second order bond extension $r\in W_{1}$. The assumption $\dim(A/N)_{1}-\dim W_{1}\geq 1$ equals the number of fully symmetric strain modes in the $\chi_{1}$ sector further tells us that there is a seconder-order motion $u\in (A/N)_{1}$ such that $C_{1}u=-r$. This deformation $v+u$ therefore does not have any bond extensions up to second order. Since the deformations $u$, $v$ and the structure $X$ are all symmetric under $G$, the new deformed structure is also symmetric under $G$. The argument above can be applied repeatedly near the starting configuration as $C_{1}$ being full rank when viewed as a map from $(A/N)_{1}$ to $W_{1}$ is an open condition. Apply the argument above repeatedly we get a nonlinear strain mode.
\end{proof}

It is worth commenting that it is enough to check the condition in Theorem~\ref{thm:nonlinearstrainmodes1} at only the configuration $X$, and the condition can be checked by computing the null space dimension of $C_{1}$ in $(A/N)_{1}$. Next we give some physical situations when the condition in Theorem~\ref{thm:nonlinearstrainmodes1} holds.

\begin{theorem}\label{thm:nonlinearstrainmodes2}
    Let $G$ be a layer group containing an inversion or a glide-$z$-reflection and $X$ be a periodic triangulated origami with layer group $G$. Let $\chi_{1}$ be the trivial representation of $G/T$. If we have the following two equations
    \begin{equation}\label{eq:thmnsmcondition}
    \dim(A/N)_{1}-\dim W_{1}\geq 1,\; \ker(C_{0})=0,
    \end{equation}
    then $X$ has a strain mode and this strain mode extends nonlinearly to at least second order.
\end{theorem}

To prove Theorem~\ref{thm:nonlinearstrainmodes2}, we first provide some lemmas.

\begin{lemma}\label{lemma:modedifferenceequals3}
    We have the following equation:
    \begin{equation}\label{eq:modedifferenceequals3}
        \dim(\ker(C_{2}))-\dim(\ker(C_{0}))=3.
    \end{equation}
\end{lemma}
\begin{proof}
    From Proposition~\ref{prop:equavarienceofthedualitymap}, we know $\dim(\ker(C_{2}))=\dim(\ker(C_{0}^{T}))$. The rank nullity theorem and the fact that $\dim(W)-\dim(V/\Delta)=3$ together give Eq.~\eqref{eq:modedifferenceequals3}.
\end{proof}

\begin{lemma}\label{lemma:C1surjectivityinthetrivialsector1}
    Let $r\in W$ be a bond extension in the $\chi_{1}$ sector. If $\ker(C_{0})=0$, the layer group $G$ contains an inversion or a glide-$z$-reflection, and $\dim(A/N)_{1}-\dim W_{1}\geq 1$, then the equation $C_{1}x=r$ has at least one solution in $(A/N)_{1}$.
\end{lemma}
\begin{proof}
    $\ker(C_{0})=0$ and Lemma~\ref{lemma:modedifferenceequals3} together give $\dim(\ker(C_{2}))=3$. Since $\dim(B/N)-\dim(W)=3$, $C_{2}$ is surjective. Hence there is a $v\in B/N$ such that $C_{2}v=r$. Since $r\in W$ lies in the $\chi_{1}$ sector, we see $v\in (B/N)_{1}$. If $v\notin A/N$, $v$ induces bending of $X$ and $v$ cannot be kept invariant by an inversion nor a glide-$z$-reflection, which contradicts the fact that $v$ lies in the $\chi_{1}$ sector. Hence $v\in A/N$. Since $r\in W_{1}$ and $C_{1}v=r$, we conclude that $v\in (A/N)_{1}$. Hence the equation $C_{1}x=r$ has at least one solution in $(A/N)_{1}$.
\end{proof}

Now we turn to the proof of Theorem~\ref{thm:nonlinearstrainmodes2}.
\begin{proof}
    The proof is almost the same as the proof for Theorem~\ref{thm:nonlinearstrainmodes1}. The only difference is the last part proving the argument can be applied repeatedly. This time, it follows from the fact that $\ker(C_{0})=0$ is also an open condition.
\end{proof}

We comment here that Lemma~\ref{lemma:C1surjectivityinthetrivialsector1} fails when the only operation in $G$ reversing the out-of-plane direction is the glide-rotation. This is because the glide-rotation preserves an off-diagonal second fundamental form. For layer groups whose only operation reversing the out-of-plane direction is the glide-rotation, Lemma~\ref{lemma:C1surjectivityinthetrivialsector1} can be replaced by Lemma~\ref{lemma:C1surjectivityinthetrivialsector2}.

\begin{lemma}\label{lemma:C1surjectivityinthetrivialsector2}
    Let $r\in W$ be a bond extension in the $\chi_{1}$ sector. Suppose $\ker(C_{0})=0$, the only operation reversing the out-of-plane direction in layer group $G$ is the glide-rotation, and $\dim(A/N)_{1}-\dim W_{1}\geq 1$. If we further assume that the structure has a zero mode $h$ in $(B/N)_{1}$ with non-zero second fundamental form or $\dim(B/N)_{1}=\dim(A/N)_{1}$, then the equation $C_{1}x=r$ has at least one solution in $(A/N)_{1}$.
\end{lemma}

\begin{proof}
    $\ker(C_{0})=0$ and Lemma~\ref{lemma:modedifferenceequals3} together give $\dim(\ker(C_{2}))=3$. Since $\dim(B/N)-\dim(W)=3$, $C_{2}$ is surjective. Hence there is a $v\in B/N$ such that $C_{2}v=r$. Since $r\in W$ lies in the $\chi_{1}$ sector, we see $v\in (B/N)_{1}$. Hence the second fundamental form of $v$ is off-diagonal. 

    If $\dim(B/N)_{1}=\dim(A/N)_{1}$, then $(B/N)_{1}=(A/N)_{1}$ and we arrive at the conclusion $v\in (A/N)_{1}$.
    
    If the structure has a zero mode $h$ in $(B/N)_{1}$ with non-zero second fundamental form, that second fundamental form must also be off-diagonal. Therefore, there is a constant $\alpha\in\mathbb{R}$ such that $v+\alpha h$ has vanishing second fundamental form and hence $v+\alpha h\in A/N$. We have $C_{1}(v+\alpha h)=C_{2}(v+\alpha h)=C_{2}v=r$. Since $r\in W_{1}$ and $C_{1}(v+\alpha h)=r$, we conclude that $(v+\alpha h)\in (A/N)_{1}$. Hence the equation $C_{1}x=r$ has at least one solution in $(A/N)_{1}$.
\end{proof}

We comment here that the extra condition in Lemma~\ref{lemma:C1surjectivityinthetrivialsector2} is very weak. If $\dim(B/N)_{1}>\dim(A/N)_{1}$, the structure has at least one bending mode in the symmetric sector, and we need to fine tune the structure parameters to ensure such a bending mode has zero second fundamental form.

\section{A Simple Proof for The Mode Compatibility Condition}
In this section we give a simpler proof for the mode compatibility condition
\begin{equation}\label{eq:EHrelations}
\begin{aligned}
    \mathrm{Tr}(H^{a}\mathrm{adj}&(E^{b})-H^{b}\mathrm{adj}(E^{a}))=\epsilon^{\alpha\mu}\epsilon^{\beta\nu}\Big(H^a_{\mu\nu}E^b_{\alpha\beta}-H^b_{\mu\nu}E^a_{\alpha\beta}\Big)=0,\\
    &\text{ for any two shape periodic zero modes }a,b.
\end{aligned}
\end{equation}

using our shape periodic ansatz Eq.~\eqref{eq:shapeperiodicansatz}, the assumption that the origami is triangulated and the fact that the mode is a zero mode.
The zero--mode (no edge extension) condition gives us
\begin{equation}\label{eq:zeromodecondition}
\Delta x_e\cdot \Delta v_e=0
\qquad\text{for every edge }e=(p\to q),\ 
\Delta x_e:=x_q-x_p,\ \Delta v_e:=v_q-v_p,
\end{equation}

\subsection{Step 0: Macroscopic \texorpdfstring{$(E,H)$}{(E,H)} From The Shape Periodic Polynomial Data.}
Let $\ell_1:=\|b_1\|$, $\ell_2:=\|b_2\|$, and define lattice differences
\[
d_1 f:=\frac{f(n_1+1,n_2)-f(n_1,n_2)}{\ell_1},
\qquad
d_2 f:=\frac{f(n_1,n_2+1)-f(n_1,n_2)}{\ell_2}.
\]
A direct finite--difference computation from Eq.~\eqref{eq:shapeperiodicansatz} yields constants
\[
d_1d_1 v=\frac{m_1\times b_1}{\ell_1^2},\qquad
d_2d_2 v=\frac{m_2\times b_2}{\ell_2^2},\qquad
d_1d_2 v=\frac{m_2\times b_1+m_1\times b_2}{2\ell_1\ell_2}.
\]
Hence, using the definition in Ref.~\cite{sun2025geometric} $H_{\alpha\beta}:=\langle d_\alpha d_\beta v,\hat z\rangle$,
\begin{equation}\label{eq:Hexpression}
H_{11}=\frac{(m_1\times b_1)\cdot \hat z}{\ell_1^2},\quad
H_{22}=\frac{(m_2\times b_2)\cdot \hat z}{\ell_2^2},\quad
H_{12}=\frac{(m_2\times b_1+m_1\times b_2)\cdot \hat z}{2\ell_1\ell_2}.
\end{equation}
Likewise, the macroscopic strain $E_{\alpha\beta}$ (defined by the paper's formula with the $+H_{\alpha\beta}x_z$
term) depends only on the translation parts $t_1,t_2$:
\begin{equation}\label{eq:Eexpression}
E_{11}=\frac{t_1\cdot b_1}{\ell_1^2},\quad
E_{22}=\frac{t_2\cdot b_2}{\ell_2^2},\quad
E_{12}=\frac12\Big(\frac{t_1\cdot b_2}{\ell_1\ell_2}+\frac{t_2\cdot b_1}{\ell_1\ell_2}\Big).
\end{equation}
Thus $(m_1,m_2)$ controls $H$ and $(t_1,t_2)$ controls $E$ at the macroscopic level.

\subsection{Step 1: Triangle-Wise Rigid Motion.}
\begin{lemma}\label{lemma:trianglerigidity}
    Let $\triangle(pqr)$ be any nondegenerate triangle in the triangulation. If Eq.~\eqref{eq:zeromodecondition} holds on its three edges,
    then there exist vectors $\Omega_\triangle,c_\triangle\in\mathbb R^3$ such that
    \begin{equation}\label{eq:trianglerigidity}
    v_p=\Omega_\triangle\times x_p+c_\triangle,\qquad
    v_q=\Omega_\triangle\times x_q+c_\triangle,\qquad
    v_r=\Omega_\triangle\times x_r+c_\triangle.
\end{equation}
\end{lemma}

\subsection{Step 2: A Closed Edge--Sum Identity for Two Zero Modes.}
Take two isometric SP modes $a,b$. On each triangle $\triangle$ define the corresponding rotation vectors
$\Omega_\triangle^a,\Omega_\triangle^b$ via Eq.~\eqref{eq:trianglerigidity}.
For each oriented edge $e=(p\to q)$ define the scalar
\begin{equation}
\eta_e^{ab}:=\Omega^a_\triangle\cdot (v_q^b-v_p^b)\;-\;\Omega^b_\triangle\cdot (v_q^a-v_p^a),
\end{equation}
where $\triangle$ is either triangle incident to $e$.

\medskip
\noindent\textbf{(i) Well-definedness.}
If $e$ is shared by triangles $\triangle,\triangle'$, the two rigid representations imply
\[
(\Omega_\triangle-\Omega_{\triangle'})\times \Delta x_e=0\ \Rightarrow\ 
(\Omega_\triangle-\Omega_{\triangle'})\parallel \Delta x_e.
\]
Adding a multiple of $\Delta x_e$ to $\Omega$ does not change $\eta_e^{ab}$ because $\Delta x_e\cdot\Delta v_e=0$ for both zero modes. Hence $\eta_e^{ab}$ is independent of the choice of incident triangle.

\medskip
\noindent\textbf{(ii) Zero curl on each triangle.}
Summing around the oriented boundary of $\triangle(pqr)$, we have
\begin{equation}\label{eq:closedness}
\oint_{\Delta(pqr)}\eta^{ab}=\eta_{p\to q}^{ab}+\eta_{q\to r}^{ab}+\eta_{r\to p}^{ab}=\Omega_\triangle^a\cdot\big[(v_q^b-v_p^b)+(v_r^b-v_q^b)+(v_p^b-v_r^b)\big]-(a\leftrightarrow b)=0.
\end{equation}
Thus $\eta^{ab}$ is a \emph{closed discrete 1-form} on the doubly periodic triangulation.

\subsection{Step 3: The Mode Compatibility Condition}
On one hand, we can pin down an orientation of this origami structure and obtain
\begin{equation}\label{eq:discretestokes}
\oint_{\partial\mathrm{cell}} \eta^{ab}=\sum_{\Delta\in\mathrm{cell}}\oint_{\Delta} \eta^{ab}=0.
\end{equation}
On the other hand, the left hand side of Eq.~\eqref{eq:discretestokes} can be computed using Eq.~\eqref{eq:shapeperiodicansatz} (compute contributions from pairs that differ by a lattice vector). Combining Eq.~\eqref{eq:Hexpression} and Eq.~\eqref{eq:Eexpression}, we arrive at the mode compatibility condition
\[
\mathrm{Tr}(H^{a}\mathrm{adj}(E^{b})-H^{b}\mathrm{adj}(E^{a}))=\epsilon^{\alpha\mu}\epsilon^{\beta\nu}\Big(H^a_{\mu\nu}E^b_{\alpha\beta}-H^b_{\mu\nu}E^a_{\alpha\beta}\Big)=0.
\]

\chapter{Reduction}
Here we introduce a useful trick for completely solving the compatibility problem between $n_{S}^{v}$, $n_{S}^{e}$, $n_{S}^{y}$ and $n_{S}^{z}$ introduced  in Section~\ref{sec:designformula}. We focus only on the origami that can be described by $(x,y,f(x,y))$ for some function $f$. Assuming the origami has some type I origami group as its space group, we first introduce some useful notation.
\section{Origami Group}
\begin{figure}[ht]\label{fig:typeIIorigamigrouplocalenvironment}
\centering
\includegraphics[width=0.5\textwidth]{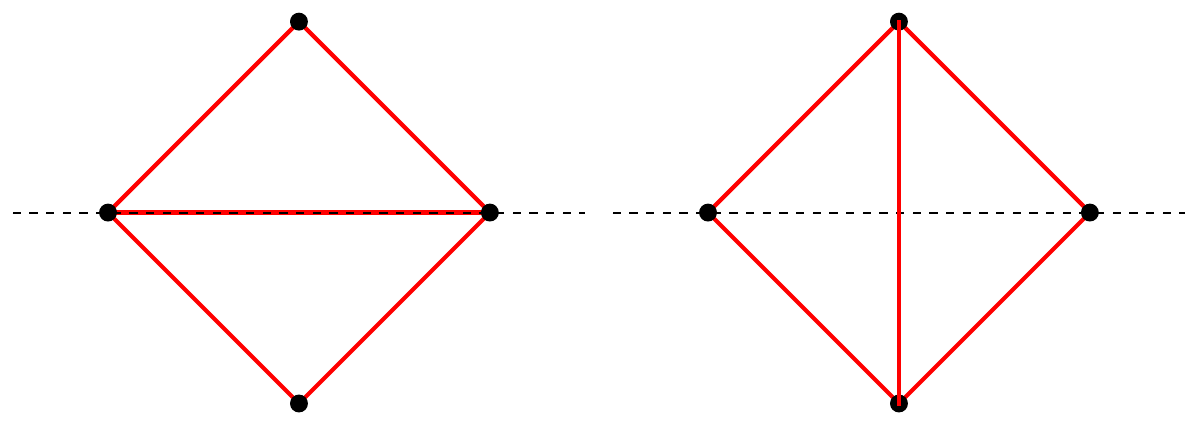}
\caption{Schematics of local connections around an in-plane rotation axis. The dashed line is the rotation axis}
\end{figure}

We now give a classification of the space groups of triangulated origami.
\begin{prop}\label{prop:oragamigroup}
    $G$ is the space group of some triangulated origami iff $G$ is a layer group with $M_{z}\notin G$.
\end{prop}
\noindent Now we give some useful definitions.
\begin{definition}
    An origami group is a layer group $G$ with $M_{z}\notin G$.
\end{definition}
\begin{definition}\label{def:typeIorigamigroup}
    An origami group $G$ is of type I if $\forall\;\hat{n}\in\mathbb{R}^{3}$ with $\hat{n}_{z}=0$, $R_{\hat{n}}(\pi)\notin G$.
\end{definition}
\begin{definition}
    An origami group $G$ is of type II if $\exists\;\hat{n}\in\mathbb{R}^{3}$ with $\hat{n}_{z}=0$ such that $R_{\hat{n}}(\pi)\in G$.
\end{definition}
\begin{prop}\label{prop:typeIIisboring}
    Let $G$ be the space group of a triangulated origami $\{X,E\}$. If $R_{\hat{n}}(\pi)\in G$ for some $\hat{n}_{z}=0$, then $X$ has a nonlinear zero mode localized around the line fixed by $R_{\hat{n}}(\pi)$.
\end{prop}
\begin{proof}
    Fig.~\ref{fig:typeIIorigamigrouplocalenvironment} shows the only possible local configurations around the rotation axis. A moment's thought reveals that there is a nonlinear zero mode localized around the rotation axis.
\end{proof}
\noindent Proposition~\ref{prop:typeIIisboring} says origami whose space group is a type II origami group can be folded nonlinearly in a trivial way. Such a folding mechanism is harmful for practical applications. \textbf{Hence, we only focus on origami whose space group is a type I origami group.} 

\section{The Relative Position of High-Symmetry Points Within an origami sheet}
\begin{definition}
    A high-symmetry point $p$ with stabilizer $S_{p}$ is said to be of discrete type if there exists some $r>0$ such that if $q$ is another point with stabilizer $S_{p}$ in the ball centered at $p$ with radius $r$, the $(x,y)$ coordinates of $p$ and $q$ are the same. Otherwise, this high-symmetry point is of continuous type.
\end{definition}
\begin{prop}
    A high-symmetry point $p$ is of continuous type if and only if its stabilizer $S_{p}$ equals $\{\mathrm{id}, M\}$, where $M$ is a mirror perpendicular to the $xy$ plane.
\end{prop}
\noindent The following Proposition gives a classification of possible positions a high-symmetry point can have in $X$.
\begin{prop}\label{prop:highsymmetrypointpositionclassification}
    Let $X$ be an origami whose space group is a type I origami group containing an operation $Q$, $q$ be a high-symmetry point fixed by at least $Q$. We have the following cases:\\
    Case 1. $Q$ is an inversion. Then $q$ must either be occupied by an end point of some edge in $X$ or the midpoint of some edge in $X$.\\
    Case 2. $Q$ equals $M_{z}R_{z}(\pi/2)$, or $M_{z}R_{z}(\pi/3)$. Then $q$ must be an end point of some edge in $X$.\\
    Case 3. $Q$ equals $R_{z}(\pi/2)$, or $R_{z}(\pi/3)$. Then there is an end point $p$ of some edge in $X$ that is invariant under $Q$ such that $p$ and $q$ have the same $(x,y)$ coordinates.\\
    Case 4. $Q=R_{z}(\pi)$. Then there is an edge in $X$ whose midpoint or end point has the same $(x,y)$ coordinates as that of $q$.\\
    Case 5. $Q=M$ is a mirror perpendicular to the $xy$ plane. If there is some point $p$ on some edge of $X$ having the same $(x,y)$ coordinates as $q$, $p$ must be an end point or the midpoint of that edge.\\
    Case 6. $Q=R_{z}(2\pi/3)$. If $q$ is not in the interior of any triangle in $X$, there is an end point $p$ of some edge in $X$ that is invariant under $Q$ such that $p$ and $q$ has the same $(x,y)$ coordinates.
\end{prop}
\begin{proof}
    For case 1, if $q$ is neither an end point of some edge nor the midpoint of some edge, we can find a triangle $T$ in $X$ containing $q$. Then $T$ has to be invariant under $Q$, which is a contradiction. The fact that the origami is described by $(x,y,f(x,y))$ for some function $f$ then implies that $q$ must be on some edge. If $q$ is not the midpoint or an end point of the edge $e$, transform edge $e$ by $Q$, we see $q$ is indeed a node, which is a contradiction. Case 2-6 can be proved similarly.
\end{proof}
\begin{prop}\label{prop:localenvironementoforder4edgestabilizer}
    Let $p$ be a high-symmetry point whose stabilizer is $S_{p}=\{\mathrm{id},M, R_{z}(\pi), MR_{z}(\pi)\}$, then $p$ must be occupied by an edge center or a vertex.
\end{prop}
\section{Edge Symmetry Classification}
\begin{prop}\label{prop:possiblemidpointsymmetries}
    Let $p$ with stabilizer $S_{p}$ be the midpoint of an edge $e$ whose stabilizer is $S_{e}$ (which also means $p$ is not an end point of any edge). Then $S_{p}=S_{e}$ and the possible non-trivial $S_{p}$ are\\
    1. $S_{p}=\{\mathrm{id},I\}$, where $I$ is an inversion.\\
    2. $S_{p}=\{\mathrm{id},M\}$, where $M$ is a mirror perpendicular to the $xy$ plane.\\
    3. $S_{p}=\{\mathrm{id}, R_{z}(\pi)\}$, where $R_{z}(\pi)$ is the rotation about the $z$-axis by $\pi$.\\
    4. $S_{p}=\{\mathrm{id},M, R_{z}(\pi), MR_{z}(\pi)\}$.
\end{prop}
\begin{proof}
    It is clear that $S_{e}\subset S_{p}$. If $S_{e}\ne S_{p}$, the extra operation would make $p$ the intersection of two edges and $p$ is then a vertex, which is a contradiction.  The resulting classifications are the consequence of the fact that the maximum possible symmetry group of an edge in $X$ must be a subgroup of the abelian group generated by three mutually perpendicular mirrors $M_{1}, M_{2}, M_{z}$, along with the fact that $M_{z}$ and $R_{\hat{n}}(\pi)$ cannot be a symmetry of our origami.
\end{proof}
\section{Reduction of The First Type: Refinement}
\subsection{Triangle Refinement}
The following theorem shows that for an origami $X$ whose space group $G$ is a layer group with $M_{z}\notin G$, if some high-symmetry point lies in the interior of some triangle in $X$, we can add them to $X$ without changing $\dim(V/\Delta)_{j}-\dim Z_{j}-\dim W_{j}+\dim Y_{j}$. Since $\dim(A/N)_{j}-\dim(V/\Delta)_{j}$ and $\dim(B/N)_{j}-\dim(V/\Delta)_{j}$ depend only on the layer group, $\dim(A/N)_{j}-\dim Z_{j}-\dim W_{j}+\dim Y_{j}$ and $\dim(B/N)_{j}-\dim Z_{j}-\dim W_{j}+\dim Y_{j}$ are thus unchanged if $\dim(V/\Delta)_{j}-\dim Z_{j}-\dim W_{j}+\dim Y_{j}$ is kept the same.
\begin{theorem}\label{thm:refinementinthebulk}
    Let $X$ be an origami with space group $G$. Suppose $G$ is a type I origami group with point group $P=G/T$. If $p$ is a point lying in the interior of some triangle $J$ in $X$, then we can refine $X$ by putting one extra node $p$ (together with its images under $G$) into $X$ and edges between $p$ and the vertices of $J$ (together with their images under $G$). This procedure keeps $P$ to be the point group of the refined structure. If the vertices of the original triangle are not inversion centers, $\dim(A/N)_{j}-\dim Z_{j}-\dim W_{j}+\dim Y_{j}$ and $\dim(V/\Delta)_{j}-\dim Z_{j}-\dim W_{j}+\dim Y_{j}$ are unchanged during this process. 
\end{theorem}
\begin{figure}[ht]
\centering
\includegraphics[width=0.5\textwidth]{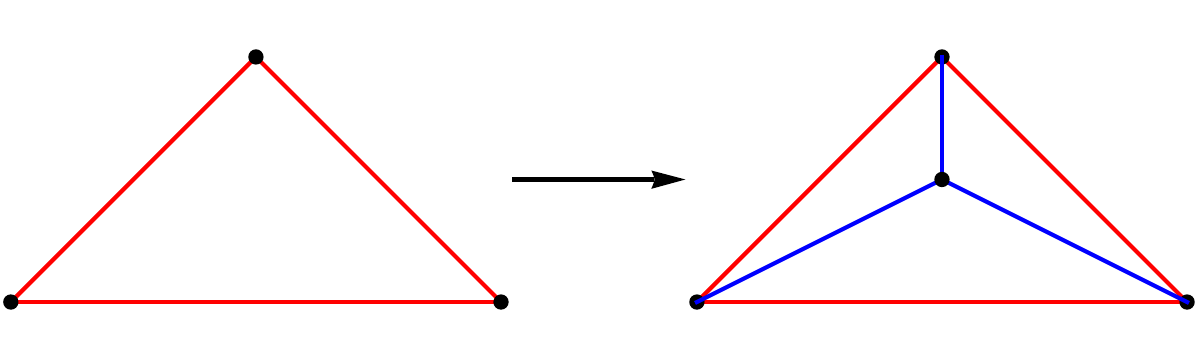}
\caption{Schematic of triangle refinement.}
\end{figure}
\begin{proof}
    Proposition~\ref{prop:highsymmetrypointpositionclassification} indicates we only have the following possibilities.\\
    Case 1: $S_{p}$ is the trivial group. Then in general we have $n_{\mathrm{id}}^{v}\to n_{\mathrm{id}}^{v}+|P|$, $n_{\mathrm{id}}^{e}\to n_{\mathrm{id}}^{e}+3|P|$. Then $\dim(A/N)_{j}\to \dim(A/N)_{j}+3\chi_{j}(\mathrm{id})\chi_{j}(\mathrm{id})^{*}$, $\dim(V/\Delta)_{j}\to \dim(V/\Delta)_{j}+3\chi_{j}(\mathrm{id})\chi_{j}(\mathrm{id})^{*}$, $\dim Z_{j}\to \dim Z_{j}$, $\dim Y_{j}\to \dim Y_{j}$, $\dim W_{j}\to \dim W_{j}+3\chi_{j}(\mathrm{id})\chi_{j}(\mathrm{id})^{*}$.\\
    Case 2: $S_{p}$ is generated by a reflection $\sigma$. Then in general we have $n_{S_{p}}^{v}\to n_{S_{p}}^{v}+|P|/2$, $n_{\mathrm{id}}^{e}\to n_{\mathrm{id}}^{e}+|P|$, $n_{S_{p}}^{e}\to n_{S_{p}}^{e}+|P|/2$. Then $\dim(A/N)_{j}\to \dim(A/N)_{j}+\chi_{j}(\mathrm{id})(3\chi_{j}(\mathrm{id})^{*}+\chi_{j}(\sigma)^{*})/2$, $\dim(V/\Delta)_{j}\to \dim(V/\Delta)_{j}+\chi_{j}(\mathrm{id})(3\chi_{j}(\mathrm{id})^{*}+\chi_{j}(\sigma)^{*})/2$, $\dim Z_{j}\to \dim Z_{j}$, $\dim Y_{j}\to \dim Y_{j}$, $\dim W_{j}\to \dim W_{j}+\chi_{j}(\mathrm{id})(3\chi_{j}(\mathrm{id})^{*}+\chi_{j}(\sigma)^{*})/2$.\\
    Case 3: $S_{p}$ is generated by a threefold rotation $\tau$. Then in general we have $n_{S_{p}}^{v}\to n_{S_{p}}^{v}+|P|/3$, $n_{\mathrm{id}}^{e}\to n_{\mathrm{id}}^{e}+|P|$. $\dim(A/N)_{j}\to \dim(A/N)_{j}+\chi_{j}(\mathrm{id})\chi_{j}(\mathrm{id})^{*}$, $\dim(V/\Delta)_{j}\to \dim(V/\Delta)_{j}+\chi_{j}(\mathrm{id})\chi_{j}(\mathrm{id})^{*}$, $\dim Z_{j}\to \dim Z_{j}$, $\dim Y_{j}\to \dim Y_{j}$, $\dim W_{j}\to \dim W_{j}+\chi_{j}(\mathrm{id})\chi_{j}(\mathrm{id})^{*}$. Hence $N_{\text{strain}}$ is unchanged.\\
    Case 4: $S_{p}$ is generated by $\sigma$ and $\tau$, where $\sigma$ is a reflection and $\tau$ is a three fold rotation about the $z$-axis. Let $S_{\sigma}$ be the group generated by $\sigma$. Then in general we have $n_{S_{p}}^{v}\to n_{S_{p}}^{v}+|P|/6$, $n_{S_{\sigma}}^{e}\to n_{S_{\sigma}}^{e}+|P|/2$. $\dim(A/N)_{j}\to \dim(A/N)_{j}+\chi_{j}(\mathrm{id})(\chi_{j}(\mathrm{id})^{*}+\chi_{j}(\sigma)^{*})/2$, $\dim(V/\Delta)_{j}\to \dim(V/\Delta)_{j}+\chi_{j}(\mathrm{id})(\chi_{j}(\mathrm{id})^{*}+\chi_{j}(\sigma)^{*})/2$, $\dim Z_{j}\to \dim Z_{j}$, $\dim Y_{j}\to \dim Y_{j}$, $\dim W_{j}\to \dim W_{j}+\chi_{j}(\mathrm{id})(\chi_{j}(\mathrm{id})^{*}+\chi_{j}(\sigma)^{*})/2$.\\
    Combine all the four possibilities, we see that $\dim(A/N)_{j}-\dim Z_{j}-\dim W_{j}+\dim Y_{j}$ and $\dim(V/\Delta)_{j}-\dim Z_{j}-\dim W_{j}+\dim Y_{j}$ are unchanged during this process. 
\end{proof}

\subsection{Edge Refinement}
\noindent We show that after certain edge refinement, the stabilizer of both end points of an edge can be made as small as possible.
\begin{theorem}\label{thm:refinementonnonsymmetricedge}
    Let $X$ be an origami with space group $G$, $e$ be an edge in $X$, the triangles containing $e$ be $J$ and $L$. Suppose $G$ is an origami group of type I, whose point group is $P=G/T$. If $e$ does not lie entirely on some mirror plane, we can find two non-symmetric points $p$ and $q$ on $e$ such that the edge $pq$ has the same symmetry as that of $e$. Join $p$ and $q$ (and their images under $G$), then join $p$ and $q$ with the other vertices of $J$ and $L$ (and their images under $G$), the new structure has point group $P$ as well. $\dim(A/N)_{j}-\dim Z_{j}-\dim W_{j}+\dim Y_{j}$ and $\dim(V/\Delta)_{j}-\dim Z_{j}-\dim W_{j}+\dim Y_{j}$ are unchanged during this process. 
\end{theorem}
\begin{figure}[ht]
\centering
\includegraphics[width=0.5\textwidth]{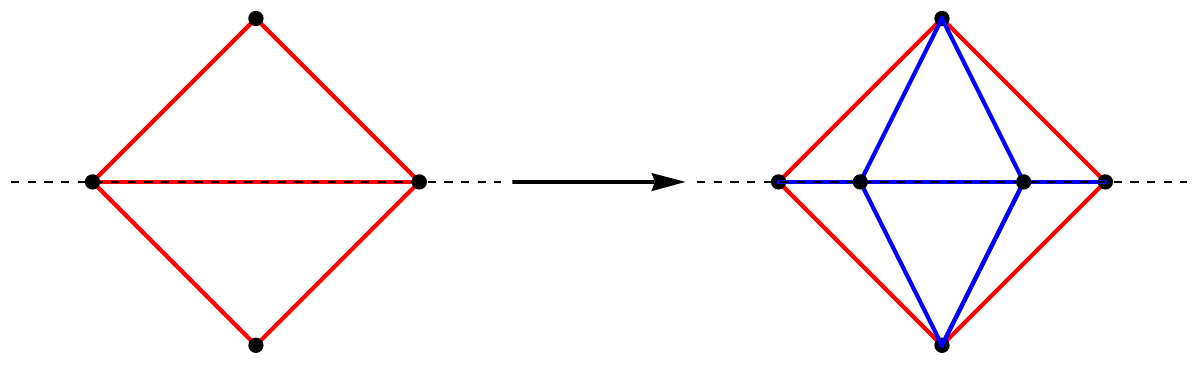}
\caption{Schematic of edge refinement for edges that do not lie entirely in any mirror and any rotation axis. The dashed line indicates that mirror plane or that rotation axis.}
\end{figure}
\begin{proof}
    Case 1. $S_{e}=\{\mathrm{id}\}$, or $S_{e}=\{\mathrm{id},M\}$, or $S_{e}=\{\mathrm{id}, R_{z}(\pi)\}$. The operations involved in these groups do not guarantee co-planarity. Hence, in general, we have $n_{\mathrm{id}}^{v}\to n_{\mathrm{id}}^{v}+2|P|$, $n_{\mathrm{id}}^{e}\to n_{\mathrm{id}}^{e}+6|P|$. $\dim(A/N)_{j}\to \dim(A/N)_{j}+6\chi_{j}(\mathrm{id})\chi_{j}(\mathrm{id})^{*}$, $\dim(V/\Delta)_{j}\to \dim(V/\Delta)_{j}+6\chi_{j}(\mathrm{id})\chi_{j}(\mathrm{id})^{*}$, $\dim W_{j}\to \dim W_{j}+6\chi_{j}(\mathrm{id})\chi_{j}(\mathrm{id})^{*}$, $\dim Z_{j}\to \dim Z_{j}$, and $\dim Y_{j}\to \dim Y_{j}$.\\
    Case 2. $S_{e}=\{\mathrm{id},I\}$. Note $I$ guarantees co-planarity, special attention is needed to this case. $\dim(A/N)_{j}\to \dim(A/N)_{j}+3\chi_{j}(\mathrm{id})\chi_{j}(\mathrm{id})^{*}$, $\dim(V/\Delta)_{j}\to \dim(V/\Delta)_{j}+3\chi_{j}(\mathrm{id})\chi_{j}(\mathrm{id})^{*}$, $\dim W_{j}\to \dim W_{j}+3\chi_{j}(\mathrm{id})\chi_{j}(\mathrm{id})^{*}$, $\dim Z_{j}\to \dim Z_{j}+1$, and $\dim Y_{j}\to \dim Y_{j}+1$.\\
    Combining the two cases above, we see that $\dim(A/N)_{j}-\dim Z_{j}-\dim W_{j}+\dim Y_{j}$ and $\dim(V/\Delta)_{j}-\dim Z_{j}-\dim W_{j}+\dim Y_{j}$ are unchanged during this process. 
\end{proof}
\begin{theorem}\label{thm:refinementonsymmetricedge}
    Let $X$ be an origami with space group $G$, $e$ be an edge in $X$, the triangles containing $e$ be $J$ and $L$. Suppose $G$ is a origami group of type I, whose point group is $P=G/T$. If $e$ lies entirely on some mirror plane generating its stabilizer $S_{e}$, we can find two symmetric points $p$ and $q$ on $e$ with $S_{p}=S_{q}=S_{e}$ such that the edge $pq$ has the same symmetry as that of $e$. Join $p$ and $q$ (and their images under $G$), then join $p$ and $q$ with the other vertices of $J$ and $L$ (and their images under $G$). The new structure has point group $P$ as well. $\dim(A/N)_{j}-\dim Z_{j}-\dim W_{j}+\dim Y_{j}$ and $\dim(V/\Delta)_{j}-\dim Z_{j}-\dim W_{j}+\dim Y_{j}$ are unchanged during this process. 
\end{theorem}
\begin{figure}[ht]
\centering
\includegraphics[width=0.8\textwidth]{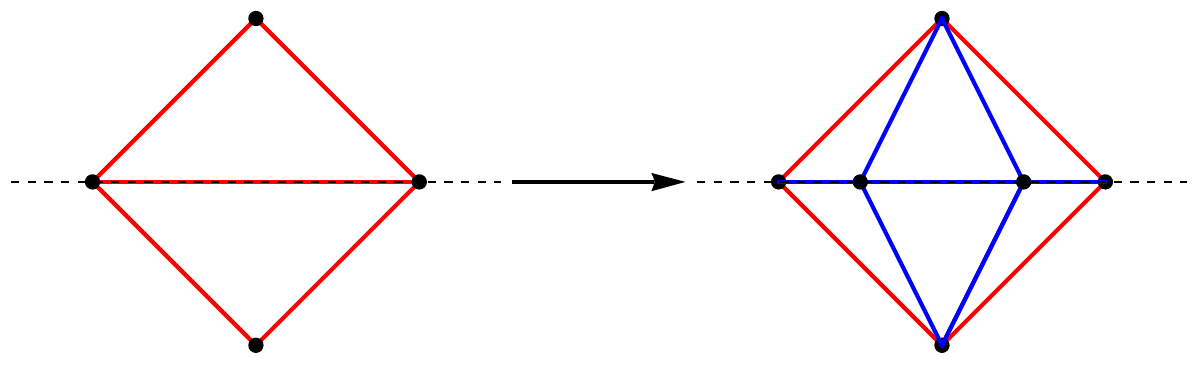}
\caption{Schematic of edge refinement for edges lie entirely in some mirror plane. The dashed line indicates that mirror plane.}
\end{figure}
\begin{proof}
    This can be proved in the same way as Theorem~\ref{thm:refinementonnonsymmetricedge}. Note inversion centers cannot lie on the mirror plane, so this proof is indeed much easier.
\end{proof}

\section{Reduction of The Second Type: Merging}
\begin{figure}[ht]\label{fig:merging}
\centering
\includegraphics[width=0.8\textwidth]{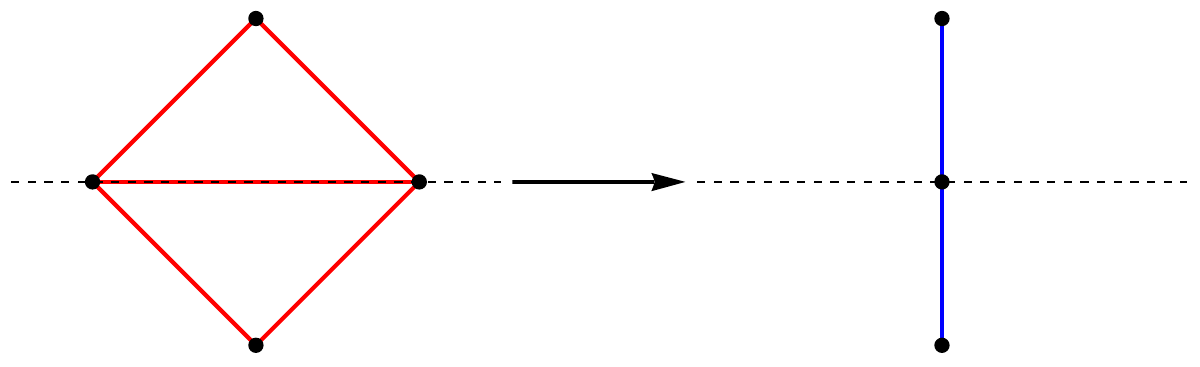}
\caption{Schematic of merging of points of continuous type. Here the point $q$ is not in the set $\{gp|\; g\in G\}$. The dashed line indicates that mirror plane or that rotation axis.}
\end{figure}

\begin{theorem}\label{thm:merging}
    Let $X$ be an origami with space group $G$, $p$, $q$ be two high-symmetry points of continuous type as shown in Fig.~\ref{fig:merging}, $r$ be the midpoint of $pq$. Suppose $G$ is a origami group of type I, whose point group is $P=G/T$. If the stabilizer $S_{r}$ of $r$ is not $\{\mathrm{id},M, R_{z}(\pi), MR_{z}(\pi)\}$, we can merge $p$ and $q$ to their midpoint, then add necessary connections to $r$ (as shown in Fig.~\ref{fig:merging}). The new structure has point group $P$ as well. $\dim(A/N)_{j}-\dim Z_{j}-\dim W_{j}+\dim Y_{j}$ and $\dim(V/\Delta)_{j}-\dim Z_{j}-\dim W_{j}+\dim Y_{j}$ are unchanged during this process.  
\end{theorem}
\begin{proof}
    This can be proved in the same way as Theorem~\ref{thm:refinementinthebulk}.
\end{proof}

\section{Reduction of The Third Type: Reconnection}
\begin{figure}[ht]\label{fig:straightlineelimination}
\centering
\includegraphics[width=0.5\textwidth]{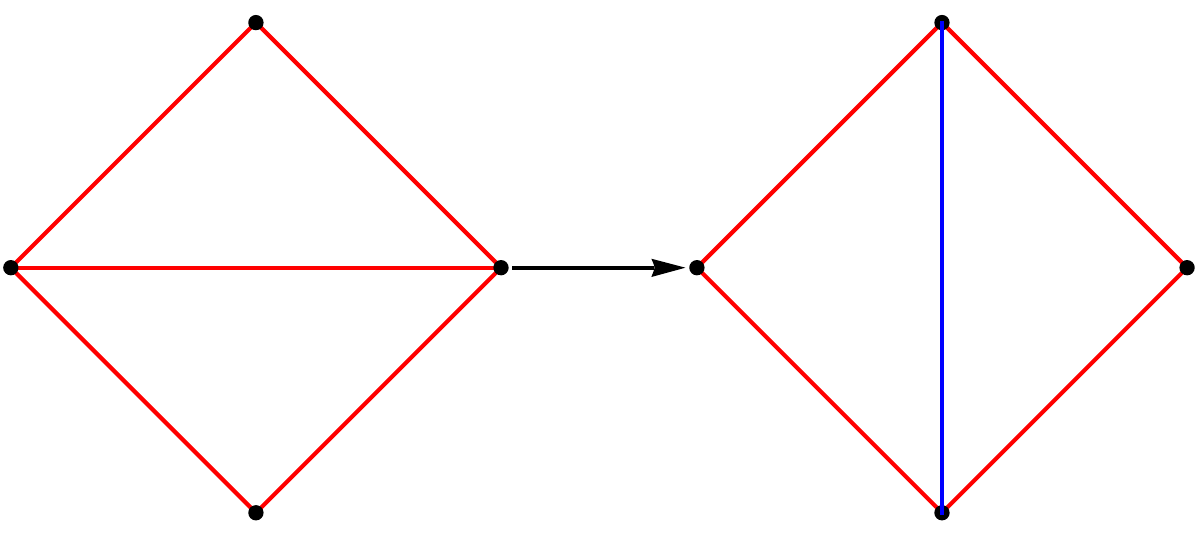}
\caption{Schematic of connection changing described in Theorem~\ref{thm:reconnection}.}
\end{figure}

\begin{theorem}\label{thm:reconnection}
    The change of connection shown in Fig.~\ref{fig:straightlineelimination} doesn't change the symmetry of the structure. Eq.~\eqref{eq:symmetryprotectedstrainmodes} gives the same $\dim(A/N)_{j}-\dim Z_{j}-\dim W_{j}+\dim Y_{j}$ and $\dim(V/\Delta)_{j}-\dim Z_{j}-\dim W_{j}+\dim Y_{j}$ if none of the four neighboring vertices are inversion centers.
\end{theorem}

\section{Reduction of The Fourth Type: Inversion Center Deletion}
\begin{figure}[ht]\label{fig:inversioncenterdeletion}
\centering
\includegraphics[width=0.5\textwidth]{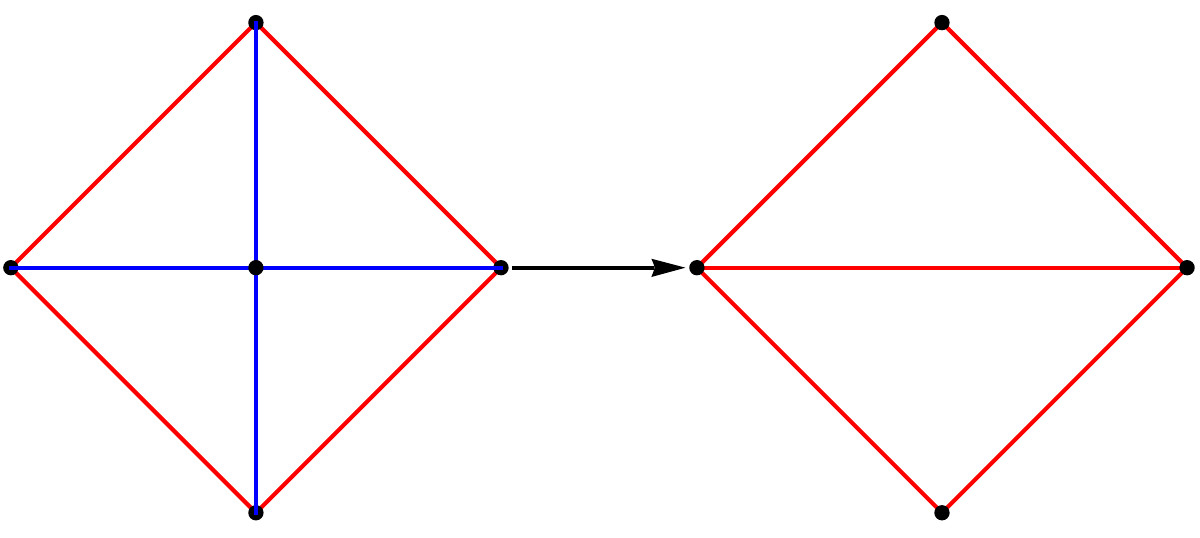}
\caption{Schematic of connection changing described in Theorem~\ref{thm:inversioncenterdeletion}.}
\end{figure}

\begin{theorem}\label{thm:inversioncenterdeletion}
    If an inversion center is occupied by a vertex and the vertex is 4-coordinated, we could remove that vertex, then reconnect vertices as shown in Fig.~\ref{fig:inversioncenterdeletion}. This process gives the same $\dim(A/N)_{j}-\dim Z_{j}-\dim W_{j}+\dim Y_{j}$ and $\dim(V/\Delta)_{j}-\dim Z_{j}-\dim W_{j}+\dim Y_{j}$ if none of the four vertices are inversion centers.
\end{theorem}

\section{The Number of Structure Classes is Finite}\label{sec:structureclasses}
\begin{definition}
    Operations described in Theorem~\ref{thm:refinementinthebulk}, Theorem~\ref{thm:refinementonnonsymmetricedge}, Theorem~\ref{thm:refinementonsymmetricedge}, Theorem~\ref{thm:merging}, Theorem~\ref{thm:reconnection} and Theorem~\ref{thm:inversioncenterdeletion} are called reversible strain-preserving operations.
\end{definition}
\begin{definition}\label{def:equivalentstructureclasses}
    Two structures $X$ and $Y$ are said to be in the same class if we can obtain $Y$ after finitely many reversible strain-preserving operations on $X$.
\end{definition}
\noindent Now we state \textbf{the most important theorem} in our note.
\begin{theorem}\label{thm:finitenessofclasses}
    Let $G$ be a type-I origami group. Then there are only finitely many classes of triangulated periodic origami whose space group is exactly $G$.
\end{theorem}
\begin{proof}
    The result follows easily from Theorem~\ref{thm:refinementinthebulk}, Theorem~\ref{thm:refinementonnonsymmetricedge}, Theorem~\ref{thm:refinementonsymmetricedge}, Theorem~\ref{thm:merging}, Theorem~\ref{thm:reconnection} and Theorem~\ref{thm:inversioncenterdeletion}.
\end{proof}

\chapter{Mode Enumeration}\label{chap:enumeration}

Here, we perform the mode enumeration for all type I origami groups defined in definition~\ref{def:typeIorigamigroup}. In this chapter, the lattice of $G$ is generated by $b_1$ and $b_2$, and we complete these vectors to a basis of $\mathbb R^3$ by setting $b_3=\hat z$.

\section{Triclinic Layer Groups}
\subsection{\texorpdfstring{$p1$}{p1}}
This group has trivial point group.
\subsection{\texorpdfstring{$p\bar{1}$}{p-1}}
\subsubsection{Group Structure}
The Bravais lattice of $G=p\bar{1}$ is generated by
\[
b_1=\begin{pmatrix}a\\0\\0\end{pmatrix},\qquad
b_2=\begin{pmatrix}b\cos\gamma\\ b\sin\gamma\\0\end{pmatrix},\qquad
b_3=\hat z=\begin{pmatrix}0\\0\\1\end{pmatrix},
\]
A convenient set of coset representatives for $G/T$ under basis $b_1,b_2,b_3$ is,
\[
g_1=(R_1|r_1),\quad R_1=
\begin{pmatrix}
1&0&0\\
0&1&0\\
0&0&1
\end{pmatrix},
\quad
\eta_1=
\begin{pmatrix}
1&0\\
0&1
\end{pmatrix},
\quad
\lambda_1=+1,
\quad
r_1=\{0,0,0\},
\]
\[
g_2=(R_2|r_2),\quad R_2=
\begin{pmatrix}
-1&0&0\\
0&-1&0\\
0&0&-1
\end{pmatrix},
\quad
\eta_2=
\begin{pmatrix}
-1&0\\
0&-1
\end{pmatrix},
\quad
\lambda_2=-1,
\quad
r_2=\{0,0,0\}.
\]
With the ordering $(g_1,g_2)$, we have:
\[
\bigl\{\mathrm{Tr}(\eta_i)\bigr\}_{i=1}^2=\{2,-2\},\qquad
\bigl\{\det(\eta_i)\bigr\}_{i=1}^2=\{1,1\},\qquad
\{\lambda_i\}_{i=1}^2=\{+1,-1\}.
\]
There are two $1$-dimensional irreducible representations of $G/T$.
The character table is
\texttt{charTable[[j,i]]} = character of irrep $j$ evaluated at element $g_i$:
\[
\texttt{charTable}=
\left\{
\{1,\ 1\},\;
\{1,\ -1\}
\right\}.
\]
Fixed points of $g_1$: every point is fixed.

Fixed points of $g_2$:
\[
\begin{aligned}
&(0,0,0),\ (\tfrac12,0,0),\ (0,\tfrac12,0),\ (\tfrac12,\tfrac12,0).
\end{aligned}
\]
\subsubsection{Mode Structure}
\begin{itemize}
    \item When all 4 inversion centers are occupied by edge centers: \textbf{one nonlinear symmetry-preserving pure $E$ mode} in the $\chi_{1}$ sector, two pure $H$ modes  in the $\chi_{2}$ sector.
    \item Otherwise, three pure $H$ modes in the $\chi_{2}$ sector.
\end{itemize}

\section{Monoclinic (Inclined) Layer Groups}
\subsection{\texorpdfstring{$p112$}{p112}}
\subsubsection{Group Structure}
The Bravais lattice of $G=p112$ is generated by
\[
b_1=\begin{pmatrix}a\\0\\0\end{pmatrix},\qquad
b_2=\begin{pmatrix}b\cos\gamma\\ b\sin\gamma\\0\end{pmatrix},\qquad
b_3=\hat z=\begin{pmatrix}0\\0\\1\end{pmatrix}.
\]
A convenient set of coset representatives for $G/T$ under basis $b_1,b_2,b_3$ is,
\[
g_1=(R_1|r_1),\quad R_1=
\begin{pmatrix}
1&0&0\\
0&1&0\\
0&0&1
\end{pmatrix},
\quad
\eta_1=
\begin{pmatrix}
1&0\\
0&1
\end{pmatrix},
\quad
\lambda_1=+1,
\quad
r_1=\{0,0,0\},
\]
\[
g_2=(R_2|r_2),\quad R_2=
\begin{pmatrix}
-1&0&0\\
0&-1&0\\
0&0&1
\end{pmatrix},
\quad
\eta_2=
\begin{pmatrix}
-1&0\\
0&-1
\end{pmatrix},
\quad
\lambda_2=+1,
\quad
r_2=\{0,0,0\}.
\]

With the ordering $(g_1,g_2)$, we have:
\[
\bigl\{\mathrm{Tr}(\eta_i)\bigr\}_{i=1}^2=\{2,-2\},\qquad
\bigl\{\det(\eta_i)\bigr\}_{i=1}^2=\{1,1\},\qquad
\{\lambda_i\}_{i=1}^2=\{+1,+1\}.
\]

There are two $1$-dimensional irreducible representations of $G/T$.
The character table is
\texttt{charTable[[j,i]]} = character of irrep $j$ evaluated at element $g_i$:
\[
\texttt{charTable}=
\left\{
\{1,\ 1\},\;
\{1,\ -1\}
\right\}.
\]

Fixed points of $g_1$: every point is fixed.

Fixed points of $g_2$: four lines parallel to $b_3$ passing through
\[
(u,v)\in\left\{(0,0),\ (\tfrac12,0),\ (0,\tfrac12),\ (\tfrac12,\tfrac12)\right\}, \qquad w\in\mathbb{R}.
\]
\subsubsection{Mode Structure}
\begin{itemize}
    \item Three $(E,H)$ modes, all of them lie in the $\chi_{1}$ sector.
\end{itemize}

\subsection{\texorpdfstring{$p11a$}{p11a}}
\subsubsection{Group Structure}
The Bravais lattice of $G=p11a$ is generated by
\[
b_1=\begin{pmatrix}a\\0\\0\end{pmatrix},\qquad
b_2=\begin{pmatrix}b\cos\gamma\\ b\sin\gamma\\0\end{pmatrix},\qquad
b_3=\hat z=\begin{pmatrix}0\\0\\1\end{pmatrix}.
\]
A convenient set of coset representatives for $G/T$ under basis $b_1,b_2,b_3$ is,
\[
g_1=(R_1|r_1),\quad R_1=
\begin{pmatrix}
1&0&0\\
0&1&0\\
0&0&1
\end{pmatrix},
\quad
\eta_1=
\begin{pmatrix}
1&0\\
0&1
\end{pmatrix},
\quad
\lambda_1=+1,
\quad
r_1=\{0,0,0\},
\]
\[
g_2=(R_2|r_2),\quad R_2=
\begin{pmatrix}
1&0&0\\
0&1&0\\
0&0&-1
\end{pmatrix},
\quad
\eta_2=
\begin{pmatrix}
1&0\\
0&1
\end{pmatrix},
\quad
\lambda_2=-1,
\quad
r_2=\{\tfrac12,0,0\}.
\]

With the ordering $(g_1,g_2)$, we have:
\[
\bigl\{\mathrm{Tr}(\eta_i)\bigr\}_{i=1}^2=\{2,2\},\qquad
\bigl\{\det(\eta_i)\bigr\}_{i=1}^2=\{1,1\},\qquad
\{\lambda_i\}_{i=1}^2=\{+1,-1\}.
\]

There are two $1$-dimensional irreducible representations of $G/T$.
The character table is
\texttt{charTable[[j,i]]} = character of irrep $j$ evaluated at element $g_i$:
\[
\texttt{charTable}=
\left\{
\{1,\ 1\},\;
\{1,\ -1\}
\right\}.
\]

Fixed points of $g_1$: every point is fixed.

Fixed points of $g_2$: no fixed points (up to lattice translations) within a unit cell.
\subsubsection{Mode Structure}
\begin{itemize}
    \item \textbf{One nonlinear symmetry-preserving pure $E$ mode} in the $\chi_{1}$ sector, two pure $H$ modes  in the $\chi_{2}$ sector.
\end{itemize}

\subsection{\texorpdfstring{$p112/a$}{p112/a}}
\subsubsection{Group Structure}
The Bravais lattice of $G=p112/a$ is generated by
\[
b_1=\begin{pmatrix}a\\0\\0\end{pmatrix},\qquad
b_2=\begin{pmatrix}b\cos\gamma\\ b\sin\gamma\\0\end{pmatrix},\qquad
b_3=\hat z=\begin{pmatrix}0\\0\\1\end{pmatrix}.
\]
A convenient set of coset representatives for $G/T$ under basis $b_1,b_2,b_3$ is,
\[
g_1=(R_1|r_1),\quad R_1=
\begin{pmatrix}
1&0&0\\
0&1&0\\
0&0&1
\end{pmatrix},
\quad
\eta_1=
\begin{pmatrix}
1&0\\
0&1
\end{pmatrix},
\quad
\lambda_1=+1,
\quad
r_1=\{0,0,0\},
\]
\[
g_2=(R_2|r_2),\quad R_2=
\begin{pmatrix}
-1&0&0\\
0&-1&0\\
0&0&1
\end{pmatrix},
\quad
\eta_2=
\begin{pmatrix}
-1&0\\
0&-1
\end{pmatrix},
\quad
\lambda_2=+1,
\quad
r_2=\{\tfrac12,0,0\},
\]
\[
g_3=(R_3|r_3),\quad R_3=
\begin{pmatrix}
-1&0&0\\
0&-1&0\\
0&0&-1
\end{pmatrix},
\quad
\eta_3=
\begin{pmatrix}
-1&0\\
0&-1
\end{pmatrix},
\quad
\lambda_3=-1,
\quad
r_3=\{0,0,0\},
\]
\[
g_4=(R_4|r_4),\quad R_4=
\begin{pmatrix}
1&0&0\\
0&1&0\\
0&0&-1
\end{pmatrix},
\quad
\eta_4=
\begin{pmatrix}
1&0\\
0&1
\end{pmatrix},
\quad
\lambda_4=-1,
\quad
r_4=\{\tfrac12,0,0\}.
\]

With the ordering $(g_1,g_2,g_3,g_4)$, we have:
\[
\bigl\{\mathrm{Tr}(\eta_i)\bigr\}_{i=1}^4=\{2,-2,-2,2\},\qquad
\bigl\{\det(\eta_i)\bigr\}_{i=1}^4=\{1,1,1,1\},\qquad
\{\lambda_i\}_{i=1}^4=\{+1,+1,-1,-1\}.
\]

There are four $1$-dimensional irreducible representations of $G/T$.
The character table is
\texttt{charTable[[j,i]]} = character of irrep $j$ evaluated at element $g_i$:
\[
\texttt{charTable}=
\left\{
\{1,\ 1,\ 1,\ 1\},\;
\{1,\ -1,\ 1,\ -1\},\;
\{1,\ 1,\ -1,\ -1\},\;
\{1,\ -1,\ -1,\ 1\}
\right\}.
\]

Fixed points of $g_1$: every point is fixed.

Fixed points of $g_2$: four lines parallel to $b_3$ passing through
\[
(u,v)\in\left\{(\tfrac14,0),\ (\tfrac34,0),\ (\tfrac14,\tfrac12),\ (\tfrac34,\tfrac12)\right\}, \qquad w\in\mathbb{R}.
\]
Fixed points of $g_3$:
\[
\begin{aligned}
&(0,0,0),\ (\tfrac12,0,0),\ (0,\tfrac12,0),\ (\tfrac12,\tfrac12,0).
\end{aligned}
\]
Fixed points of $g_4$: no fixed points (up to lattice translations) within a unit cell.
\subsubsection{Mode Structure}
\begin{itemize}
    \item When all 4 inversion centers are occupied by edge centers: \textbf{one nonlinear symmetry-preserving pure $E$ mode} in the $\chi_{1}$ sector, two pure $H$ modes  in the $\chi_{3}$ sector.
    \item Otherwise, three pure $H$ modes in the $\chi_{3}$ sector.
\end{itemize}

\section{Monoclinic (Orthogonal) Layer Groups}
\subsection{\texorpdfstring{$p2_{1}11$}{p2-111}}
\subsubsection{Group Structure}
The Bravais lattice of $G=p2_{1}11$ is generated by
\[
b_1=\begin{pmatrix}a\\0\\0\end{pmatrix},\qquad
b_2=\begin{pmatrix}0\\ b\\0\end{pmatrix},\qquad
b_3=\hat z=\begin{pmatrix}0\\0\\1\end{pmatrix}.
\]
A convenient set of coset representatives for $G/T$ under basis $b_1,b_2,b_3$ is,
\[
g_1=(R_1|r_1),\quad R_1=
\begin{pmatrix}
1&0&0\\
0&1&0\\
0&0&1
\end{pmatrix},
\quad
\eta_1=
\begin{pmatrix}
1&0\\
0&1
\end{pmatrix},
\quad
\lambda_1=+1,
\quad
r_1=\{0,0,0\},
\]
\[
g_2=(R_2|r_2),\quad R_2=
\begin{pmatrix}
1&0&0\\
0&-1&0\\
0&0&-1
\end{pmatrix},
\quad
\eta_2=
\begin{pmatrix}
1&0\\
0&-1
\end{pmatrix},
\quad
\lambda_2=-1,
\quad
r_2=\{\tfrac12,0,0\}.
\]

With the ordering $(g_1,g_2)$, we have:
\[
\bigl\{\mathrm{Tr}(\eta_i)\bigr\}_{i=1}^2=\{2,0\},\qquad
\bigl\{\det(\eta_i)\bigr\}_{i=1}^2=\{1,-1\},\qquad
\{\lambda_i\}_{i=1}^2=\{+1,-1\}.
\]

There are two $1$-dimensional irreducible representations of $G/T$.
The character table is
\texttt{charTable[[j,i]]} = character of irrep $j$ evaluated at element $g_i$:
\[
\texttt{charTable}=
\left\{
\{1,\ 1\},\;
\{1,\ -1\}
\right\}.
\]

Fixed points of $g_1$: every point is fixed.

Fixed points of $g_2$: no fixed points (up to lattice translations) within a unit cell.
\subsubsection{Mode Structure}
\begin{itemize}
    \item \textbf{one nonlinear symmetry-preserving pure $E$ mode} in the $\chi_{1}$ sector, $E$ is diagonal. One $(E,H)$ mode in sector $\chi_{1}$, with diagonal $E$ and off-diagonal $H$. One $(E,H)$ mode in sector $\chi_{2}$, with off-diagonal $E$ and diagonal $H$. 
\end{itemize}

\subsection{\texorpdfstring{$pm11$}{pm11}}
\subsubsection{Group Structure}
The Bravais lattice of $G=pm11$ is generated by
\[
b_1=\begin{pmatrix}a\\0\\0\end{pmatrix},\qquad
b_2=\begin{pmatrix}0\\ b\\0\end{pmatrix},\qquad
b_3=\hat z=\begin{pmatrix}0\\0\\1\end{pmatrix}.
\]
A convenient set of coset representatives for $G/T$ under basis $b_1,b_2,b_3$ is,
\[
g_1=(R_1|r_1),\quad R_1=
\begin{pmatrix}
1&0&0\\
0&1&0\\
0&0&1
\end{pmatrix},
\quad
\eta_1=
\begin{pmatrix}
1&0\\
0&1
\end{pmatrix},
\quad
\lambda_1=+1,
\quad
r_1=\{0,0,0\},
\]
\[
g_2=(R_2|r_2),\quad R_2=
\begin{pmatrix}
-1&0&0\\
0&1&0\\
0&0&1
\end{pmatrix},
\quad
\eta_2=
\begin{pmatrix}
-1&0\\
0&1
\end{pmatrix},
\quad
\lambda_2=+1,
\quad
r_2=\{0,0,0\}.
\]

With the ordering $(g_1,g_2)$, we have:
\[
\bigl\{\mathrm{Tr}(\eta_i)\bigr\}_{i=1}^2=\{2,0\},\qquad
\bigl\{\det(\eta_i)\bigr\}_{i=1}^2=\{1,-1\},\qquad
\{\lambda_i\}_{i=1}^2=\{+1,+1\}.
\]

There are two $1$-dimensional irreducible representations of $G/T$.
The character table is
\texttt{charTable[[j,i]]} = character of irrep $j$ evaluated at element $g_i$:
\[
\texttt{charTable}=
\left\{
\{1,\ 1\},\;
\{1,\ -1\}
\right\}.
\]

Fixed points of $g_1$: every point is fixed.

Fixed points of $g_2$: two planes parallel to $\mathrm{span}\{b_2,b_3\}$ with
\[
u\in\left\{0,\ \tfrac12\right\},\qquad v\in\mathbb R, \qquad w\in\mathbb{R}.
\]
\subsubsection{Mode Structure}
\begin{itemize}
    \item Two linear $(E,H)$ modes in sector $\chi_{1}$ with diagonal $E$ and diagonal $H$. One $(E,H)$ mode in sector $\chi_{2}$, with off-diagonal $E$ and off-diagonal $H$. 
\end{itemize}

\subsection{\texorpdfstring{$pb11$}{pb11}}
\subsubsection{Group Structure}
The Bravais lattice of $G=pb11$ is generated by
\[
b_1=\begin{pmatrix}a\\0\\0\end{pmatrix},\qquad
b_2=\begin{pmatrix}0\\ b\\0\end{pmatrix},\qquad
b_3=\hat z=\begin{pmatrix}0\\0\\1\end{pmatrix}.
\]
A convenient set of coset representatives for $G/T$ under basis $b_1,b_2,b_3$ is,
\[
g_1=(R_1|r_1),\quad R_1=
\begin{pmatrix}
1&0&0\\
0&1&0\\
0&0&1
\end{pmatrix},
\quad
\eta_1=
\begin{pmatrix}
1&0\\
0&1
\end{pmatrix},
\quad
\lambda_1=+1,
\quad
r_1=\{0,0,0\},
\]
\[
g_2=(R_2|r_2),\quad R_2=
\begin{pmatrix}
-1&0&0\\
0&1&0\\
0&0&1
\end{pmatrix},
\quad
\eta_2=
\begin{pmatrix}
-1&0\\
0&1
\end{pmatrix},
\quad
\lambda_2=+1,
\quad
r_2=\{0,\tfrac12,0\}.
\]

With the ordering $(g_1,g_2)$, we have:
\[
\bigl\{\mathrm{Tr}(\eta_i)\bigr\}_{i=1}^2=\{2,0\},\qquad
\bigl\{\det(\eta_i)\bigr\}_{i=1}^2=\{1,-1\},\qquad
\{\lambda_i\}_{i=1}^2=\{+1,+1\}.
\]

There are two $1$-dimensional irreducible representations of $G/T$.
The character table is
\texttt{charTable[[j,i]]} = character of irrep $j$ evaluated at element $g_i$:
\[
\texttt{charTable}=
\left\{
\{1,\ 1\},\;
\{1,\ -1\}
\right\}.
\]

Fixed points of $g_1$: every point is fixed.

Fixed points of $g_2$: no fixed points (up to lattice translations) within a unit cell.

\subsubsection{Mode Structure}
\begin{itemize}
    \item \item Two linear $(E,H)$ modes in the $\chi_{1}$ sector with diagonal $E$ and diagonal $H$. One bending mode $(E,H)$ in sector $\chi_{2}$ with off-diagonal $E$ and off-diagonal $H$. 
\end{itemize}

\subsection{\texorpdfstring{$cm11$}{cm11}}
Same as $pm11$.

\subsection{\texorpdfstring{$p2_{1}/m11$}{p2-1/mm11}}
\subsubsection{Group Structure}
The Bravais lattice of $G=p2_{1}/m11$ is generated by
\[
b_1=\begin{pmatrix}a\\0\\0\end{pmatrix},\qquad
b_2=\begin{pmatrix}0\\ b\\0\end{pmatrix},\qquad
b_3=\hat z=\begin{pmatrix}0\\0\\1\end{pmatrix}.
\]
A convenient set of coset representatives for $G/T$ under basis $b_1,b_2,b_3$ is,
\[
g_1=(R_1|r_1),\quad R_1=
\begin{pmatrix}
1&0&0\\
0&1&0\\
0&0&1
\end{pmatrix},
\quad
\eta_1=
\begin{pmatrix}
1&0\\
0&1
\end{pmatrix},
\quad
\lambda_1=+1,
\quad
r_1=\{0,0,0\},
\]
\[
g_2=(R_2|r_2),\quad R_2=
\begin{pmatrix}
1&0&0\\
0&-1&0\\
0&0&-1
\end{pmatrix},
\quad
\eta_2=
\begin{pmatrix}
1&0\\
0&-1
\end{pmatrix},
\quad
\lambda_2=-1,
\quad
r_2=\{\tfrac12,0,0\},
\]
\[
g_3=(R_3|r_3),\quad R_3=
\begin{pmatrix}
-1&0&0\\
0&-1&0\\
0&0&-1
\end{pmatrix},
\quad
\eta_3=
\begin{pmatrix}
-1&0\\
0&-1
\end{pmatrix},
\quad
\lambda_3=-1,
\quad
r_3=\{0,0,0\},
\]
\[
g_4=(R_4|r_4),\quad R_4=
\begin{pmatrix}
-1&0&0\\
0&1&0\\
0&0&1
\end{pmatrix},
\quad
\eta_4=
\begin{pmatrix}
-1&0\\
0&1
\end{pmatrix},
\quad
\lambda_4=+1,
\quad
r_4=\{\tfrac12,0,0\}.
\]

With the ordering $(g_1,g_2,g_3,g_4)$, we have:
\[
\bigl\{\mathrm{Tr}(\eta_i)\bigr\}_{i=1}^4=\{2,0,-2,0\},\qquad
\bigl\{\det(\eta_i)\bigr\}_{i=1}^4=\{1,-1,1,-1\},\qquad
\{\lambda_i\}_{i=1}^4=\{+1,-1,-1,+1\}.
\]

There are four $1$-dimensional irreducible representations of $G/T$.
The character table is
\texttt{charTable[[j,i]]} = character of irrep $j$ evaluated at element $g_i$:
\[
\texttt{charTable}=
\left\{
\{1,\ 1,\ 1,\ 1\},\;
\{1,\ -1,\ 1,\ -1\},\;
\{1,\ 1,\ -1,\ -1\},\;
\{1,\ -1,\ -1,\ 1\}
\right\}.
\]

Fixed points of $g_1$: every point is fixed.

Fixed points of $g_2$: no fixed points (up to lattice translations) within a unit cell.

Fixed points of $g_3$:
\[
\begin{aligned}
&(0,0,0),\ (\tfrac12,0,0),\ (0,\tfrac12,0),\ (\tfrac12,\tfrac12,0).
\end{aligned}
\]

Fixed points of $g_4$: two planes parallel to $\mathrm{span}\{b_2,b_3\}$ with
\[
u\in\left\{\tfrac14,\ \tfrac34\right\},\qquad v\in\mathbb R, \qquad w\in\mathbb{R}.
\]
\subsubsection{Mode Structure}
\begin{itemize}
    \item When all 4 inversion centers are occupied by edge centers: \textbf{one nonlinear symmetry-preserving pure $E$ mode} in the $\chi_{1}$ sector with diagonal $E$. One pure $H$ mode in the $\chi_{3}$ sector with off-diagonal $H$. One pure $H$ mode  in the $\chi_{4}$ sector, with diagonal $H$.
    \item Otherwise, one pure $H$ mode in the $\chi_{3}$ sector with off-diagonal $H$. Two pure $H$ modes in the $\chi_{4}$ sector with diagonal $H$. 
\end{itemize}

\subsection{\texorpdfstring{$p2_{1}/b11$}{p2-1/b11}}
\subsubsection{Group Structure}
The translation lattice of $G=p2_{1}/b11$ is generated by
\[
b_1=\begin{pmatrix}a\\0\\0\end{pmatrix},\qquad
b_2=\begin{pmatrix}0\\ b\\0\end{pmatrix}.
\]
We complete these vectors to a basis of $\mathbb R^3$ by setting
\[
b_3=\hat z=\begin{pmatrix}0\\0\\1\end{pmatrix}.
\]
A convenient set of coset representatives for $G/T$ under basis $b_1,b_2,b_3$ is,
\[
g_1=(R_1|r_1),\quad R_1=
\begin{pmatrix}
1&0&0\\
0&1&0\\
0&0&1
\end{pmatrix},
\quad
\eta_1=
\begin{pmatrix}
1&0\\
0&1
\end{pmatrix},
\quad
\lambda_1=+1,
\quad
r_1=\{0,0,0\},
\]
\[
g_2=(R_2|r_2),\quad R_2=
\begin{pmatrix}
1&0&0\\
0&-1&0\\
0&0&-1
\end{pmatrix},
\quad
\eta_2=
\begin{pmatrix}
1&0\\
0&-1
\end{pmatrix},
\quad
\lambda_2=-1,
\quad
r_2=\{\tfrac12,\tfrac12,0\},
\]
\[
g_3=(R_3|r_3),\quad R_3=
\begin{pmatrix}
-1&0&0\\
0&-1&0\\
0&0&-1
\end{pmatrix},
\quad
\eta_3=
\begin{pmatrix}
-1&0\\
0&-1
\end{pmatrix},
\quad
\lambda_3=-1,
\quad
r_3=\{0,0,0\},
\]
\[
g_4=(R_4|r_4),\quad R_4=
\begin{pmatrix}
-1&0&0\\
0&1&0\\
0&0&1
\end{pmatrix},
\quad
\eta_4=
\begin{pmatrix}
-1&0\\
0&1
\end{pmatrix},
\quad
\lambda_4=+1,
\quad
r_4=\{\tfrac12,\tfrac12,0\}.
\]

With the ordering $(g_1,g_2,g_3,g_4)$, we have:
\[
\bigl\{\mathrm{Tr}(\eta_i)\bigr\}_{i=1}^4
=\{2,0,-2,0\},\qquad
\bigl\{\det(\eta_i)\bigr\}_{i=1}^4
=\{1,-1,1,-1\},\qquad
\{\lambda_i\}_{i=1}^4
=\{+1,-1,-1,+1\}.
\]

There are four $1$-dimensional irreducible representations of $G/T$.
The character table is
\texttt{charTable[[j,i]]} = character of irrep $j$ evaluated at element $g_i$:
\[
\texttt{charTable}=
\left\{
\{1,\ 1,\ 1,\ 1\},\;
\{1,\ -1,\ 1,\ -1\},\;
\{1,\ 1,\ -1,\ -1\},\;
\{1,\ -1,\ -1,\ 1\}
\right\}.
\]

Fixed points of $g_1$: every point is fixed.

Fixed points of $g_2$: no fixed points (up to lattice translations) within a unit cell.

Fixed points of $g_3$:
\[
\begin{aligned}
&(0,0,0),\quad
(\tfrac12,0,0),\quad
(0,\tfrac12,0),\quad
(\tfrac12,\tfrac12,0).
\end{aligned}
\]

Fixed points of $g_4$: no fixed points (up to lattice translations) within a unit cell.

\subsubsection{Mode Structure}
\begin{itemize}
    \item When all 4 inversion centers are occupied by edge centers:
    \textbf{one nonlinear symmetry-preserving pure $E$ mode} in the $\chi_{1}$ sector with diagonal $E$. One pure $H$ mode in the $\chi_{3}$ sector with off-diagonal $H$. One pure $H$ mode in the $\chi_{4}$ sector with diagonal $H$.
    
    \item Otherwise, one pure $H$ mode in the $\chi_{3}$ sector with off-diagonal $H$. Two pure $H$ modes in the $\chi_{4}$ sector with diagonal $H$.
\end{itemize}

\section{Orthorhombic Layer Groups}
\subsection{\texorpdfstring{$p2_{1}2_{1}2$}{p2-12-12}}
\subsubsection{Group Structure}
The Bravais lattice of $G=p2_{1}2_{1}2$ is generated by
\[
b_1=\begin{pmatrix}a\\0\\0\end{pmatrix},\qquad
b_2=\begin{pmatrix}0\\ b\\0\end{pmatrix},\qquad
b_3=\hat z=\begin{pmatrix}0\\0\\1\end{pmatrix}.
\]
A convenient set of coset representatives for $G/T$ under basis $b_1,b_2,b_3$ is,
\[
g_1=(R_1|r_1),\quad R_1=
\begin{pmatrix}
1&0&0\\
0&1&0\\
0&0&1
\end{pmatrix},
\quad
\eta_1=
\begin{pmatrix}
1&0\\
0&1
\end{pmatrix},
\quad
\lambda_1=+1,
\quad
r_1=\{0,0,0\},
\]
\[
g_2=(R_2|r_2),\quad R_2=
\begin{pmatrix}
1&0&0\\
0&-1&0\\
0&0&-1
\end{pmatrix},
\quad
\eta_2=
\begin{pmatrix}
1&0\\
0&-1
\end{pmatrix},
\quad
\lambda_2=-1,
\quad
r_2=\{\tfrac12,0,0\},
\]
\[
g_3=(R_3|r_3),\quad R_3=
\begin{pmatrix}
-1&0&0\\
0&1&0\\
0&0&-1
\end{pmatrix},
\quad
\eta_3=
\begin{pmatrix}
-1&0\\
0&1
\end{pmatrix},
\quad
\lambda_3=-1,
\quad
r_3=\{0,\tfrac12,0\},
\]
\[
g_4=(R_4|r_4),\quad R_4=
\begin{pmatrix}
-1&0&0\\
0&-1&0\\
0&0&1
\end{pmatrix},
\quad
\eta_4=
\begin{pmatrix}
-1&0\\
0&-1
\end{pmatrix},
\quad
\lambda_4=+1,
\quad
r_4=\{\tfrac12,\tfrac12,0\}.
\]

With the ordering $(g_1,g_2,g_3,g_4)$, we have:
\[
\bigl\{\mathrm{Tr}(\eta_i)\bigr\}_{i=1}^4=\{2,0,0,-2\},\qquad
\bigl\{\det(\eta_i)\bigr\}_{i=1}^4=\{1,-1,-1,1\},\qquad
\{\lambda_i\}_{i=1}^4=\{+1,-1,-1,+1\}.
\]

There are four $1$-dimensional irreducible representations of $G/T$.
The character table is
\texttt{charTable[[j,i]]} = character of irrep $j$ evaluated at element $g_i$:
\[
\texttt{charTable}=
\left\{
\{1,\ 1,\ 1,\ 1\},\;
\{1,\ -1,\ 1,\ -1\},\;
\{1,\ 1,\ -1,\ -1\},\;
\{1,\ -1,\ -1,\ 1\}
\right\}.
\]

Fixed points of $g_1$: every point is fixed.

Fixed points of $g_2$: no fixed points (up to lattice translations) within a unit cell.

Fixed points of $g_3$: no fixed points (up to lattice translations) within a unit cell.

Fixed points of $g_4$: four lines parallel to $b_3$ passing through
\[
(u,v)\in\left\{(\tfrac14,\tfrac14),\ (\tfrac14,\tfrac34),\ (\tfrac34,\tfrac14),\ (\tfrac34,\tfrac34)\right\},\qquad w\in\mathbb{R}.
\]
\subsubsection{Mode Structure}
\begin{itemize}
    \item\textbf{One nonlinear symmetry-preserving pure $E$ mode} in the $\chi_{1}$ sector with diagonal $E$, one $(E,H)$ mode in the $\chi_{1}$ sector with diagonal $E$ and off-diagonal $H$. One $(E,H)$ mode in the $\chi_{4}$ sector with off-diagonal $E$ and diagonal $H$.
\end{itemize}

\subsection{\texorpdfstring{$pmm2$}{pmm2}}
\subsubsection{Group Structure}
The Bravais lattice of $G=pmm2$ is generated by
\[
b_1=\begin{pmatrix}a\\0\\0\end{pmatrix},\qquad
b_2=\begin{pmatrix} 0\\ b\\0\end{pmatrix},\qquad
b_3=\hat z=\begin{pmatrix}0\\0\\1\end{pmatrix}.
\]
A convenient set of coset representatives for $G/T$ under basis $b_1,b_2,b_3$ is,
\[
g_1=(R_1|r_1),\quad R_1=
\begin{pmatrix}
1&0&0\\
0&1&0\\
0&0&1
\end{pmatrix},
\quad
\eta_1=
\begin{pmatrix}
1&0\\
0&1
\end{pmatrix},
\quad
\lambda_1=+1,
\quad
r_1=\{0,0,0\},
\]
\[
g_2=(R_2|r_2),\quad R_2=
\begin{pmatrix}
-1&0&0\\
0&1&0\\
0&0&1
\end{pmatrix},
\quad
\eta_2=
\begin{pmatrix}
-1&0\\
0&1
\end{pmatrix},
\quad
\lambda_2=+1,
\quad
r_2=\{0,0,0\},
\]
\[
g_3=(R_3|r_3),\quad R_3=
\begin{pmatrix}
1&0&0\\
0&-1&0\\
0&0&1
\end{pmatrix},
\quad
\eta_3=
\begin{pmatrix}
1&0\\
0&-1
\end{pmatrix},
\quad
\lambda_3=+1,
\quad
r_3=\{0,0,0\},
\]
\[
g_4=(R_4|r_4),\quad R_4=
\begin{pmatrix}
-1&0&0\\
0&-1&0\\
0&0&1
\end{pmatrix},
\quad
\eta_4=
\begin{pmatrix}
-1&0\\
0&-1
\end{pmatrix},
\quad
\lambda_4=+1,
\quad
r_4=\{0,0,0\}.
\]

With the ordering $(g_1,g_2,g_3,g_4)$, we have:
\[
\bigl\{\mathrm{Tr}(\eta_i)\bigr\}_{i=1}^4=\{2,0,0,-2\},\qquad
\bigl\{\det(\eta_i)\bigr\}_{i=1}^4=\{1,-1,-1,1\},\qquad
\{\lambda_i\}_{i=1}^4=\{+1,+1,+1,+1\}.
\]

There are four $1$-dimensional irreducible representations of $G/T$.
The character table is
\texttt{charTable[[j,i]]} = character of irrep $j$ evaluated at element $g_i$:
\[
\texttt{charTable}=
\left\{
\{1,\ 1,\ 1,\ 1\},\;
\{1,\ -1,\ 1,\ -1\},\;
\{1,\ 1,\ -1,\ -1\},\;
\{1,\ -1,\ -1,\ 1\}
\right\}.
\]

Fixed points of $g_1$: every point is fixed.

Fixed points of $g_2$: two planes parallel to $\mathrm{span}\{b_2,b_3\}$ with
\[
u\in\left\{0,\ \tfrac12\right\},\qquad v\in\mathbb R,\qquad w\in\mathbb R.
\]

Fixed points of $g_3$: two planes parallel to $\mathrm{span}\{b_1,b_3\}$ with
\[
v\in\left\{0,\ \tfrac12\right\},\qquad u\in\mathbb R,\qquad w\in\mathbb R.
\]

Fixed points of $g_4$: four lines parallel to $b_3$ passing through
\[
(u,v)\in\left\{(0,0),\ (\tfrac12,0),\ (0,\tfrac12),\ (\tfrac12,\tfrac12)\right\}.
\]
\subsubsection{Mode Structure}
\begin{itemize}
    \item Two linear $(E,H)$ modes in the $\chi_{1}$ sector with diagonal $E$ and diagonal $H$. One $(E,H)$ mode with off-diagonal $E$, off-diagonal $H$ in the $\chi_{4}$ sector.
\end{itemize}

\subsection{\texorpdfstring{$cmm2$}{cmm2}}
This is the same as $pmm2$.

\subsection{\texorpdfstring{$pba2$}{pba2}}
\subsubsection{Group Structure}
The translation lattice of $G=pba2$ is generated by
\[
b_1=\begin{pmatrix}a\\0\\0\end{pmatrix},\qquad
b_2=\begin{pmatrix}0\\ b\\0\end{pmatrix}.
\]
We complete these vectors to a basis of $\mathbb R^3$ by setting
\[
b_3=\hat z=\begin{pmatrix}0\\0\\1\end{pmatrix}.
\]
A convenient set of coset representatives for $G/T$ under basis $b_1,b_2,b_3$ is,
\[
g_1=(R_1|r_1),\quad R_1=
\begin{pmatrix}
1&0&0\\
0&1&0\\
0&0&1
\end{pmatrix},
\quad
\eta_1=
\begin{pmatrix}
1&0\\
0&1
\end{pmatrix},
\quad
\lambda_1=+1,
\quad
r_1=\{0,0,0\},
\]
\[
g_2=(R_2|r_2),\quad R_2=
\begin{pmatrix}
-1&0&0\\
0&-1&0\\
0&0&1
\end{pmatrix},
\quad
\eta_2=
\begin{pmatrix}
-1&0\\
0&-1
\end{pmatrix},
\quad
\lambda_2=+1,
\quad
r_2=\{0,0,0\},
\]
\[
g_3=(R_3|r_3),\quad R_3=
\begin{pmatrix}
1&0&0\\
0&-1&0\\
0&0&1
\end{pmatrix},
\quad
\eta_3=
\begin{pmatrix}
1&0\\
0&-1
\end{pmatrix},
\quad
\lambda_3=+1,
\quad
r_3=\{\tfrac12,\tfrac12,0\},
\]
\[
g_4=(R_4|r_4),\quad R_4=
\begin{pmatrix}
-1&0&0\\
0&1&0\\
0&0&1
\end{pmatrix},
\quad
\eta_4=
\begin{pmatrix}
-1&0\\
0&1
\end{pmatrix},
\quad
\lambda_4=+1,
\quad
r_4=\{\tfrac12,\tfrac12,0\}.
\]

With the ordering $(g_1,g_2,g_3,g_4)$, we have:
\[
\bigl\{\mathrm{Tr}(\eta_i)\bigr\}_{i=1}^4
=\{2,-2,0,0\},\qquad
\bigl\{\det(\eta_i)\bigr\}_{i=1}^4
=\{1,1,-1,-1\},\qquad
\{\lambda_i\}_{i=1}^4
=\{+1,+1,+1,+1\}.
\]

There are four $1$-dimensional irreducible representations of $G/T$.
The character table is
\texttt{charTable[[j,i]]} = character of irrep $j$ evaluated at element $g_i$:
\[
\texttt{charTable}=
\left\{
\{1,\ 1,\ 1,\ 1\},\;
\{1,\ -1,\ 1,\ -1\},\;
\{1,\ 1,\ -1,\ -1\},\;
\{1,\ -1,\ -1,\ 1\}
\right\}.
\]

Fixed points of $g_1$: every point is fixed.

Fixed points of $g_2$: four lines parallel to $b_3$ passing through
\[
(u,v)\in
\left\{
(0,0),\quad
(\tfrac12,0),\quad
(0,\tfrac12),\quad
(\tfrac12,\tfrac12)
\right\},
\qquad w\in\mathbb R.
\]

Fixed points of $g_3$: no fixed points (up to lattice translations) within a unit cell.

Fixed points of $g_4$: no fixed points (up to lattice translations) within a unit cell.

\subsubsection{Mode Structure}
\begin{itemize}
    \item Two linear $(E,H)$ modes in the $\chi_{1}$ sector, with diagonal $E$ and diagonal $H$. One $(E,H)$ mode in the $\chi_{3}$ sector, with off-diagonal $E$ and off-diagonal $H$.
\end{itemize}

\subsection{\texorpdfstring{$pma2$}{pma2}}
\subsubsection{Group Structure}
The Bravais lattice of $G=pma2$ is generated by
\[
b_1=\begin{pmatrix}a\\0\\0\end{pmatrix},\qquad
b_2=\begin{pmatrix}0\\ b\\0\end{pmatrix},\qquad
b_3=\hat z=\begin{pmatrix}0\\0\\1\end{pmatrix}.
\]
A convenient set of coset representatives for $G/T$ under basis $b_1,b_2,b_3$ is,
\[
g_1=(R_1|r_1),\quad R_1=
\begin{pmatrix}
1&0&0\\
0&1&0\\
0&0&1
\end{pmatrix},
\quad
\eta_1=
\begin{pmatrix}
1&0\\
0&1
\end{pmatrix},
\quad
\lambda_1=+1,
\quad
r_1=\{0,0,0\},
\]
\[
g_2=(R_2|r_2),\quad R_2=
\begin{pmatrix}
-1&0&0\\
0&-1&0\\
0&0&1
\end{pmatrix},
\quad
\eta_2=
\begin{pmatrix}
-1&0\\
0&-1
\end{pmatrix},
\quad
\lambda_2=+1,
\quad
r_2=\{0,0,0\},
\]
\[
g_3=(R_3|r_3),\quad R_3=
\begin{pmatrix}
1&0&0\\
0&-1&0\\
0&0&1
\end{pmatrix},
\quad
\eta_3=
\begin{pmatrix}
1&0\\
0&-1
\end{pmatrix},
\quad
\lambda_3=+1,
\quad
r_3=\{\tfrac12,0,0\},
\]
\[
g_4=(R_4|r_4),\quad R_4=
\begin{pmatrix}
-1&0&0\\
0&1&0\\
0&0&1
\end{pmatrix},
\quad
\eta_4=
\begin{pmatrix}
-1&0\\
0&1
\end{pmatrix},
\quad
\lambda_4=+1,
\quad
r_4=\{\tfrac12,0,0\}.
\]

With the ordering $(g_1,g_2,g_3,g_4)$, we have:
\[
\bigl\{\mathrm{Tr}(\eta_i)\bigr\}_{i=1}^4=\{2,-2,0,0\},\qquad
\bigl\{\det(\eta_i)\bigr\}_{i=1}^4=\{1,1,-1,-1\},\qquad
\{\lambda_i\}_{i=1}^4=\{+1,+1,+1,+1\}.
\]

There are four $1$-dimensional irreducible representations of $G/T$.
The character table is
\texttt{charTable[[j,i]]} = character of irrep $j$ evaluated at element $g_i$:
\[
\texttt{charTable}=
\left\{
\{1,\ 1,\ 1,\ 1\},\;
\{1,\ -1,\ 1,\ -1\},\;
\{1,\ 1,\ -1,\ -1\},\;
\{1,\ -1,\ -1,\ 1\}
\right\}.
\]

Fixed points of $g_1$: every point is fixed.

Fixed points of $g_2$: four lines parallel to $b_3$ passing through
\[
(u,v)\in\left\{(0,0),\ (\tfrac12,0),\ (0,\tfrac12),\ (\tfrac12,\tfrac12)\right\},
\qquad w\in\mathbb R.
\]

Fixed points of $g_3$: no fixed points (up to lattice translations) within a unit cell.

Fixed points of $g_4$: two planes parallel to $\mathrm{span}\{b_2,b_3\}$ with
\[
u\in\left\{\tfrac14,\ \tfrac34\right\},\qquad v\in\mathbb R,\qquad w\in\mathbb R.
\]
\subsubsection{Mode Structure}
\begin{itemize}
    \item Two linear $(E,H)$ modes in the $\chi_{1}$ sector with diagonal $E$ and diagonal $H$. One $(E,H)$ mode with off-diagonal $E$, off-diagonal $H$ in the $\chi_{3}$ sector.
\end{itemize}

\subsection{\texorpdfstring{$pm2_{1}b$}{pm2-1b}}
\subsubsection{Group Structure}
The Bravais lattice of $G=pm2_{1}b$ is generated by
\[
b_1=\begin{pmatrix}a\\0\\0\end{pmatrix},\qquad
b_2=\begin{pmatrix}0\\ b\\0\end{pmatrix},\qquad
b_3=\hat z=\begin{pmatrix}0\\0\\1\end{pmatrix}.
\]
A convenient set of coset representatives for $G/T$ under basis $b_1,b_2,b_3$ is,
\[
g_1=(R_1|r_1),\quad R_1=
\begin{pmatrix}
1&0&0\\
0&1&0\\
0&0&1
\end{pmatrix},
\quad
\eta_1=
\begin{pmatrix}
1&0\\
0&1
\end{pmatrix},
\quad
\lambda_1=+1,
\quad
r_1=\{0,0,0\},
\]
\[
g_2=(R_2|r_2),\quad R_2=
\begin{pmatrix}
-1&0&0\\
0&1&0\\
0&0&1
\end{pmatrix},
\quad
\eta_2=
\begin{pmatrix}
-1&0\\
0&1
\end{pmatrix},
\quad
\lambda_2=+1,
\quad
r_2=\{0,0,0\},
\]
\[
g_3=(R_3|r_3),\quad R_3=
\begin{pmatrix}
-1&0&0\\
0&1&0\\
0&0&-1
\end{pmatrix},
\quad
\eta_3=
\begin{pmatrix}
-1&0\\
0&1
\end{pmatrix},
\quad
\lambda_3=-1,
\quad
r_3=\{0,\tfrac12,0\},
\]
\[
g_4=(R_4|r_4),\quad R_4=
\begin{pmatrix}
1&0&0\\
0&1&0\\
0&0&-1
\end{pmatrix},
\quad
\eta_4=
\begin{pmatrix}
1&0\\
0&1
\end{pmatrix},
\quad
\lambda_4=-1,
\quad
r_4=\{0,\tfrac12,0\}.
\]

With the ordering $(g_1,g_2,g_3,g_4)$, we have:
\[
\bigl\{\mathrm{Tr}(\eta_i)\bigr\}_{i=1}^4=\{2,0,0,2\},\qquad
\bigl\{\det(\eta_i)\bigr\}_{i=1}^4=\{1,-1,-1,1\},\qquad
\{\lambda_i\}_{i=1}^4=\{+1,+1,-1,-1\}.
\]

There are four $1$-dimensional irreducible representations of $G/T$.
The character table is
\texttt{charTable[[j,i]]} = character of irrep $j$ evaluated at element $g_i$:
\[
\texttt{charTable}=
\left\{
\{1,\ 1,\ 1,\ 1\},\;
\{1,\ -1,\ 1,\ -1\},\;
\{1,\ 1,\ -1,\ -1\},\;
\{1,\ -1,\ -1,\ 1\}
\right\}.
\]

Fixed points of $g_1$: every point is fixed.

Fixed points of $g_2$: two planes parallel to $\mathrm{span}\{b_2,b_3\}$ with
\[
u\in\left\{0,\ \tfrac12\right\},\qquad v\in\mathbb R,\qquad w\in\mathbb R.
\]

Fixed points of $g_3$: no fixed points (up to lattice translations) within a unit cell.

Fixed points of $g_4$: no fixed points (up to lattice translations) within a unit cell.
\subsubsection{Mode Structure}
\begin{itemize}
    \item\textbf{One  nonlinear symmetry-preserving pure $E$ mode} with diagonal $E$ in the $\chi_{1}$ sector. One pure $H$ mode with off-diagonal $H$ in the $\chi_{2}$ sector. One pure $H$ mode with diagonal $H$ in the $\chi_{3}$ sector.
\end{itemize}

\subsection{\texorpdfstring{$pm2_{1}n$}{pm2-1n}}
\subsubsection{Group Structure}
The Bravais lattice of $G=pm2_{1}n$ is generated by
\[
b_1=\begin{pmatrix}a\\0\\0\end{pmatrix},\qquad
b_2=\begin{pmatrix}0\\ b\\0\end{pmatrix},\qquad
b_3=\hat z=\begin{pmatrix}0\\0\\1\end{pmatrix}.
\]
A convenient set of coset representatives for $G/T$ under basis $b_1,b_2,b_3$ is,
\[
g_1=(R_1|r_1),\quad R_1=
\begin{pmatrix}
1&0&0\\
0&1&0\\
0&0&1
\end{pmatrix},
\quad
\eta_1=
\begin{pmatrix}
1&0\\
0&1
\end{pmatrix},
\quad
\lambda_1=+1,
\quad
r_1=\{0,0,0\},
\]
\[
g_2=(R_2|r_2),\quad R_2=
\begin{pmatrix}
-1&0&0\\
0&1&0\\
0&0&1
\end{pmatrix},
\quad
\eta_2=
\begin{pmatrix}
-1&0\\
0&1
\end{pmatrix},
\quad
\lambda_2=+1,
\quad
r_2=\{0,0,0\},
\]
\[
g_3=(R_3|r_3),\quad R_3=
\begin{pmatrix}
-1&0&0\\
0&1&0\\
0&0&-1
\end{pmatrix},
\quad
\eta_3=
\begin{pmatrix}
-1&0\\
0&1
\end{pmatrix},
\quad
\lambda_3=-1,
\quad
r_3=\{\tfrac12,\tfrac12,0\},
\]
\[
g_4=(R_4|r_4),\quad R_4=
\begin{pmatrix}
1&0&0\\
0&1&0\\
0&0&-1
\end{pmatrix},
\quad
\eta_4=
\begin{pmatrix}
1&0\\
0&1
\end{pmatrix},
\quad
\lambda_4=-1,
\quad
r_4=\{\tfrac12,\tfrac12,0\}.
\]

With the ordering $(g_1,g_2,g_3,g_4)$, we have:
\[
\bigl\{\mathrm{Tr}(\eta_i)\bigr\}_{i=1}^4=\{2,0,0,2\},\qquad
\bigl\{\det(\eta_i)\bigr\}_{i=1}^4=\{1,-1,-1,1\},\qquad
\{\lambda_i\}_{i=1}^4=\{+1,+1,-1,-1\}.
\]

There are four $1$-dimensional irreducible representations of $G/T$.
The character table is
\texttt{charTable[[j,i]]} = character of irrep $j$ evaluated at element $g_i$:
\[
\texttt{charTable}=
\left\{
\{1,\ 1,\ 1,\ 1\},\;
\{1,\ -1,\ 1,\ -1\},\;
\{1,\ 1,\ -1,\ -1\},\;
\{1,\ -1,\ -1,\ 1\}
\right\}.
\]

Fixed points of $g_1$: every point is fixed.

Fixed points of $g_2$: two planes parallel to $\mathrm{span}\{b_2,b_3\}$ with
\[
u\in\left\{0,\ \tfrac12\right\},\qquad v\in\mathbb R,\qquad w\in\mathbb R.
\]

Fixed points of $g_3$: no fixed points (up to lattice translations) within a unit cell.

Fixed points of $g_4$: no fixed points (up to lattice translations) within a unit cell.
\subsubsection{Mode Structure}
\begin{itemize}
    \item\textbf{One nonlinear symmetry-preserving pure $E$ mode} with diagonal $E$ in the $\chi_{1}$ sector. One pure $H$ mode with off-diagonal $H$ in the $\chi_{2}$ sector. One pure $H$ mode with diagonal $H$ in the $\chi_{3}$ sector.
\end{itemize}

\subsection{\texorpdfstring{$pb2_{1}a$}{pb2-1a}}
\subsubsection{Group Structure}
The translation lattice of $G=pb2_{1}a$ is generated by
\[
b_1=\begin{pmatrix}a\\0\\0\end{pmatrix},\qquad
b_2=\begin{pmatrix}0\\ b\\0\end{pmatrix}.
\]
We complete these vectors to a basis of $\mathbb R^3$ by setting
\[
b_3=\hat z=\begin{pmatrix}0\\0\\1\end{pmatrix}.
\]
A convenient set of coset representatives for $G/T$ under basis $b_1,b_2,b_3$ is,
\[
g_1=(R_1|r_1),\quad R_1=
\begin{pmatrix}
1&0&0\\
0&1&0\\
0&0&1
\end{pmatrix},
\quad
\eta_1=
\begin{pmatrix}
1&0\\
0&1
\end{pmatrix},
\quad
\lambda_1=+1,
\quad
r_1=\{0,0,0\},
\]
\[
g_2=(R_2|r_2),\quad R_2=
\begin{pmatrix}
-1&0&0\\
0&1&0\\
0&0&1
\end{pmatrix},
\quad
\eta_2=
\begin{pmatrix}
-1&0\\
0&1
\end{pmatrix},
\quad
\lambda_2=+1,
\quad
r_2=\{\tfrac12,\tfrac12,0\},
\]
\[
g_3=(R_3|r_3),\quad R_3=
\begin{pmatrix}
-1&0&0\\
0&1&0\\
0&0&-1
\end{pmatrix},
\quad
\eta_3=
\begin{pmatrix}
-1&0\\
0&1
\end{pmatrix},
\quad
\lambda_3=-1,
\quad
r_3=\{0,\tfrac12,0\},
\]
\[
g_4=(R_4|r_4),\quad R_4=
\begin{pmatrix}
1&0&0\\
0&1&0\\
0&0&-1
\end{pmatrix},
\quad
\eta_4=
\begin{pmatrix}
1&0\\
0&1
\end{pmatrix},
\quad
\lambda_4=-1,
\quad
r_4=\{\tfrac12,0,0\}.
\]

With the ordering $(g_1,g_2,g_3,g_4)$, we have:
\[
\bigl\{\mathrm{Tr}(\eta_i)\bigr\}_{i=1}^4
=\{2,0,0,2\},\qquad
\bigl\{\det(\eta_i)\bigr\}_{i=1}^4
=\{1,-1,-1,1\},\qquad
\{\lambda_i\}_{i=1}^4
=\{+1,+1,-1,-1\}.
\]

There are four $1$-dimensional irreducible representations of $G/T$.
The character table is
\texttt{charTable[[j,i]]} = character of irrep $j$ evaluated at element $g_i$:
\[
\texttt{charTable}=
\left\{
\{1,\ 1,\ 1,\ 1\},\;
\{1,\ -1,\ 1,\ -1\},\;
\{1,\ 1,\ -1,\ -1\},\;
\{1,\ -1,\ -1,\ 1\}
\right\}.
\]

Fixed points of $g_1$: every point is fixed.

Fixed points of $g_2$: no fixed points (up to lattice translations) within a unit cell.

Fixed points of $g_3$: no fixed points (up to lattice translations) within a unit cell.

Fixed points of $g_4$: no fixed points (up to lattice translations) within a unit cell.

\subsubsection{Mode Structure}
\begin{itemize}
    \item \textbf{One nonlinear symmetry-preserving pure $E$ mode} in the $\chi_{1}$ sector with diagonal $E$. One pure $H$ mode in the $\chi_{2}$ sector with off-diagonal $H$. One pure $H$ mode in the $\chi_{3}$ sector with diagonal $H$.
\end{itemize}

\subsection{\texorpdfstring{$pbma$}{pbma}}
\subsubsection{Group Structure}
The Bravais lattice of $G=pbma$ is generated by
\[
b_1=\begin{pmatrix}a\\0\\0\end{pmatrix},\qquad
b_2=\begin{pmatrix}0\\ b\\0\end{pmatrix},\qquad
b_3=\hat z=\begin{pmatrix}0\\0\\1\end{pmatrix}.
\]
A convenient set of coset representatives for $G/T$ under basis $b_1,b_2,b_3$ is,
\[
g_1=(R_1|r_1),\quad R_1=
\begin{pmatrix}
1&0&0\\
0&1&0\\
0&0&1
\end{pmatrix},
\quad
\eta_1=
\begin{pmatrix}
1&0\\
0&1
\end{pmatrix},
\quad
\lambda_1=+1,
\quad
r_1=\{0,0,0\},
\]
\[
g_2=(R_2|r_2),\quad R_2=
\begin{pmatrix}
-1&0&0\\
0&1&0\\
0&0&1
\end{pmatrix},
\quad
\eta_2=
\begin{pmatrix}
-1&0\\
0&1
\end{pmatrix},
\quad
\lambda_2=+1,
\quad
r_2=\{0,\tfrac12,0\},
\]
\[
g_3=(R_3|r_3),\quad R_3=
\begin{pmatrix}
1&0&0\\
0&-1&0\\
0&0&1
\end{pmatrix},
\quad
\eta_3=
\begin{pmatrix}
1&0\\
0&-1
\end{pmatrix},
\quad
\lambda_3=+1,
\quad
r_3=\{0,0,0\},
\]
\[
g_4=(R_4|r_4),\quad R_4=
\begin{pmatrix}
-1&0&0\\
0&-1&0\\
0&0&1
\end{pmatrix},
\quad
\eta_4=
\begin{pmatrix}
-1&0\\
0&-1
\end{pmatrix},
\quad
\lambda_4=+1,
\quad
r_4=\{0,\tfrac12,0\},
\]
\[
g_5=(R_5|r_5),\quad R_5=
\begin{pmatrix}
1&0&0\\
0&1&0\\
0&0&-1
\end{pmatrix},
\quad
\eta_5=
\begin{pmatrix}
1&0\\
0&1
\end{pmatrix},
\quad
\lambda_5=-1,
\quad
r_5=\{\tfrac12,0,0\},
\]
\[
g_6=(R_6|r_6),\quad R_6=
\begin{pmatrix}
-1&0&0\\
0&1&0\\
0&0&-1
\end{pmatrix},
\quad
\eta_6=
\begin{pmatrix}
-1&0\\
0&1
\end{pmatrix},
\quad
\lambda_6=-1,
\quad
r_6=\{\tfrac12,\tfrac12,0\},
\]
\[
g_7=(R_7|r_7),\quad R_7=
\begin{pmatrix}
1&0&0\\
0&-1&0\\
0&0&-1
\end{pmatrix},
\quad
\eta_7=
\begin{pmatrix}
1&0\\
0&-1
\end{pmatrix},
\quad
\lambda_7=-1,
\quad
r_7=\{\tfrac12,0,0\},
\]
\[
g_8=(R_8|r_8),\quad R_8=
\begin{pmatrix}
-1&0&0\\
0&-1&0\\
0&0&-1
\end{pmatrix},
\quad
\eta_8=
\begin{pmatrix}
-1&0\\
0&-1
\end{pmatrix},
\quad
\lambda_8=-1,
\quad
r_8=\{\tfrac12,\tfrac12,0\}.
\]

With the ordering $(g_1,g_2,g_3,g_4,g_5,g_6,g_7,g_8)$, we have:
\[
\begin{aligned}
&\bigl\{\mathrm{Tr}(\eta_i)\bigr\}_{i=1}^8=\{2,0,0,-2,2,0,0,-2\},\\
&\bigl\{\det(\eta_i)\bigr\}_{i=1}^8=\{1,-1,-1,1,1,-1,-1,1\},\\
&\{\lambda_i\}_{i=1}^8=\{+1,+1,+1,+1,-1,-1,-1,-1\}.
\end{aligned}
\]

There are eight $1$-dimensional irreducible representations of $G/T$.
The character table is
\texttt{charTable[[j,i]]} = character of irrep $j$ evaluated at element $g_i$:
\[
\begin{aligned}
\texttt{charTable}=
\{&
\{1,\ 1,\ 1,\ 1,\ 1,\ 1,\ 1,\ 1\},\\
&\{1,\ -1,\ 1,\ -1,\ 1,\ -1,\ 1,\ -1\},\\
&\{1,\ 1,\ -1,\ -1,\ 1,\ 1,\ -1,\ -1\},\\
&\{1,\ -1,\ -1,\ 1,\ 1,\ -1,\ -1,\ 1\},\\
&\{1,\ 1,\ 1,\ 1,\ -1,\ -1,\ -1,\ -1\},\\
&\{1,\ -1,\ 1,\ -1,\ -1,\ 1,\ -1,\ 1\},\\
&\{1,\ 1,\ -1,\ -1,\ -1,\ -1,\ 1,\ 1\},\\
&\{1,\ -1,\ -1,\ 1,\ -1,\ 1,\ 1,\ -1\}
\}.
\end{aligned}
\]

Fixed points of $g_1$: every point is fixed.

Fixed points of $g_2$: no fixed points (up to lattice translations) within a unit cell.

Fixed points of $g_3$: two planes parallel to $\mathrm{span}\{b_1,b_3\}$ given by
\[
v\in\left\{0,\ \tfrac12\right\},\qquad u\in\mathbb R,\qquad w\in\mathbb R.
\]

Fixed points of $g_4$: four lines parallel to $b_3$ passing through
\[
(u,v)\in\left\{(0,\tfrac14),\ (\tfrac12,\tfrac14),\ (0,\tfrac34),\ (\tfrac12,\tfrac34)\right\},
\qquad w\in\mathbb R.
\]

Fixed points of $g_5$: no fixed points (up to lattice translations) within a unit cell.

Fixed points of $g_6$: no fixed points (up to lattice translations) within a unit cell.

Fixed points of $g_7$: no fixed points (up to lattice translations) within a unit cell.

Fixed points of $g_8$:
\[
\begin{aligned}
&(\tfrac14,\tfrac14,0),\ (\tfrac34,\tfrac14,0),\ (\tfrac14,\tfrac34,0),\ (\tfrac34,\tfrac34,0).
\end{aligned}
\]
\subsubsection{Mode Structure}
\begin{itemize}
    \item When all 4 inversion centers are occupied by edge centers: \textbf{one nonlinear symmetry-preserving pure $E$ mode} in the $\chi_{1}$ sector with diagonal $E$. One pure $H$ mode  in the $\chi_{5}$ sector with diagonal $H$. One pure $H$ mode in the $\chi_{8}$ sector, with off-diagonal $H$.
    \item Otherwise, two pure $H$ mode  in the $\chi_{5}$ sector with diagonal $H$. One pure $H$ mode  in the $\chi_{8}$ sector, with off-diagonal $H$. 
\end{itemize}

\subsection{\texorpdfstring{$pmmn$}{pmmn}}
\subsubsection{Group Structure}
The Bravais lattice of $G=pmmn$ is generated by
\[
b_1=\begin{pmatrix}a\\0\\0\end{pmatrix},\qquad
b_2=\begin{pmatrix}0\\ b\\0\end{pmatrix},\qquad
b_3=\hat z=\begin{pmatrix}0\\0\\1\end{pmatrix}.
\]
A convenient set of coset representatives for $G/T$ under basis $b_1,b_2,b_3$ is,
\[
g_1=(R_1|r_1),\quad R_1=
\begin{pmatrix}
1&0&0\\
0&1&0\\
0&0&1
\end{pmatrix},
\quad
\eta_1=
\begin{pmatrix}
1&0\\
0&1
\end{pmatrix},
\quad
\lambda_1=+1,
\quad
r_1=\{0,0,0\},
\]
\[
g_2=(R_2|r_2),\quad R_2=
\begin{pmatrix}
-1&0&0\\
0&1&0\\
0&0&1
\end{pmatrix},
\quad
\eta_2=
\begin{pmatrix}
-1&0\\
0&1
\end{pmatrix},
\quad
\lambda_2=+1,
\quad
r_2=\{0,0,0\},
\]
\[
g_3=(R_3|r_3),\quad R_3=
\begin{pmatrix}
1&0&0\\
0&-1&0\\
0&0&1
\end{pmatrix},
\quad
\eta_3=
\begin{pmatrix}
1&0\\
0&-1
\end{pmatrix},
\quad
\lambda_3=+1,
\quad
r_3=\{0,0,0\},
\]
\[
g_4=(R_4|r_4),\quad R_4=
\begin{pmatrix}
-1&0&0\\
0&-1&0\\
0&0&1
\end{pmatrix},
\quad
\eta_4=
\begin{pmatrix}
-1&0\\
0&-1
\end{pmatrix},
\quad
\lambda_4=+1,
\quad
r_4=\{0,0,0\},
\]
\[
g_5=(R_5|r_5),\quad R_5=
\begin{pmatrix}
1&0&0\\
0&1&0\\
0&0&-1
\end{pmatrix},
\quad
\eta_5=
\begin{pmatrix}
1&0\\
0&1
\end{pmatrix},
\quad
\lambda_5=-1,
\quad
r_5=\{\tfrac12,\tfrac12,0\},
\]
\[
g_6=(R_6|r_6),\quad R_6=
\begin{pmatrix}
-1&0&0\\
0&1&0\\
0&0&-1
\end{pmatrix},
\quad
\eta_6=
\begin{pmatrix}
-1&0\\
0&1
\end{pmatrix},
\quad
\lambda_6=-1,
\quad
r_6=\{\tfrac12,\tfrac12,0\},
\]
\[
g_7=(R_7|r_7),\quad R_7=
\begin{pmatrix}
1&0&0\\
0&-1&0\\
0&0&-1
\end{pmatrix},
\quad
\eta_7=
\begin{pmatrix}
1&0\\
0&-1
\end{pmatrix},
\quad
\lambda_7=-1,
\quad
r_7=\{\tfrac12,\tfrac12,0\},
\]
\[
g_8=(R_8|r_8),\quad R_8=
\begin{pmatrix}
-1&0&0\\
0&-1&0\\
0&0&-1
\end{pmatrix},
\quad
\eta_8=
\begin{pmatrix}
-1&0\\
0&-1
\end{pmatrix},
\quad
\lambda_8=-1,
\quad
r_8=\{\tfrac12,\tfrac12,0\}.
\]

With the ordering $(g_1,g_2,g_3,g_4,g_5,g_6,g_7,g_8)$, we have:
\[
\begin{aligned}
&\bigl\{\mathrm{Tr}(\eta_i)\bigr\}_{i=1}^8=\{2,0,0,-2,2,0,0,-2\},\\
&\bigl\{\det(\eta_i)\bigr\}_{i=1}^8=\{1,-1,-1,1,1,-1,-1,1\},\\
&\{\lambda_i\}_{i=1}^8=\{+1,+1,+1,+1,-1,-1,-1,-1\}.
\end{aligned}
\]

There are eight $1$-dimensional irreducible representations of $G/T$.
The character table is
\texttt{charTable[[j,i]]} = character of irrep $j$ evaluated at element $g_i$:
\[
\begin{aligned}
\texttt{charTable}=
\{&
\{1,\ 1,\ 1,\ 1,\ 1,\ 1,\ 1,\ 1\},\\
&\{1,\ -1,\ 1,\ -1,\ 1,\ -1,\ 1,\ -1\},\\
&\{1,\ 1,\ -1,\ -1,\ 1,\ 1,\ -1,\ -1\},\\
&\{1,\ -1,\ -1,\ 1,\ 1,\ -1,\ -1,\ 1\},\\
&\{1,\ 1,\ 1,\ 1,\ -1,\ -1,\ -1,\ -1\},\\
&\{1,\ -1,\ 1,\ -1,\ -1,\ 1,\ -1,\ 1\},\\
&\{1,\ 1,\ -1,\ -1,\ -1,\ -1,\ 1,\ 1\},\\
&\{1,\ -1,\ -1,\ 1,\ -1,\ 1,\ 1,\ -1\}
\}.
\end{aligned}
\]

Fixed points of $g_1$: every point is fixed.

Fixed points of $g_2$: two planes parallel to $\mathrm{span}\{b_2,b_3\}$ given by
\[
u\in\left\{0,\ \tfrac12\right\},\qquad v\in\mathbb R,\qquad w\in\mathbb R.
\]

Fixed points of $g_3$: two planes parallel to $\mathrm{span}\{b_1,b_3\}$ given by
\[
v\in\left\{0,\ \tfrac12\right\},\qquad u\in\mathbb R,\qquad w\in\mathbb R.
\]

Fixed points of $g_4$: four lines parallel to $b_3$ passing through
\[
(u,v)\in\left\{(0,0),\ (\tfrac12,0),\ (0,\tfrac12),\ (\tfrac12,\tfrac12)\right\},
\qquad w\in\mathbb R.
\]

Fixed points of $g_5$: no fixed points (up to lattice translations) within a unit cell.

Fixed points of $g_6$: no fixed points (up to lattice translations) within a unit cell.

Fixed points of $g_7$: no fixed points (up to lattice translations) within a unit cell.

Fixed points of $g_8$:
\[
\begin{aligned}
&(\tfrac14,\tfrac14,0),\ (\tfrac34,\tfrac14,0),\ (\tfrac14,\tfrac34,0),\ (\tfrac34,\tfrac34,0).
\end{aligned}
\]
\subsubsection{Mode Structure}
\begin{itemize}
    \item When all 4 inversion centers are occupied by edge centers: \textbf{one nonlinear symmetry-preserving pure $E$ mode} in the $\chi_{1}$ sector with diagonal $E$. One pure $H$ mode in the $\chi_{5}$ sector with diagonal $H$. One pure $H$ mode  in the $\chi_{8}$ sector, with off-diagonal $H$.
    \item Otherwise, two pure $H$ mode  in the $\chi_{5}$ sector with diagonal $H$. One pure $H$ mode in the $\chi_{8}$ sector, with off-diagonal $H$. 
\end{itemize}

\section{Tetragonal Layer Groups}
\subsection{\texorpdfstring{$p4$}{p4}}
\subsubsection{Group Structure}
The Bravais lattice of $G=p4$ is generated by
\[
b_1=\begin{pmatrix}a\\0\\0\end{pmatrix},\qquad
b_2=\begin{pmatrix}0\\ a\\0\end{pmatrix},\qquad
b_3=\hat z=\begin{pmatrix}0\\0\\1\end{pmatrix}.
\]
A convenient set of coset representatives for $G/T$ under basis $b_1,b_2,b_3$ is,
\[
g_1=(R_1|r_1),\quad R_1=
\begin{pmatrix}
1&0&0\\
0&1&0\\
0&0&1
\end{pmatrix},
\quad
\eta_1=
\begin{pmatrix}
1&0\\
0&1
\end{pmatrix},
\quad
\lambda_1=+1,
\quad
r_1=\{0,0,0\},
\]
\[
g_2=(R_2|r_2),\quad R_2=
\begin{pmatrix}
0&-1&0\\
1&0&0\\
0&0&1
\end{pmatrix},
\quad
\eta_2=
\begin{pmatrix}
0&-1\\
1&0
\end{pmatrix},
\quad
\lambda_2=+1,
\quad
r_2=\{0,0,0\},
\]
\[
g_3=(R_3|r_3),\quad R_3=
\begin{pmatrix}
-1&0&0\\
0&-1&0\\
0&0&1
\end{pmatrix},
\quad
\eta_3=
\begin{pmatrix}
-1&0\\
0&-1
\end{pmatrix},
\quad
\lambda_3=+1,
\quad
r_3=\{0,0,0\},
\]
\[
g_4=(R_4|r_4),\quad R_4=
\begin{pmatrix}
0&1&0\\
-1&0&0\\
0&0&1
\end{pmatrix},
\quad
\eta_4=
\begin{pmatrix}
0&1\\
-1&0
\end{pmatrix},
\quad
\lambda_4=+1,
\quad
r_4=\{0,0,0\}.
\]

With the ordering $(g_1,g_2,g_3,g_4)$, we have:
\[
\bigl\{\mathrm{Tr}(\eta_i)\bigr\}_{i=1}^4=\{2,0,-2,0\},\qquad
\bigl\{\det(\eta_i)\bigr\}_{i=1}^4=\{1,1,1,1\},\qquad
\{\lambda_i\}_{i=1}^4=\{+1,+1,+1,+1\}.
\]

There are four $1$-dimensional irreducible representations of $G/T$.
The character table is
\texttt{charTable[[j,i]]} = character of irrep $j$ evaluated at element $g_i$:
\[
\texttt{charTable}=
\left\{
\{1,\ 1,\ 1,\ 1\},\;
\{1,\ \mathrm{i},\ -1,\ -\mathrm{i}\},\;
\{1,\ -1,\ 1,\ -1\},\;
\{1,\ -\mathrm{i},\ -1,\ \mathrm{i}\}
\right\}.
\]

Fixed points of $g_1$: every point is fixed.

Fixed points of $g_2$: two lines parallel to $b_3$ passing through
\[
(u,v)\in\left\{(0,0),\ (\tfrac12,\tfrac12)\right\},
\qquad w\in\mathbb R.
\]

Fixed points of $g_3$: four lines parallel to $b_3$ passing through
\[
(u,v)\in\left\{(0,0),\ (\tfrac12,0),\ (0,\tfrac12),\ (\tfrac12,\tfrac12)\right\},
\qquad w\in\mathbb R.
\]

Fixed points of $g_4$: two lines parallel to $b_3$ passing through
\[
(u,v)\in\left\{(0,0),\ (\tfrac12,\tfrac12)\right\},
\qquad w\in\mathbb R.
\]
\subsubsection{Mode Structure}
\begin{itemize}
    \item One $(E,H)$ mode in the $\chi_{1}$ sector, with $E_{11}-E_{22}=E_{12}=0$ and $H_{11}-H_{22}=H_{12}=0$. Two $(E,H)$ modes in the $\chi_{3}$ sector, with $E_{11}+E_{22}=0$ and $ H_{11}+H_{22}=0$.
\end{itemize}

\subsection{\texorpdfstring{$p\bar{4}$}{p-4}}
\subsubsection{Group Structure}
The Bravais lattice of $G=p\overline{4}$ is generated by
\[
b_1=\begin{pmatrix}a\\0\\0\end{pmatrix},\qquad
b_2=\begin{pmatrix}0\\ a\\0\end{pmatrix},\qquad
b_3=\hat z=\begin{pmatrix}0\\0\\1\end{pmatrix}.
\]
A convenient set of coset representatives for $G/T$ under basis $b_1,b_2,b_3$ is,
\[
g_1=(R_1|r_1),\quad R_1=
\begin{pmatrix}
1&0&0\\
0&1&0\\
0&0&1
\end{pmatrix},
\quad
\eta_1=
\begin{pmatrix}
1&0\\
0&1
\end{pmatrix},
\quad
\lambda_1=+1,
\quad
r_1=\{0,0,0\},
\]
\[
g_2=(R_2|r_2),\quad R_2=
\begin{pmatrix}
0&1&0\\
-1&0&0\\
0&0&-1
\end{pmatrix},
\quad
\eta_2=
\begin{pmatrix}
0&1\\
-1&0
\end{pmatrix},
\quad
\lambda_2=-1,
\quad
r_2=\{0,0,0\},
\]
\[
g_3=(R_3|r_3),\quad R_3=
\begin{pmatrix}
-1&0&0\\
0&-1&0\\
0&0&1
\end{pmatrix},
\quad
\eta_3=
\begin{pmatrix}
-1&0\\
0&-1
\end{pmatrix},
\quad
\lambda_3=+1,
\quad
r_3=\{0,0,0\},
\]
\[
g_4=(R_4|r_4),\quad R_4=
\begin{pmatrix}
0&-1&0\\
1&0&0\\
0&0&-1
\end{pmatrix},
\quad
\eta_4=
\begin{pmatrix}
0&-1\\
1&0
\end{pmatrix},
\quad
\lambda_4=-1,
\quad
r_4=\{0,0,0\}.
\]

With the ordering $(g_1,g_2,g_3,g_4)$, we have:
\[
\bigl\{\mathrm{Tr}(\eta_i)\bigr\}_{i=1}^4=\{2,0,-2,0\},\qquad
\bigl\{\det(\eta_i)\bigr\}_{i=1}^4=\{1,1,1,1\},\qquad
\{\lambda_i\}_{i=1}^4=\{+1,-1,+1,-1\}.
\]

There are four $1$-dimensional irreducible representations of $G/T$.
The character table is
\texttt{charTable[[j,i]]} = character of irrep $j$ evaluated at element $g_i$:
\[
\texttt{charTable}=
\left\{
\{1,\ 1,\ 1,\ 1\},\;
\{1,\ \mathrm{i},\ -1,\ -\mathrm{i}\},\;
\{1,\ -1,\ 1,\ -1\},\;
\{1,\ -\mathrm{i},\ -1,\ \mathrm{i}\}
\right\}.
\]

Fixed points of $g_1$: every point is fixed.

Fixed points of $g_2$:
\[
\begin{aligned}
&(0,0,0),\ (\tfrac12,\tfrac12,0).
\end{aligned}
\]

Fixed points of $g_3$: four lines parallel to $b_3$ passing through
\[
(u,v)\in\left\{(0,0),\ (\tfrac12,0),\ (0,\tfrac12),\ (\tfrac12,\tfrac12)\right\},
\qquad w\in\mathbb R.
\]

Fixed points of $g_4$:
\[
\begin{aligned}
&(0,0,0),\ (\tfrac12,\tfrac12,0).
\end{aligned}
\]
\subsubsection{Mode Structure}
\begin{itemize}
    \item One pure $E$ mode in the $\chi_{3}$ sector, with $E_{11}+E_{22}=0$. One $(E,H)$ mode in the $\chi_{1}$ sector, with $E_{11}-E_{22}=E_{12}=0$ and $H_{11}+H_{22}=0$. One $(E,H)$ modes in the $\chi_{3}$ sector, with $E_{11}+E_{22}=0$ and $H_{11}-H_{22}=H_{12}=0$.
\end{itemize}

\subsection{\texorpdfstring{$p4/n$}{p4/n}}
\subsubsection{Group Structure}
The Bravais lattice of $G=p4/n$ is generated by
\[
b_1=\begin{pmatrix}a\\0\\0\end{pmatrix},\qquad
b_2=\begin{pmatrix}0\\ a\\0\end{pmatrix},\qquad
b_3=\hat z=\begin{pmatrix}0\\0\\1\end{pmatrix}.
\]
A convenient set of coset representatives for $G/T$ under basis $b_1,b_2,b_3$ is,
\[
g_1=(R_1|r_1),\quad R_1=
\begin{pmatrix}
1&0&0\\
0&1&0\\
0&0&1
\end{pmatrix},
\quad
\eta_1=
\begin{pmatrix}
1&0\\
0&1
\end{pmatrix},
\quad
\lambda_1=+1,
\quad
r_1=\{0,0,0\},
\]
\[
g_2=(R_2|r_2),\quad R_2=
\begin{pmatrix}
-1&0&0\\
0&-1&0\\
0&0&1
\end{pmatrix},
\quad
\eta_2=
\begin{pmatrix}
-1&0\\
0&-1
\end{pmatrix},
\quad
\lambda_2=+1,
\quad
r_2=\{0,0,0\},
\]
\[
g_3=(R_3|r_3),\quad R_3=
\begin{pmatrix}
0&-1&0\\
1&0&0\\
0&0&1
\end{pmatrix},
\quad
\eta_3=
\begin{pmatrix}
0&-1\\
1&0
\end{pmatrix},
\quad
\lambda_3=+1,
\quad
r_3=\{\tfrac12,\tfrac12,0\},
\]
\[
g_4=(R_4|r_4),\quad R_4=
\begin{pmatrix}
0&1&0\\
-1&0&0\\
0&0&1
\end{pmatrix},
\quad
\eta_4=
\begin{pmatrix}
0&1\\
-1&0
\end{pmatrix},
\quad
\lambda_4=+1,
\quad
r_4=\{\tfrac12,\tfrac12,0\},
\]
\[
g_5=(R_5|r_5),\quad R_5=
\begin{pmatrix}
-1&0&0\\
0&-1&0\\
0&0&-1
\end{pmatrix},
\quad
\eta_5=
\begin{pmatrix}
-1&0\\
0&-1
\end{pmatrix},
\quad
\lambda_5=-1,
\quad
r_5=\{\tfrac12,\tfrac12,0\},
\]
\[
g_6=(R_6|r_6),\quad R_6=
\begin{pmatrix}
1&0&0\\
0&1&0\\
0&0&-1
\end{pmatrix},
\quad
\eta_6=
\begin{pmatrix}
1&0\\
0&1
\end{pmatrix},
\quad
\lambda_6=-1,
\quad
r_6=\{\tfrac12,\tfrac12,0\},
\]
\[
g_7=(R_7|r_7),\quad R_7=
\begin{pmatrix}
0&1&0\\
-1&0&0\\
0&0&-1
\end{pmatrix},
\quad
\eta_7=
\begin{pmatrix}
0&1\\
-1&0
\end{pmatrix},
\quad
\lambda_7=-1,
\quad
r_7=\{0,0,0\},
\]
\[
g_8=(R_8|r_8),\quad R_8=
\begin{pmatrix}
0&-1&0\\
1&0&0\\
0&0&-1
\end{pmatrix},
\quad
\eta_8=
\begin{pmatrix}
0&-1\\
1&0
\end{pmatrix},
\quad
\lambda_8=-1,
\quad
r_8=\{0,0,0\}.
\]

With the ordering $(g_1,g_2,g_3,g_4,g_5,g_6,g_7,g_8)$, we have:
\[
\begin{aligned}
&\bigl\{\mathrm{Tr}(\eta_i)\bigr\}_{i=1}^8=\{2,-2,0,0,-2,2,0,0\},\\
&\bigl\{\det(\eta_i)\bigr\}_{i=1}^8=\{1,1,1,1,1,1,1,1\},\\
&\{\lambda_i\}_{i=1}^8=\{+1,+1,+1,+1,-1,-1,-1,-1\}.
\end{aligned}
\]

There are eight $1$-dimensional irreducible representations of $G/T$.
The character table is
\texttt{charTable[[j,i]]} = character of irrep $j$ evaluated at element $g_i$:
\[
\begin{aligned}
\texttt{charTable}=
\{&
\{1,\ 1,\ 1,\ 1,\ 1,\ 1,\ 1,\ 1\},\\
&\{1,\ -1,\ \mathrm{i},\ -\mathrm{i},\ -1,\ 1,\ -\mathrm{i},\ \mathrm{i}\},\\
&\{1,\ 1,\ -1,\ -1,\ 1,\ 1,\ -1,\ -1\},\\
&\{1,\ -1,\ -\mathrm{i},\ \mathrm{i},\ -1,\ 1,\ \mathrm{i},\ -\mathrm{i}\},\\
&\{1,\ 1,\ 1,\ 1,\ -1,\ -1,\ -1,\ -1\},\\
&\{1,\ -1,\ \mathrm{i},\ -\mathrm{i},\ 1,\ -1,\ \mathrm{i},\ -\mathrm{i}\},\\
&\{1,\ 1,\ -1,\ -1,\ -1,\ -1,\ 1,\ 1\},\\
&\{1,\ -1,\ -\mathrm{i},\ \mathrm{i},\ 1,\ -1,\ -\mathrm{i},\ \mathrm{i}\}
\}.
\end{aligned}
\]

Fixed points of $g_1$: every point is fixed.

Fixed points of $g_2$: four lines parallel to $b_3$ passing through
\[
(u,v)\in\left\{(0,0),\ (\tfrac12,0),\ (0,\tfrac12),\ (\tfrac12,\tfrac12)\right\},
\qquad w\in\mathbb R.
\]

Fixed points of $g_3$: two lines parallel to $b_3$ passing through
\[
(u,v)\in\left\{(0,\tfrac12),\ (\tfrac12,0)\right\},
\qquad w\in\mathbb R.
\]

Fixed points of $g_4$: two lines parallel to $b_3$ passing through
\[
(u,v)\in\left\{(0,\tfrac12),\ (\tfrac12,0)\right\},
\qquad w\in\mathbb R.
\]

Fixed points of $g_5$:
\[
(\tfrac14,\tfrac14,0),\ (\tfrac34,\tfrac14,0),\ (\tfrac14,\tfrac34,0),\ (\tfrac34,\tfrac34,0),
\]

Fixed points of $g_6$: no fixed points (up to lattice translations) within a unit cell.

Fixed points of $g_7$:
\[
\begin{aligned}
&(0,0,0),\ (\tfrac12,\tfrac12,0).
\end{aligned}
\]

Fixed points of $g_8$:
\[
\begin{aligned}
&(0,0,0),\ (\tfrac12,\tfrac12,0).
\end{aligned}
\]
\subsubsection{Mode Structure}
\begin{itemize}
    \item When all 4 inversion centers are occupied by edge centers: one pure $E$ mode in the $\chi_{3}$ sector with $E_{11}+E_{22}=0$. One pure $H$ mode in the $\chi_{5}$ sector with $H_{11}-H_{22}=H_{12}=0$. One pure $H$ mode in the $\chi_{7}$ sector, with $H_{11}+H_{22}=0$.
    \item Otherwise, one pure $H$ mode in the $\chi_{5}$ sector with $H_{11}-H_{22}=H_{12}=0$. Two pure $H$ modes in the $\chi_{7}$ sector, with $H_{11}+H_{22}=0$.
\end{itemize}

\subsection{\texorpdfstring{$p4mm$}{p4mm}}
\subsubsection{Group Structure}
The Bravais lattice of $G=p4mm$ is generated by
\[
b_1=\begin{pmatrix}a\\0\\0\end{pmatrix},\qquad
b_2=\begin{pmatrix}0\\ a\\0\end{pmatrix},\qquad
b_3=\hat z=\begin{pmatrix}0\\0\\1\end{pmatrix}.
\]
A convenient set of coset representatives for $G/T$ under basis $b_1,b_2,b_3$ is,
\[
g_1=(R_1|r_1),\quad R_1=
\begin{pmatrix}
1&0&0\\
0&1&0\\
0&0&1
\end{pmatrix},
\quad
\eta_1=
\begin{pmatrix}
1&0\\
0&1
\end{pmatrix},
\quad
\lambda_1=+1,
\quad
r_1=\{0,0,0\},
\]
\[
g_2=(R_2|r_2),\quad R_2=
\begin{pmatrix}
0&-1&0\\
1&0&0\\
0&0&1
\end{pmatrix},
\quad
\eta_2=
\begin{pmatrix}
0&-1\\
1&0
\end{pmatrix},
\quad
\lambda_2=+1,
\quad
r_2=\{0,0,0\},
\]
\[
g_3=(R_3|r_3),\quad R_3=
\begin{pmatrix}
-1&0&0\\
0&-1&0\\
0&0&1
\end{pmatrix},
\quad
\eta_3=
\begin{pmatrix}
-1&0\\
0&-1
\end{pmatrix},
\quad
\lambda_3=+1,
\quad
r_3=\{0,0,0\},
\]
\[
g_4=(R_4|r_4),\quad R_4=
\begin{pmatrix}
0&1&0\\
-1&0&0\\
0&0&1
\end{pmatrix},
\quad
\eta_4=
\begin{pmatrix}
0&1\\
-1&0
\end{pmatrix},
\quad
\lambda_4=+1,
\quad
r_4=\{0,0,0\},
\]
\[
g_5=(R_5|r_5),\quad R_5=
\begin{pmatrix}
-1&0&0\\
0&1&0\\
0&0&1
\end{pmatrix},
\quad
\eta_5=
\begin{pmatrix}
-1&0\\
0&1
\end{pmatrix},
\quad
\lambda_5=+1,
\quad
r_5=\{0,0,0\},
\]
\[
g_6=(R_6|r_6),\quad R_6=
\begin{pmatrix}
1&0&0\\
0&-1&0\\
0&0&1
\end{pmatrix},
\quad
\eta_6=
\begin{pmatrix}
1&0\\
0&-1
\end{pmatrix},
\quad
\lambda_6=+1,
\quad
r_6=\{0,0,0\},
\]
\[
g_7=(R_7|r_7),\quad R_7=
\begin{pmatrix}
0&1&0\\
1&0&0\\
0&0&1
\end{pmatrix},
\quad
\eta_7=
\begin{pmatrix}
0&1\\
1&0
\end{pmatrix},
\quad
\lambda_7=+1,
\quad
r_7=\{0,0,0\},
\]
\[
g_8=(R_8|r_8),\quad R_8=
\begin{pmatrix}
0&-1&0\\
-1&0&0\\
0&0&1
\end{pmatrix},
\quad
\eta_8=
\begin{pmatrix}
0&-1\\
-1&0
\end{pmatrix},
\quad
\lambda_8=+1,
\quad
r_8=\{0,0,0\}.
\]

With the ordering $(g_1,g_2,g_3,g_4,g_5,g_6,g_7,g_8)$, we have:
\[
\begin{aligned}
&\bigl\{\mathrm{Tr}(\eta_i)\bigr\}_{i=1}^8=\{2,0,-2,0,0,0,0,0\},\\
&\bigl\{\det(\eta_i)\bigr\}_{i=1}^8=\{1,1,1,1,-1,-1,-1,-1\},\\
&\{\lambda_i\}_{i=1}^8=\{+1,+1,+1,+1,+1,+1,+1,+1\}.
\end{aligned}
\]

There are five irreducible representations of $G/T$ (four $1$-dimensional and one $2$-dimensional).
The character table is
\texttt{charTable[[j,i]]} = character of irrep $j$ evaluated at element $g_i$:
\[
\begin{aligned}
\texttt{charTable}=
\{&
\{1,\ 1,\ 1,\ 1,\ 1,\ 1,\ 1,\ 1\},\\
&\{1,\ 1,\ 1,\ 1,\ -1,\ -1,\ -1,\ -1\},\\
&\{1,\ -1,\ 1,\ -1,\ 1,\ 1,\ -1,\ -1\},\\
&\{1,\ -1,\ 1,\ -1,\ -1,\ -1,\ 1,\ 1\},\\
&\{2,\ 0,\ -2,\ 0,\ 0,\ 0,\ 0,\ 0\}
\}.
\end{aligned}
\]

Fixed points of $g_1$: every point is fixed.

Fixed points of $g_2$: two lines parallel to $b_3$ passing through
\[
(u,v)\in\left\{(0,0),\ (\tfrac12,\tfrac12)\right\},
\qquad w\in\mathbb R.
\]

Fixed points of $g_3$: four lines parallel to $b_3$ passing through
\[
(u,v)\in\left\{(0,0),\ (\tfrac12,0),\ (0,\tfrac12),\ (\tfrac12,\tfrac12)\right\},
\qquad w\in\mathbb R.
\]

Fixed points of $g_4$: two lines parallel to $b_3$ passing through
\[
(u,v)\in\left\{(0,0),\ (\tfrac12,\tfrac12)\right\},
\qquad w\in\mathbb R.
\]

Fixed points of $g_5$: two planes parallel to $\mathrm{span}\{b_2,b_3\}$ given by
\[
u\in\left\{0,\ \tfrac12\right\},\qquad v\in\mathbb R,\qquad w\in\mathbb R.
\]

Fixed points of $g_6$: two planes parallel to $\mathrm{span}\{b_1,b_3\}$ given by
\[
v\in\left\{0,\ \tfrac12\right\},\qquad u\in\mathbb R,\qquad w\in\mathbb R.
\]

Fixed points of $g_7$: one plane parallel to $\mathrm{span}\{b_1+b_2,\ b_3\}$ given by
\[
u=v,\qquad w\in\mathbb R.
\]

Fixed points of $g_8$: two planes parallel to $\mathrm{span}\{b_1-b_2,\ b_3\}$ given by
\[
u+v\in\left\{0,\ 1\right\},\qquad w\in\mathbb R.
\]
\subsubsection{Mode Structure}
\begin{itemize}
    \item One $(E,H)$ mode in the $\chi_{1}$ sector with $E_{11}-E_{22}=E_{12}=0$ and $H_{11}-H_{22}=H_{12}=0$. One $(E,H)$ mode in the $\chi_{3}$ sector with $E_{11}+E_{22}=E_{12}=0$ and $H_{11}+H_{22}=H_{12}=0$. One $(E,H)$ mode in the $\chi_{4}$ sector with $E_{11}=E_{22}=0$ and $H_{11}=H_{22}=0$.
\end{itemize}
\subsection{\texorpdfstring{$p4bm$}{p4bm}}
\subsubsection{Group Structure}
The Bravais lattice of $G=p4bm$ is generated by
\[
b_1=\begin{pmatrix}a\\0\\0\end{pmatrix},\qquad
b_2=\begin{pmatrix}0\\ a\\0\end{pmatrix},\qquad
b_3=\hat z=\begin{pmatrix}0\\0\\1\end{pmatrix}.
\]
A convenient set of coset representatives for $G/T$ under basis $b_1,b_2,b_3$ is,
\[
g_1=(R_1|r_1),\quad R_1=
\begin{pmatrix}
1&0&0\\
0&1&0\\
0&0&1
\end{pmatrix},
\quad
\eta_1=
\begin{pmatrix}
1&0\\
0&1
\end{pmatrix},
\quad
\lambda_1=+1,
\quad
r_1=\{0,0,0\},
\]
\[
g_2=(R_2|r_2),\quad R_2=
\begin{pmatrix}
-1&0&0\\
0&-1&0\\
0&0&1
\end{pmatrix},
\quad
\eta_2=
\begin{pmatrix}
-1&0\\
0&-1
\end{pmatrix},
\quad
\lambda_2=+1,
\quad
r_2=\{0,0,0\},
\]
\[
g_3=(R_3|r_3),\quad R_3=
\begin{pmatrix}
0&-1&0\\
1&0&0\\
0&0&1
\end{pmatrix},
\quad
\eta_3=
\begin{pmatrix}
0&-1\\
1&0
\end{pmatrix},
\quad
\lambda_3=+1,
\quad
r_3=\{0,0,0\},
\]
\[
g_4=(R_4|r_4),\quad R_4=
\begin{pmatrix}
0&1&0\\
-1&0&0\\
0&0&1
\end{pmatrix},
\quad
\eta_4=
\begin{pmatrix}
0&1\\
-1&0
\end{pmatrix},
\quad
\lambda_4=+1,
\quad
r_4=\{0,0,0\},
\]
\[
g_5=(R_5|r_5),\quad R_5=
\begin{pmatrix}
1&0&0\\
0&-1&0\\
0&0&1
\end{pmatrix},
\quad
\eta_5=
\begin{pmatrix}
1&0\\
0&-1
\end{pmatrix},
\quad
\lambda_5=+1,
\quad
r_5=\{\tfrac12,\tfrac12,0\},
\]
\[
g_6=(R_6|r_6),\quad R_6=
\begin{pmatrix}
-1&0&0\\
0&1&0\\
0&0&1
\end{pmatrix},
\quad
\eta_6=
\begin{pmatrix}
-1&0\\
0&1
\end{pmatrix},
\quad
\lambda_6=+1,
\quad
r_6=\{\tfrac12,\tfrac12,0\},
\]
\[
g_7=(R_7|r_7),\quad R_7=
\begin{pmatrix}
0&-1&0\\
-1&0&0\\
0&0&1
\end{pmatrix},
\quad
\eta_7=
\begin{pmatrix}
0&-1\\
-1&0
\end{pmatrix},
\quad
\lambda_7=+1,
\quad
r_7=\{\tfrac12,\tfrac12,0\},
\]
\[
g_8=(R_8|r_8),\quad R_8=
\begin{pmatrix}
0&1&0\\
1&0&0\\
0&0&1
\end{pmatrix},
\quad
\eta_8=
\begin{pmatrix}
0&1\\
1&0
\end{pmatrix},
\quad
\lambda_8=+1,
\quad
r_8=\{\tfrac12,\tfrac12,0\}.
\]

With the ordering $(g_1,g_2,g_3,g_4,g_5,g_6,g_7,g_8)$, we have:
\[
\begin{aligned}
&\bigl\{\mathrm{Tr}(\eta_i)\bigr\}_{i=1}^8=\{2,-2,0,0,0,0,0,0\},\\
&\bigl\{\det(\eta_i)\bigr\}_{i=1}^8=\{1,1,1,1,-1,-1,-1,-1\},\\
&\{\lambda_i\}_{i=1}^8=\{+1,+1,+1,+1,+1,+1,+1,+1\}.
\end{aligned}
\]

There are five irreducible representations of $G/T$ (four $1$-dimensional and one $2$-dimensional).
The character table is
\texttt{charTable[[j,i]]} = character of irrep $j$ evaluated at element $g_i$:
\[
\begin{aligned}
\texttt{charTable}=
\{&
\{1,\ 1,\ 1,\ 1,\ 1,\ 1,\ 1,\ 1\},\\
&\{1,\ 1,\ 1,\ 1,\ -1,\ -1,\ -1,\ -1\},\\
&\{1,\ 1,\ -1,\ -1,\ 1,\ 1,\ -1,\ -1\},\\
&\{1,\ 1,\ -1,\ -1,\ -1,\ -1,\ 1,\ 1\},\\
&\{2,\ -2,\ 0,\ 0,\ 0,\ 0,\ 0,\ 0\}
\}.
\end{aligned}
\]

Fixed points of $g_1$: every point is fixed.

Fixed points of $g_2$: four lines parallel to $b_3$ passing through
\[
(u,v)\in\left\{(0,0),\ (\tfrac12,0),\ (0,\tfrac12),\ (\tfrac12,\tfrac12)\right\},
\qquad w\in\mathbb R.
\]

Fixed points of $g_3$: two lines parallel to $b_3$ passing through
\[
(u,v)\in\left\{(0,0),\ (\tfrac12,\tfrac12)\right\},
\qquad w\in\mathbb R.
\]

Fixed points of $g_4$: two lines parallel to $b_3$ passing through
\[
(u,v)\in\left\{(0,0),\ (\tfrac12,\tfrac12)\right\},
\qquad w\in\mathbb R.
\]

Fixed points of $g_5$: no fixed points (up to lattice translations) within a unit cell.

Fixed points of $g_6$: no fixed points (up to lattice translations) within a unit cell.

Fixed points of $g_7$: two planes parallel to $\mathrm{span}\{b_1-b_2,\ b_3\}$ given by
\[
u+v\in\left\{\tfrac12,\ \tfrac32\right\},\qquad w\in\mathbb R.
\]

Fixed points of $g_8$: two planes parallel to $\mathrm{span}\{b_1+b_2,\ b_3\}$ given by
\[
u-v\in\left\{\tfrac12,\ -\tfrac12\right\},\qquad w\in\mathbb R.
\]
\subsubsection{Mode Structure}
\begin{itemize}
    \item One $(E,H)$ mode in the $\chi_{1}$ sector with $E_{11}-E_{22}=E_{12}=0$ and $H_{11}-H_{22}=H_{12}=0$. One $(E,H)$ mode in the $\chi_{3}$ sector with $E_{11}+E_{22}=E_{12}=0$ and $H_{11}+H_{22}=H_{12}=0$. One $(E,H)$ mode in the $\chi_{4}$ sector with $E_{11}=E_{22}=0$ and $H_{11}=H_{22}=0$.
\end{itemize}

\subsection{\texorpdfstring{$p\bar{4}2_{1}m$}{p-42-1m}}
\subsubsection{Group Structure}
The Bravais lattice of $G=p\bar{4}2_{1}m$ is generated by
\[
b_1=\begin{pmatrix}a\\0\\0\end{pmatrix},\qquad
b_2=\begin{pmatrix}0\\ a\\0\end{pmatrix},\qquad
b_3=\hat z=\begin{pmatrix}0\\0\\1\end{pmatrix}.
\]
A convenient set of coset representatives for $G/T$ under basis $b_1,b_2,b_3$ is,
\[
g_1=(R_1|r_1),\quad R_1=
\begin{pmatrix}
1&0&0\\
0&1&0\\
0&0&1
\end{pmatrix},
\quad
\eta_1=
\begin{pmatrix}
1&0\\
0&1
\end{pmatrix},
\quad
\lambda_1=+1,
\quad
r_1=\{0,0,0\},
\]
\[
g_2=(R_2|r_2),\quad R_2=
\begin{pmatrix}
0&1&0\\
-1&0&0\\
0&0&-1
\end{pmatrix},
\quad
\eta_2=
\begin{pmatrix}
0&1\\
-1&0
\end{pmatrix},
\quad
\lambda_2=-1,
\quad
r_2=\{0,0,0\},
\]
\[
g_3=(R_3|r_3),\quad R_3=
\begin{pmatrix}
-1&0&0\\
0&-1&0\\
0&0&1
\end{pmatrix},
\quad
\eta_3=
\begin{pmatrix}
-1&0\\
0&-1
\end{pmatrix},
\quad
\lambda_3=+1,
\quad
r_3=\{0,0,0\},
\]
\[
g_4=(R_4|r_4),\quad R_4=
\begin{pmatrix}
0&-1&0\\
1&0&0\\
0&0&-1
\end{pmatrix},
\quad
\eta_4=
\begin{pmatrix}
0&-1\\
1&0
\end{pmatrix},
\quad
\lambda_4=-1,
\quad
r_4=\{0,0,0\},
\]
\[
g_5=(R_5|r_5),\quad R_5=
\begin{pmatrix}
1&0&0\\
0&-1&0\\
0&0&-1
\end{pmatrix},
\quad
\eta_5=
\begin{pmatrix}
1&0\\
0&-1
\end{pmatrix},
\quad
\lambda_5=-1,
\quad
r_5=\{\tfrac12,\tfrac12,0\},
\]
\[
g_6=(R_6|r_6),\quad R_6=
\begin{pmatrix}
0&-1&0\\
-1&0&0\\
0&0&1
\end{pmatrix},
\quad
\eta_6=
\begin{pmatrix}
0&-1\\
-1&0
\end{pmatrix},
\quad
\lambda_6=+1,
\quad
r_6=\{\tfrac12,\tfrac12,0\},
\]
\[
g_7=(R_7|r_7),\quad R_7=
\begin{pmatrix}
-1&0&0\\
0&1&0\\
0&0&-1
\end{pmatrix},
\quad
\eta_7=
\begin{pmatrix}
-1&0\\
0&1
\end{pmatrix},
\quad
\lambda_7=-1,
\quad
r_7=\{\tfrac12,\tfrac12,0\},
\]
\[
g_8=(R_8|r_8),\quad R_8=
\begin{pmatrix}
0&1&0\\
1&0&0\\
0&0&1
\end{pmatrix},
\quad
\eta_8=
\begin{pmatrix}
0&1\\
1&0
\end{pmatrix},
\quad
\lambda_8=+1,
\quad
r_8=\{\tfrac12,\tfrac12,0\}.
\]

With the ordering $(g_1,g_2,g_3,g_4,g_5,g_6,g_7,g_8)$, we have:
\[
\begin{aligned}
&\bigl\{\mathrm{Tr}(\eta_i)\bigr\}_{i=1}^8=\{2,0,-2,0,0,0,0,0\},\\
&\bigl\{\det(\eta_i)\bigr\}_{i=1}^8=\{1,1,1,1,-1,-1,-1,-1\},\\
&\{\lambda_i\}_{i=1}^8=\{+1,-1,+1,-1,-1,+1,-1,+1\}.
\end{aligned}
\]

There are five irreducible representations of $G/T$ (four $1$-dimensional and one $2$-dimensional).
The character table is
\texttt{charTable[[j,i]]} = character of irrep $j$ evaluated at element $g_i$:
\[
\begin{aligned}
\texttt{charTable}=
\{&
\{1,\ 1,\ 1,\ 1,\ 1,\ 1,\ 1,\ 1\},\\
&\{1,\ 1,\ 1,\ 1,\ -1,\ -1,\ -1,\ -1\},\\
&\{1,\ -1,\ 1,\ -1,\ 1,\ -1,\ 1,\ -1\},\\
&\{1,\ -1,\ 1,\ -1,\ -1,\ 1,\ -1,\ 1\},\\
&\{2,\ 0,\ -2,\ 0,\ 0,\ 0,\ 0,\ 0\}
\}.
\end{aligned}
\]

Fixed points of $g_1$: every point is fixed.

Fixed points of $g_2$:
\[
\begin{aligned}
&(0,0,0),\ (\tfrac12,\tfrac12,0).
\end{aligned}
\]

Fixed points of $g_3$: four lines parallel to $b_3$ passing through
\[
(u,v)\in\left\{(0,0),\ (\tfrac12,0),\ (0,\tfrac12),\ (\tfrac12,\tfrac12)\right\},
\qquad w\in\mathbb R.
\]

Fixed points of $g_4$:
\[
\begin{aligned}
&(0,0,0),\ (\tfrac12,\tfrac12,0).
\end{aligned}
\]

Fixed points of $g_5$: no fixed points (up to lattice translations) within a unit cell.

Fixed points of $g_6$: two planes parallel to $\mathrm{span}\{b_1-b_2,\ b_3\}$ given by
\[
u+v\in\left\{\tfrac12,\ \tfrac32\right\},\qquad w\in\mathbb R.
\]

Fixed points of $g_7$: no fixed points (up to lattice translations) within a unit cell.

Fixed points of $g_8$: two planes parallel to $\mathrm{span}\{b_1+b_2,\ b_3\}$ given by
\[
u-v\in\left\{\tfrac12,\ -\tfrac12\right\},\qquad w\in\mathbb R.
\]
\subsubsection{Mode Structure}
\begin{itemize}
    \item One pure $E$ mode in the $\chi_{4}$ sector with $E_{11}=E_{22}=0$. One pure $H$ mode in the $\chi_{2}$ sector with $H_{11}+H_{22}=H_{12}=0$. One $(E,H)$ mode in the $\chi_{4}$ sector with $E_{11}=E_{22}=0$ and $H_{11}-H_{22}=H_{12}=0$.
\end{itemize}

\section{Trigonal Layer Groups}
\subsection{\texorpdfstring{$p3$}{p3}}
\subsubsection{Group Structure}
The Bravais lattice of $G=p3$ is generated by
\[
b_1=\begin{pmatrix}a\\0\\0\end{pmatrix},\qquad
b_2=\begin{pmatrix}\tfrac{a}{2}\\ \tfrac{\sqrt3\,a}{2}\\0\end{pmatrix},\qquad
b_3=\hat z=\begin{pmatrix}0\\0\\1\end{pmatrix}.
\]
A convenient set of coset representatives for $G/T$ under basis $b_1,b_2,b_3$ is,
\[
g_1=(R_1|r_1),\quad R_1=
\begin{pmatrix}
1&0&0\\
0&1&0\\
0&0&1
\end{pmatrix},
\quad
\eta_1=
\begin{pmatrix}
1&0\\
0&1
\end{pmatrix},
\quad
\lambda_1=+1,
\quad
r_1=\{0,0,0\},
\]
\[
g_2=(R_2|r_2),\quad R_2=
\begin{pmatrix}
-1&-1&0\\
1&0&0\\
0&0&1
\end{pmatrix},
\quad
\eta_2=
\begin{pmatrix}
-1&-1\\
1&0
\end{pmatrix},
\quad
\lambda_2=+1,
\quad
r_2=\{0,0,0\},
\]
\[
g_3=(R_3|r_3),\quad R_3=
\begin{pmatrix}
0&1&0\\
-1&-1&0\\
0&0&1
\end{pmatrix},
\quad
\eta_3=
\begin{pmatrix}
0&1\\
-1&-1
\end{pmatrix},
\quad
\lambda_3=+1,
\quad
r_3=\{0,0,0\}.
\]

With the ordering $(g_1,g_2,g_3)$, we have:
\[
\begin{aligned}
&\bigl\{\mathrm{Tr}(\eta_i)\bigr\}_{i=1}^3=\{2,-1,-1\},\\
&\bigl\{\det(\eta_i)\bigr\}_{i=1}^3=\{1,1,1\},\\
&\{\lambda_i\}_{i=1}^3=\{+1,+1,+1\}.
\end{aligned}
\]

There are three $1$-dimensional irreducible representations of $G/T$.
Let $\omega=\exp\!\bigl(\tfrac{2\pi\,\mathrm{i}}{3}\bigr)$.
The character table is
\texttt{charTable[[j,i]]} = character of irrep $j$ evaluated at element $g_i$:
\[
\begin{aligned}
\texttt{charTable}=
\{&
\{1,\ 1,\ 1\},\\
&\{1,\ \omega,\ \omega^{2}\},\\
&\{1,\ \omega^{2},\ \omega\}
\}.
\end{aligned}
\]

Fixed points of $g_1$: every point is fixed.

Fixed points of $g_2$: three lines parallel to $b_3$ passing through
\[
(u,v)\in\left\{(0,0),\ \left(\tfrac13,\tfrac13\right),\ \left(\tfrac23,\tfrac23\right)\right\},
\qquad w\in\mathbb R.
\]

Fixed points of $g_3$: three lines parallel to $b_3$ passing through
\[
(u,v)\in\left\{(0,0),\ \left(\tfrac13,\tfrac13\right),\ \left(\tfrac23,\tfrac23\right)\right\},
\qquad w\in\mathbb R.
\]
\subsubsection{Mode Structure}
\begin{itemize}
    \item One $(E,H)$ mode in the $\chi_{1}$ sector with $2E_{12}=E_{11}=E_{22}$ and $2H_{12}=H_{11}=H_{22}$. One $(E,H)$ mode in the $\chi_{2}$ sector. One $(E,H)$ mode in the $\chi_{3}$ sector. They together can be recombined to a mode with $E_{11}=-E_{12},\;E_{22}=2E_{12}$ and $H_{11}=-H_{12},\;H_{22}=2H_{12}$, and a mode with $E_{11}=E_{12},\;E_{22}=0$  and $H_{11}=H_{12},\;H_{22}=0$.
\end{itemize}

\subsection{\texorpdfstring{$p\bar{3}$}{p-3}}
\subsubsection{Group Structure}
The translation lattice of $G=p\bar{3}$ is generated by
\[
b_1=\begin{pmatrix}a\\0\\0\end{pmatrix},\qquad
b_2=\begin{pmatrix}
\tfrac{a}{2}\\
\tfrac{\sqrt3\,a}{2}\\
0
\end{pmatrix}.
\]
We complete these vectors to a basis of $\mathbb R^3$ by setting
\[
b_3=\hat z=\begin{pmatrix}0\\0\\1\end{pmatrix}.
\]
A convenient set of coset representatives for $G/T$ under basis $b_1,b_2,b_3$ is,
\[
g_1=(R_1|r_1),\quad R_1=
\begin{pmatrix}
1&0&0\\
0&1&0\\
0&0&1
\end{pmatrix},
\quad
\eta_1=
\begin{pmatrix}
1&0\\
0&1
\end{pmatrix},
\quad
\lambda_1=+1,
\quad
r_1=\{0,0,0\},
\]
\[
g_2=(R_2|r_2),\quad R_2=
\begin{pmatrix}
-1&-1&0\\
1&0&0\\
0&0&1
\end{pmatrix},
\quad
\eta_2=
\begin{pmatrix}
-1&-1\\
1&0
\end{pmatrix},
\quad
\lambda_2=+1,
\quad
r_2=\{0,0,0\},
\]
\[
g_3=(R_3|r_3),\quad R_3=
\begin{pmatrix}
0&1&0\\
-1&-1&0\\
0&0&1
\end{pmatrix},
\quad
\eta_3=
\begin{pmatrix}
0&1\\
-1&-1
\end{pmatrix},
\quad
\lambda_3=+1,
\quad
r_3=\{0,0,0\},
\]
\[
g_4=(R_4|r_4),\quad R_4=
\begin{pmatrix}
-1&0&0\\
0&-1&0\\
0&0&-1
\end{pmatrix},
\quad
\eta_4=
\begin{pmatrix}
-1&0\\
0&-1
\end{pmatrix},
\quad
\lambda_4=-1,
\quad
r_4=\{0,0,0\},
\]
\[
g_5=(R_5|r_5),\quad R_5=
\begin{pmatrix}
1&1&0\\
-1&0&0\\
0&0&-1
\end{pmatrix},
\quad
\eta_5=
\begin{pmatrix}
1&1\\
-1&0
\end{pmatrix},
\quad
\lambda_5=-1,
\quad
r_5=\{0,0,0\},
\]
\[
g_6=(R_6|r_6),\quad R_6=
\begin{pmatrix}
0&-1&0\\
1&1&0\\
0&0&-1
\end{pmatrix},
\quad
\eta_6=
\begin{pmatrix}
0&-1\\
1&1
\end{pmatrix},
\quad
\lambda_6=-1,
\quad
r_6=\{0,0,0\}.
\]

With the ordering $(g_1,g_2,g_3,g_4,g_5,g_6)$, we have:
\[
\begin{aligned}
&\bigl\{\mathrm{Tr}(\eta_i)\bigr\}_{i=1}^6
=\{2,-1,-1,-2,1,1\},\\
&\bigl\{\det(\eta_i)\bigr\}_{i=1}^6
=\{1,1,1,1,1,1\},\\
&\{\lambda_i\}_{i=1}^6
=\{+1,+1,+1,-1,-1,-1\}.
\end{aligned}
\]

There are six $1$-dimensional irreducible representations of $G/T$.
Let
\[
\omega=\exp\!\left(\frac{2\pi\,\mathrm{i}}{3}\right).
\]
The character table is
\texttt{charTable[[j,i]]} = character of irrep $j$ evaluated at element $g_i$:
\[
\begin{aligned}
\texttt{charTable}=
\{&
\{1,\ 1,\ 1,\ 1,\ 1,\ 1\},\\
&\{1,\ \omega,\ \omega^{2},\ 1,\ \omega,\ \omega^{2}\},\\
&\{1,\ \omega^{2},\ \omega,\ 1,\ \omega^{2},\ \omega\},\\
&\{1,\ 1,\ 1,\ -1,\ -1,\ -1\},\\
&\{1,\ \omega,\ \omega^{2},\ -1,\ -\omega,\ -\omega^{2}\},\\
&\{1,\ \omega^{2},\ \omega,\ -1,\ -\omega^{2},\ -\omega\}
\}.
\end{aligned}
\]

Fixed points of $g_1$: every point is fixed.

Fixed points of $g_2$: three lines parallel to $b_3$ passing through
\[
(u,v)\in
\left\{
(0,0),\quad
\left(\tfrac13,\tfrac13\right),\quad
\left(\tfrac23,\tfrac23\right)
\right\},
\qquad w\in\mathbb R.
\]

Fixed points of $g_3$: three lines parallel to $b_3$ passing through
\[
(u,v)\in
\left\{
(0,0),\quad
\left(\tfrac13,\tfrac13\right),\quad
\left(\tfrac23,\tfrac23\right)
\right\},
\qquad w\in\mathbb R.
\]

Fixed points of $g_4$:
\[
\begin{aligned}
&(0,0,0),\quad
(\tfrac12,0,0),\quad
(0,\tfrac12,0),\quad
(\tfrac12,\tfrac12,0).
\end{aligned}
\]

Fixed points of $g_5$:
\[
(0,0,0).
\]

Fixed points of $g_6$:
\[
(0,0,0).
\]

\subsubsection{Mode Structure}
\begin{itemize}
    \item The inversion center at $(0,0,0)$ is also fixed by $g_5$ and $g_6$, and therefore must be occupied by a vertex. Consequently, there is one pure $H$ mode in the $\chi_{4}$ sector with $2H_{12}=H_{11}=H_{22}$, one pure $H$ mode in the $\chi_{5}$ sector, and one pure $H$ mode in the $\chi_{6}$ sector. The $\chi_{5}$ and $\chi_{6}$ modes can be recombined into two real modes: one with $H_{11}=-H_{12},\;  H_{22}=2H_{12}$, and one with $H_{11}=H_{12},\; H_{22}=0$.
\end{itemize}

\subsection{\texorpdfstring{$p3m1$}{p3m1}}
\subsubsection{Group Structure}
The Bravais lattice of $G=p3m1$ is generated by
\[
b_1=\begin{pmatrix}a\\0\\0\end{pmatrix},\qquad
b_2=\begin{pmatrix}\tfrac{a}{2}\\ \tfrac{\sqrt3\,a}{2}\\0\end{pmatrix},\qquad
b_3=\hat z=\begin{pmatrix}0\\0\\1\end{pmatrix}.
\]
A convenient set of coset representatives for $G/T$ under basis $b_1,b_2,b_3$ is,
\[
g_1=(R_1|r_1),\quad R_1=
\begin{pmatrix}
1&0&0\\
0&1&0\\
0&0&1
\end{pmatrix},
\quad
\eta_1=
\begin{pmatrix}
1&0\\
0&1
\end{pmatrix},
\quad
\lambda_1=+1,
\quad
r_1=\{0,0,0\},
\]
\[
g_2=(R_2|r_2),\quad R_2=
\begin{pmatrix}
-1&-1&0\\
1&0&0\\
0&0&1
\end{pmatrix},
\quad
\eta_2=
\begin{pmatrix}
-1&-1\\
1&0
\end{pmatrix},
\quad
\lambda_2=+1,
\quad
r_2=\{0,0,0\},
\]
\[
g_3=(R_3|r_3),\quad R_3=
\begin{pmatrix}
0&1&0\\
-1&-1&0\\
0&0&1
\end{pmatrix},
\quad
\eta_3=
\begin{pmatrix}
0&1\\
-1&-1
\end{pmatrix},
\quad
\lambda_3=+1,
\quad
r_3=\{0,0,0\},
\]
\[
g_4=(R_4|r_4),\quad R_4=
\begin{pmatrix}
1&1&0\\
0&-1&0\\
0&0&1
\end{pmatrix},
\quad
\eta_4=
\begin{pmatrix}
1&1\\
0&-1
\end{pmatrix},
\quad
\lambda_4=+1,
\quad
r_4=\{0,0,0\},
\]
\[
g_5=(R_5|r_5),\quad R_5=
\begin{pmatrix}
-1&0&0\\
1&1&0\\
0&0&1
\end{pmatrix},
\quad
\eta_5=
\begin{pmatrix}
-1&0\\
1&1
\end{pmatrix},
\quad
\lambda_5=+1,
\quad
r_5=\{0,0,0\},
\]
\[
g_6=(R_6|r_6),\quad R_6=
\begin{pmatrix}
0&-1&0\\
-1&0&0\\
0&0&1
\end{pmatrix},
\quad
\eta_6=
\begin{pmatrix}
0&-1\\
-1&0
\end{pmatrix},
\quad
\lambda_6=+1,
\quad
r_6=\{0,0,0\}.
\]

With the ordering $(g_1,g_2,g_3,g_4,g_5,g_6)$, we have:
\[
\begin{aligned}
&\bigl\{\mathrm{Tr}(\eta_i)\bigr\}_{i=1}^6=\{2,-1,-1,0,0,0\},\\
&\bigl\{\det(\eta_i)\bigr\}_{i=1}^6=\{1,1,1,-1,-1,-1\},\\
&\{\lambda_i\}_{i=1}^6=\{+1,+1,+1,+1,+1,+1\}.
\end{aligned}
\]

There are three irreducible representations of $G/T$ (two $1$-dimensional and one $2$-dimensional).
The character table is
\texttt{charTable[[j,i]]} = character of irrep $j$ evaluated at element $g_i$:
\[
\begin{aligned}
\texttt{charTable}=
\{&
\{1,\ 1,\ 1,\ 1,\ 1,\ 1\},\\
&\{1,\ 1,\ 1,\ -1,\ -1,\ -1\},\\
&\{2,\ -1,\ -1,\ 0,\ 0,\ 0\}
\}.
\end{aligned}
\]

Fixed points of $g_1$: every point is fixed.

Fixed points of $g_2$: three lines parallel to $b_3$ passing through
\[
(u,v)\in\left\{(0,0),\ \left(\tfrac13,\tfrac13\right),\ \left(\tfrac23,\tfrac23\right)\right\},
\qquad w\in\mathbb R.
\]

Fixed points of $g_3$: three lines parallel to $b_3$ passing through
\[
(u,v)\in\left\{(0,0),\ \left(\tfrac13,\tfrac13\right),\ \left(\tfrac23,\tfrac23\right)\right\},
\qquad w\in\mathbb R.
\]

Fixed points of $g_4$: one plane parallel to $\mathrm{span}\{b_1,b_3\}$ given by
\[
v=0,\qquad u\in\mathbb R,\qquad w\in\mathbb R.
\]

Fixed points of $g_5$: one plane parallel to $\mathrm{span}\{b_2,b_3\}$ given by
\[
u=0,\qquad v\in\mathbb R,\qquad w\in\mathbb R.
\]

Fixed points of $g_6$: two planes parallel to $\mathrm{span}\{b_1-b_2,\ b_3\}$ given by
\[
u+v\in\left\{0,\ 1\right\},\qquad w\in\mathbb R.
\]
\subsubsection{Mode Structure}
\begin{itemize}
    \item One $(E,H)$ mode in the $\chi_{1}$ sector with $2E_{12}=E_{11}=E_{22}$ and $2H_{12}=H_{11}=H_{22}$. Two $(E,H)$ modes in the $\chi_{3}$ sector. These two modes together can be recombined to a mode with $E_{11}=-E_{12},\;E_{22}=2E_{12}$ and $H_{11}=-H_{12},\;H_{22}=2H_{12}$, and a mode with $E_{11}=E_{12},\;E_{22}=0$  and $H_{11}=H_{12},\;H_{22}=0$.
\end{itemize}

\section{Hexagonal Layer Groups}
\subsection{\texorpdfstring{$p6$}{p6}}
\subsubsection{Group Structure}
The Bravais lattice of $G=p6$ is generated by
\[
b_1=\begin{pmatrix}a\\0\\0\end{pmatrix},\qquad
b_2=\begin{pmatrix}\tfrac{a}{2}\\ \tfrac{\sqrt3\,a}{2}\\0\end{pmatrix},\qquad
b_3=\hat z=\begin{pmatrix}0\\0\\1\end{pmatrix}.
\]
A convenient set of coset representatives for $G/T$ under basis $b_1,b_2,b_3$ is,
\[
g_1=(R_1|r_1),\quad R_1=
\begin{pmatrix}
1&0&0\\
0&1&0\\
0&0&1
\end{pmatrix},
\quad
\eta_1=
\begin{pmatrix}
1&0\\
0&1
\end{pmatrix},
\quad
\lambda_1=+1,
\quad
r_1=\{0,0,0\},
\]
\[
g_2=(R_2|r_2),\quad R_2=
\begin{pmatrix}
0&-1&0\\
1&1&0\\
0&0&1
\end{pmatrix},
\quad
\eta_2=
\begin{pmatrix}
0&-1\\
1&1
\end{pmatrix},
\quad
\lambda_2=+1,
\quad
r_2=\{0,0,0\},
\]
\[
g_3=(R_3|r_3),\quad R_3=
\begin{pmatrix}
-1&-1&0\\
1&0&0\\
0&0&1
\end{pmatrix},
\quad
\eta_3=
\begin{pmatrix}
-1&-1\\
1&0
\end{pmatrix},
\quad
\lambda_3=+1,
\quad
r_3=\{0,0,0\},
\]
\[
g_4=(R_4|r_4),\quad R_4=
\begin{pmatrix}
-1&0&0\\
0&-1&0\\
0&0&1
\end{pmatrix},
\quad
\eta_4=
\begin{pmatrix}
-1&0\\
0&-1
\end{pmatrix},
\quad
\lambda_4=+1,
\quad
r_4=\{0,0,0\},
\]
\[
g_5=(R_5|r_5),\quad R_5=
\begin{pmatrix}
0&1&0\\
-1&-1&0\\
0&0&1
\end{pmatrix},
\quad
\eta_5=
\begin{pmatrix}
0&1\\
-1&-1
\end{pmatrix},
\quad
\lambda_5=+1,
\quad
r_5=\{0,0,0\},
\]
\[
g_6=(R_6|r_6),\quad R_6=
\begin{pmatrix}
1&1&0\\
-1&0&0\\
0&0&1
\end{pmatrix},
\quad
\eta_6=
\begin{pmatrix}
1&1\\
-1&0
\end{pmatrix},
\quad
\lambda_6=+1,
\quad
r_6=\{0,0,0\}.
\]

With the ordering $(g_1,g_2,g_3,g_4,g_5,g_6)$, we have:
\[
\begin{aligned}
&\bigl\{\mathrm{Tr}(\eta_i)\bigr\}_{i=1}^6=\{2,1,-1,-2,-1,1\},\\
&\bigl\{\det(\eta_i)\bigr\}_{i=1}^6=\{1,1,1,1,1,1\},\\
&\{\lambda_i\}_{i=1}^6=\{+1,+1,+1,+1,+1,+1\}.
\end{aligned}
\]

There are six $1$-dimensional irreducible representations of $G/T$.
Let $\omega=\exp\bigl(\tfrac{2\pi\,\mathrm{i}}{6}\bigr)$.
The character table is
\texttt{charTable[[j,i]]} = character of irrep $j$ evaluated at element $g_i$:
\[
\begin{aligned}
\texttt{charTable}=
\{&
\{1,\ 1,\ 1,\ 1,\ 1,\ 1\},\\
&\{1,\ \omega,\ \omega^{2},\ \omega^{3},\ \omega^{4},\ \omega^{5}\},\\
&\{1,\ \omega^{2},\ \omega^{4},\ 1,\ \omega^{2},\ \omega^{4}\},\\
&\{1,\ \omega^{3},\ 1,\ \omega^{3},\ 1,\ \omega^{3}\},\\
&\{1,\ \omega^{4},\ \omega^{2},\ 1,\ \omega^{4},\ \omega^{2}\},\\
&\{1,\ \omega^{5},\ \omega^{4},\ \omega^{3},\ \omega^{2},\ \omega\}
\}.
\end{aligned}
\]

Fixed points of $g_1$: every point is fixed.

Fixed points of $g_2$: one line parallel to $b_3$ passing through
\[
(u,v)=(0,0),\qquad w\in\mathbb R.
\]

Fixed points of $g_3$: three lines parallel to $b_3$ passing through
\[
(u,v)\in\left\{(0,0),\ \left(\tfrac13,\tfrac13\right),\ \left(\tfrac23,\tfrac23\right)\right\},
\qquad w\in\mathbb R.
\]

Fixed points of $g_4$: four lines parallel to $b_3$ passing through
\[
(u,v)\in\left\{(0,0),\ \left(\tfrac12,0\right),\ \left(0,\tfrac12\right),\ \left(\tfrac12,\tfrac12\right)\right\},
\qquad w\in\mathbb R.
\]

Fixed points of $g_5$: three lines parallel to $b_3$ passing through
\[
(u,v)\in\left\{(0,0),\ \left(\tfrac13,\tfrac13\right),\ \left(\tfrac23,\tfrac23\right)\right\},
\qquad w\in\mathbb R.
\]

Fixed points of $g_6$: one line parallel to $b_3$ passing through
\[
(u,v)=(0,0),\qquad w\in\mathbb R.
\]
\subsubsection{Mode Structure}
\begin{itemize}
    \item One $(E,H)$ mode in the $\chi_{1}$ sector with $2E_{12}=E_{11}=E_{22}$ and $2H_{12}=H_{11}=H_{22}$. One $(E,H)$ mode in the $\chi_{3}$ sector. One $(E,H)$ mode in the $\chi_{5}$ sector. They together can be recombined to a mode with $E_{11}=-E_{12},\;E_{22}=2E_{12}$ and $H_{11}=-H_{12},\;H_{22}=2H_{12}$, and a mode with $E_{11}=E_{12},\;E_{22}=0$  and $H_{11}=H_{12},\;H_{22}=0$.
\end{itemize}

\subsection{\texorpdfstring{$p6mm$}{p6mm}}
\subsubsection{Group Structure}
The Bravais lattice of $G=p6mm$ is generated by
\[
b_1=\begin{pmatrix}a\\0\\0\end{pmatrix},\qquad
b_2=\begin{pmatrix}\tfrac{a}{2}\\ \tfrac{\sqrt3\,a}{2}\\0\end{pmatrix},\qquad
b_3=\hat z=\begin{pmatrix}0\\0\\1\end{pmatrix}.
\]
A convenient set of coset representatives for $G/T$ under basis $b_1,b_2,b_3$ is,
\[
g_1=(R_1|r_1),\quad R_1=
\begin{pmatrix}
1&0&0\\
0&1&0\\
0&0&1
\end{pmatrix},
\quad
\eta_1=
\begin{pmatrix}
1&0\\
0&1
\end{pmatrix},
\quad
\lambda_1=+1,
\quad
r_1=\{0,0,0\},
\]
\[
g_2=(R_2|r_2),\quad R_2=
\begin{pmatrix}
0&-1&0\\
1&1&0\\
0&0&1
\end{pmatrix},
\quad
\eta_2=
\begin{pmatrix}
0&-1\\
1&1
\end{pmatrix},
\quad
\lambda_2=+1,
\quad
r_2=\{0,0,0\},
\]
\[
g_3=(R_3|r_3),\quad R_3=
\begin{pmatrix}
-1&-1&0\\
1&0&0\\
0&0&1
\end{pmatrix},
\quad
\eta_3=
\begin{pmatrix}
-1&-1\\
1&0
\end{pmatrix},
\quad
\lambda_3=+1,
\quad
r_3=\{0,0,0\},
\]
\[
g_4=(R_4|r_4),\quad R_4=
\begin{pmatrix}
-1&0&0\\
0&-1&0\\
0&0&1
\end{pmatrix},
\quad
\eta_4=
\begin{pmatrix}
-1&0\\
0&-1
\end{pmatrix},
\quad
\lambda_4=+1,
\quad
r_4=\{0,0,0\},
\]
\[
g_5=(R_5|r_5),\quad R_5=
\begin{pmatrix}
0&1&0\\
-1&-1&0\\
0&0&1
\end{pmatrix},
\quad
\eta_5=
\begin{pmatrix}
0&1\\
-1&-1
\end{pmatrix},
\quad
\lambda_5=+1,
\quad
r_5=\{0,0,0\},
\]
\[
g_6=(R_6|r_6),\quad R_6=
\begin{pmatrix}
1&1&0\\
-1&0&0\\
0&0&1
\end{pmatrix},
\quad
\eta_6=
\begin{pmatrix}
1&1\\
-1&0
\end{pmatrix},
\quad
\lambda_6=+1,
\quad
r_6=\{0,0,0\},
\]
\[
g_7=(R_7|r_7),\quad R_7=
\begin{pmatrix}
1&1&0\\
0&-1&0\\
0&0&1
\end{pmatrix},
\quad
\eta_7=
\begin{pmatrix}
1&1\\
0&-1
\end{pmatrix},
\quad
\lambda_7=+1,
\quad
r_7=\{0,0,0\},
\]
\[
g_8=(R_8|r_8),\quad R_8=
\begin{pmatrix}
1&0&0\\
-1&-1&0\\
0&0&1
\end{pmatrix},
\quad
\eta_8=
\begin{pmatrix}
1&0\\
-1&-1
\end{pmatrix},
\quad
\lambda_8=+1,
\quad
r_8=\{0,0,0\},
\]
\[
g_9=(R_9|r_9),\quad R_9=
\begin{pmatrix}
0&-1&0\\
-1&0&0\\
0&0&1
\end{pmatrix},
\quad
\eta_9=
\begin{pmatrix}
0&-1\\
-1&0
\end{pmatrix},
\quad
\lambda_9=+1,
\quad
r_9=\{0,0,0\},
\]
\[
g_{10}=(R_{10}|r_{10}),\quad R_{10}=
\begin{pmatrix}
-1&-1&0\\
0&1&0\\
0&0&1
\end{pmatrix},
\quad
\eta_{10}=
\begin{pmatrix}
-1&-1\\
0&1
\end{pmatrix},
\quad
\lambda_{10}=+1,
\quad
r_{10}=\{0,0,0\},
\]
\[
g_{11}=(R_{11}|r_{11}),\quad R_{11}=
\begin{pmatrix}
-1&0&0\\
1&1&0\\
0&0&1
\end{pmatrix},
\quad
\eta_{11}=
\begin{pmatrix}
-1&0\\
1&1
\end{pmatrix},
\quad
\lambda_{11}=+1,
\quad
r_{11}=\{0,0,0\},
\]
\[
g_{12}=(R_{12}|r_{12}),\quad R_{12}=
\begin{pmatrix}
0&1&0\\
1&0&0\\
0&0&1
\end{pmatrix},
\quad
\eta_{12}=
\begin{pmatrix}
0&1\\
1&0
\end{pmatrix},
\quad
\lambda_{12}=+1,
\quad
r_{12}=\{0,0,0\}.
\]

With the ordering $(g_1,g_2,g_3,g_4,g_5,g_6,g_7,g_8,g_9,g_{10},g_{11},g_{12})$, we have:
\[
\begin{aligned}
&\bigl\{\mathrm{Tr}(\eta_i)\bigr\}_{i=1}^{12}=\{2,1,-1,-2,-1,1,0,0,0,0,0,0\},\\
&\bigl\{\det(\eta_i)\bigr\}_{i=1}^{12}=\{1,1,1,1,1,1,-1,-1,-1,-1,-1,-1\},\\
&\{\lambda_i\}_{i=1}^{12}=\{+1,+1,+1,+1,+1,+1,+1,+1,+1,+1,+1,+1\}.
\end{aligned}
\]

There are six irreducible representations of $G/T$ (four $1$-dimensional and two $2$-dimensional).
The character table is
\texttt{charTable[[j,i]]} = character of irrep $j$ evaluated at element $g_i$:
\[
\begin{aligned}
\texttt{charTable}=
\{&
\{1,\ 1,\ 1,\ 1,\ 1,\ 1,\ 1,\ 1,\ 1,\ 1,\ 1,\ 1\},\\
&\{1,\ 1,\ 1,\ 1,\ 1,\ 1,\ -1,\ -1,\ -1,\ -1,\ -1,\ -1\},\\
&\{1,\ -1,\ 1,\ -1,\ 1,\ -1,\ 1,\ -1,\ 1,\ -1,\ 1,\ -1\},\\
&\{1,\ -1,\ 1,\ -1,\ 1,\ -1,\ -1,\ 1,\ -1,\ 1,\ -1,\ 1\},\\
&\{2,\ 1,\ -1,\ -2,\ -1,\ 1,\ 0,\ 0,\ 0,\ 0,\ 0,\ 0\},\\
&\{2,\ -1,\ -1,\ 2,\ -1,\ -1,\ 0,\ 0,\ 0,\ 0,\ 0,\ 0\}
\}.
\end{aligned}
\]

Fixed points of $g_1$: every point is fixed.

Fixed points of $g_2$: one line parallel to $b_3$ passing through
\[
(u,v)=(0,0),\qquad w\in\mathbb R.
\]

Fixed points of $g_3$: three lines parallel to $b_3$ passing through
\[
(u,v)\in\left\{(0,0),\ \left(\tfrac13,\tfrac13\right),\ \left(\tfrac23,\tfrac23\right)\right\},
\qquad w\in\mathbb R.
\]

Fixed points of $g_4$: four lines parallel to $b_3$ passing through
\[
(u,v)\in\left\{(0,0),\ \left(\tfrac12,0\right),\ \left(0,\tfrac12\right),\ \left(\tfrac12,\tfrac12\right)\right\},
\qquad w\in\mathbb R.
\]

Fixed points of $g_5$: three lines parallel to $b_3$ passing through
\[
(u,v)\in\left\{(0,0),\ \left(\tfrac13,\tfrac13\right),\ \left(\tfrac23,\tfrac23\right)\right\},
\qquad w\in\mathbb R.
\]

Fixed points of $g_6$: one line parallel to $b_3$ passing through
\[
(u,v)=(0,0),\qquad w\in\mathbb R.
\]

Fixed points of $g_7$: one plane parallel to $\mathrm{span}\{b_1,b_3\}$ given by
\[
v=0,\qquad u\in\mathbb R,\qquad w\in\mathbb R.
\]

Fixed points of $g_8$: three planes parallel to $\mathrm{span}\{2b_1-b_2,\ b_3\}$ given by
\[
u+2v\in\left\{0,\ 1,\ 2\right\},\qquad w\in\mathbb R.
\]

Fixed points of $g_9$: two planes parallel to $\mathrm{span}\{b_1-b_2,\ b_3\}$ given by
\[
u+v\in\left\{0,\ 1\right\},\qquad w\in\mathbb R.
\]

Fixed points of $g_{10}$: three planes parallel to $\mathrm{span}\{b_1-2b_2,\ b_3\}$ given by
\[
2u+v\in\left\{0,\ 1,\ 2\right\},\qquad w\in\mathbb R.
\]

Fixed points of $g_{11}$: one plane parallel to $\mathrm{span}\{b_2,b_3\}$ given by
\[
u=0,\qquad v\in\mathbb R,\qquad w\in\mathbb R.
\]

Fixed points of $g_{12}$: one plane parallel to $\mathrm{span}\{b_1+b_2,\ b_3\}$ given by
\[
u=v,\qquad w\in\mathbb R.
\]
\subsubsection{Mode Structure}
\begin{itemize}
    \item One $(E,H)$ mode in the $\chi_{1}$ sector with $2E_{12}=E_{11}=E_{22}$ and $2H_{12}=H_{11}=H_{22}$. Two $(E,H)$ modes in the $\chi_{6}$ sector. They together can be recombined to a mode with $E_{11}=-E_{12},\;E_{22}=2E_{12}$, $H_{11}=-H_{12},\;H_{22}=2H_{12}$, and a mode with $E_{11}=E_{12},\;E_{22}=0$, $H_{11}=H_{12},\;H_{22}=0$.
\end{itemize}

\chapter{Experimental validation}
To validate the theory presented in this work we fabricate three physical sheets based on the glide-rotation symmetry discussed in Fig.~3 (a) of the main text, the inversion symmetry in Fig.~4 (c), and broken inversion symmetry in Fig.~4 (e) of the main text. The following sections will discuss both the fabrication and test setup used to generate the shown results.

\section{Fabrication}

\begin{figure}[ht]\label{fig:SIFabrication}
\centering
\includegraphics[width=\textwidth]{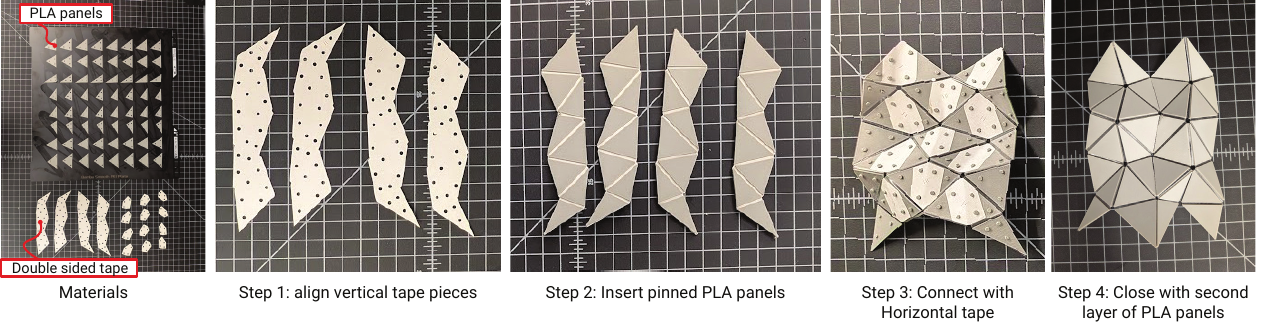}
\caption{Fabrication process of a sample.}
\end{figure}

The physical origami sheets are assembled from 3D-printed PLA panels connected by a continuous layer of tape that forms the flexible hinges between adjacent panels. Adjacent panels are offset by $1.7\mathrm{mm}$ to accommodate the thickness of the PLA panels, providing enough space for two panels to fold flat on top of each other without contact.
It is not possible to lasercut the double-sided tape in one continuous piece, as the sheets presented in this work are not flat-foldable. Therefore, we employ a strategy with two layers of tape which is visualized in Fig.~\ref{fig:SIFabrication}. First, we lasercut continuous strips connecting all the panels in the vertical direction of the sheet and attach the bottom PLA panels. Next, we use a second layer of tape to connect all the horizontal gaps and push the sheet into a 3D-shape. The sheets are finalized by attaching the second layer of PLA panels on top. 
To facilitate assembly and accurate alignment, each bottom panel has a set of alignment pins of diameter $2.6\mathrm{mm}$ and a height of $1 \mathrm{mm}$, while the tape and top panels have matching holes of diameter $2.4\mathrm{mm}$. This creates a press fit that secures the panels and tape together and ensures that all panels stay accurately spaced during assembly.

\section{Experimental setup}

\begin{figure}[ht]\label{fig:SIExpSetup}
\centering
\includegraphics[width=0.5\textwidth]{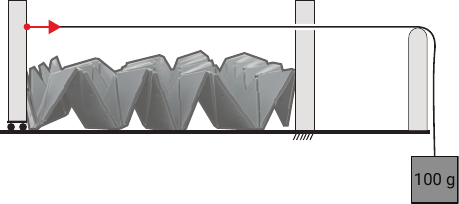}
\caption{Schematic of the experimental setup.}
\end{figure}

To test the fabricated samples we places them between two acrylic plates with a width of 300 mm, height of 50 mm and thickness of 6 mm. One of the acrylic plates is fixed to to the substrate while the other is free to slide, so that the plate can compress the samples from the boundary. For the experiments shown in Fig.~3 (d) of the main text we manually set a fixed displacement between the plates and see if the structure can fold without bending out of plane. Furthermore, to ensure equal loading conditions for the experiments shown in Fig.~3 (f) of the main text, we attach a string to a mass of 100 grams to the sliding plates. Using a pulley the mass pulls the plates together until the origami can no longer fold. At this positions we record the the configuration of the sheets to compare their response. A schematic of this setup is shown in Fig.~\ref{fig:SIExpSetup}.

\printbibliography[heading=sourceline]